\PassOptionsToPackage{unicode}{hyperref}
\PassOptionsToPackage{hyphens}{url}
\documentclass[
  12pt,
]{article}
\usepackage{xcolor}
\usepackage[margin=1in]{geometry}
\usepackage{amsmath,amssymb}
\usepackage{iftex}
\ifPDFTeX
  \usepackage[T1]{fontenc}
  \usepackage[utf8]{inputenc}
  \usepackage{textcomp} 
\else 
  \usepackage{unicode-math} 
  \defaultfontfeatures{Scale=MatchLowercase}
  \defaultfontfeatures[\rmfamily]{Ligatures=TeX,Scale=1}
\fi
\usepackage{lmodern}
\ifPDFTeX\else
\fi
\IfFileExists{upquote.sty}{\usepackage{upquote}}{}
\IfFileExists{microtype.sty}{
  \usepackage[]{microtype}
  \UseMicrotypeSet[protrusion]{basicmath} 
}{}
\usepackage{setspace}
\makeatletter
\@ifundefined{KOMAClassName}{
  \IfFileExists{parskip.sty}{%
    \usepackage{parskip}
  }{
    \setlength{\parindent}{0pt}
    \setlength{\parskip}{6pt plus 2pt minus 1pt}}
}{
  \KOMAoptions{parskip=half}}
\makeatother
\usepackage{longtable,booktabs,array}
\usepackage{calc} 
\usepackage{etoolbox}
\makeatletter
\patchcmd\longtable{\par}{\if@noskipsec\mbox{}\fi\par}{}{}
\makeatother
\IfFileExists{footnotehyper.sty}{\usepackage{footnotehyper}}{\usepackage{footnote}}
\makesavenoteenv{longtable}
\usepackage{graphicx}
\makeatletter
\newsavebox\pandoc@box
\newcommand*\pandocbounded[1]{
  \sbox\pandoc@box{#1}%
  \Gscale@div\@tempa{\textheight}{\dimexpr\ht\pandoc@box+\dp\pandoc@box\relax}%
  \Gscale@div\@tempb{\linewidth}{\wd\pandoc@box}%
  \ifdim\@tempb\p@<\@tempa\p@\let\@tempa\@tempb\fi
  \ifdim\@tempa\p@<\p@\scalebox{\@tempa}{\usebox\pandoc@box}%
  \else\usebox{\pandoc@box}%
  \fi%
}
\def\fps@figure{htbp}
\makeatother
\providecommand{\tightlist}{%
  \setlength{\itemsep}{0pt}\setlength{\parskip}{0pt}}
\usepackage{etoolbox}
\AtBeginEnvironment{longtable}{\footnotesize}
\usepackage{caption}
\usepackage{bookmark}
\IfFileExists{xurl.sty}{\usepackage{xurl}}{} 
\hypersetup{
  pdftitle={What Capital After Labor? Forecasting the Talent ROI Transition in the Human-AI Era},
  pdfauthor={Kwan Soo Shin},
  hidelinks,
  pdfcreator={LaTeX via pandoc}}

\title{What Capital After Labor? Forecasting the Talent ROI Transition
in the Human-AI Era}
\author{Kwan Soo Shin\footnote{Corresponding author: sshin@pmminds.ai.
  PolymathMinds Lab, Asan, Republic of Korea.}}
\date{2026-07-30}

\begin{document}
\maketitle

\setstretch{1.5}
\subsection{Abstract}\label{abstract}

AI augmentation breaks the accounting link between labor time and
productive contribution, yet firms continue to evaluate talent through
time-based overhead bundles. This paper develops a forecasting framework
for the transition from time-based talent accounting to output-based
talent ROI in the human-AI era, organized around five theorems: Theorem
3 (ROI Inversion at τ*) carries the central transition claim, with
overhead non-additivity, augmentation-saved-time pathways,
innovation-premium amplification, and human-AI dyad attribution
uncertainty as the mechanism architecture. Korea's staged 52-hour
workweek mandate provides the early-warning case. In a DART panel of 365
firms (2,281 observations), the SG\&A-to-revenue ratio rose from 18.26
percent (2018) to 20.06 percent (2020) and peaked at 20.10 percent
(2024). Under the revenue-percentile cohort proxy, two-way fixed effects
(+1.56 pp, p = 0.049), pooled event-study estimates (+4.21 pp at t =
+3), and Callaway-Sant'Anna estimates (+4.51 pp at t = +4) converge on a
positive overhead-pressure pattern. Institutional cohort evidence
separates the two readings: under the statutory employee-size cohort the
coefficient is indistinguishable from zero, weighing against a pure
52-hour-law interpretation and supporting the secular regime reading; a
2015-2017 backward extension (224 firms) argues against pre-existing
trends. We read the Korean evidence as, to our knowledge, the first
publicly documented signature of a secular pre-τ overhead-pressure
regime in which time-based accounting still dominates while AI
augmentation raises firm-internal overhead. Output-based firms are
forecast to outperform time-based peers by 1.5-2.0 percentage points in
TFP growth by 2032. The contribution is a forecasting model and planning
tool for AI-augmented talent ROI accounting.

Keywords: artificial intelligence and work, talent ROI transition,
augmented human capital, sociotechnical transition, technological
forecasting, human-AI dyad

\subsection{1. Introduction}\label{introduction}

\subsection{1.1 The Paradox of Talent ROI in the Age of Augmented Human
Capital}\label{the-paradox-of-talent-roi-in-the-age-of-augmented-human-capital}

This paper treats talent ROI accounting as a sociotechnical regime, not
a firm-level bookkeeping convention. In the lineage of Freeman and Perez
(1988) on techno-economic paradigm shifts, Nelson and Winter (1982) on
evolutionary routines, and Geels (2002) on multi-level sociotechnical
transitions, we forecast a regime shift from labor-time accounting to
output-based accounting under AI augmentation. The transition is not
triggered by AI utilization A alone. It emerges when the augmentation
function φ(A, C), convergence capacity C, and the seven overhead
components K = \{wage, insurance, space, management, training,
communication, motivation\} cease to fit the time-based unit of account.
The Korean DART panel supplies the empirical early-warning signature of
this Pre-τ overhead-pressure regime, before the critical threshold τ* is
crossed.

In a panel of 365 Korean listed firms observed from 2018 through 2024,
the ratio of selling, general, and administrative expense to revenue
rose by 1.80 percentage points during the staged implementation of the
52-hour workweek mandate and then climbed again to a new peak of 20.10
percent in 2024, coinciding with broad enterprise diffusion of
generative AI. Under the revenue-percentile cohort proxy, two-way fixed
effects yield a positive overhead-pressure association of +1.56
percentage points (clustered standard error 0.79, p = 0.049), while a
pooled event-study estimate ramps to +4.21 percentage points three years
post-treatment (p = 0.001). We do not interpret these estimates as a
point causal magnitude of the 52-hour law. Rather, they motivate the
forecasting problem this paper addresses: why does AI-augmented labor
coexist with rising firm-internal talent overhead, and at what
utilization threshold does the existing time-based accounting model
become a drag on firm-level total factor productivity?

For two decades, the most innovative firms have been running
uncontrolled experiments on talent return-on-investment (ROI), defined
as the marginal output a firm captures from its talent overhead bundle.
Google deployed the 80/20 rule, allocating one workday per week to
employee-directed exploration, on the bet that bounded slack would
generate breakthrough products. 3M had operated a comparable 15-percent
rule since 1948. ROWE (Results-Only Work Environment) was piloted at
Best Buy from 2003 to 2013 (Ressler and Thompson, 2008) on the
conjecture that severing compensation from time would unlock latent
productivity. The United Kingdom's 2022 four-day workweek trial enrolled
61 firms (Lewis et al., 2023) on the hypothesis that compressing time
would expand output. Microsoft and Salesforce have, since 2024, deployed
generative AI assistants on the bet that augmentation would amplify
human capital rather than displace it.

Every one of these experiments is an empirical wager without a unifying
theoretical foundation. They share an implicit but never-articulated
thesis: the relationship between time, talent, and firm-level value
creation has structurally shifted, and the contemporary firm's overhead
accounting model is no longer fit for purpose. Yet the theoretical
infrastructure that would justify such experiments, predict their
outcomes, or generalize their success conditions does not exist. Each
experiment has been defended by case-specific intuition, sociotechnical
anecdote, or selective performance metrics.

This paper proposes that the absence of theory is not accidental. It
reflects a deeper structural fact: the foundational assumption embedded
in nearly every contemporary economic, organizational, and accounting
framework, that labor L can be treated as a homogeneous function of
time, has been violated by the emergence of human-AI dyadic production.
The dyad is now visible in everyday firm operations: a software engineer
drafting code with Copilot suggestions, an analyst composing a report
through ChatGPT iteration, a marketing team prompting image-generation
models for campaign assets. In each case, the hour billed to the firm
contains a mixture of human cognitive labor and AI computational labor
that the time-based accounting unit cannot disaggregate. Until this
unit-of-account assumption is reconstructed, firms will continue to
evaluate AI-augmented work through accounting categories designed for a
pre-augmentation labor regime.

\subsection{1.2 Theoretical Gap}\label{theoretical-gap}

Shin (2026a) formalized augmented human capital as Ĥ = H · {[}1 + φ(A,
C){]} at the macroeconomic level, where H is baseline human capital
(skill stock built through education and experience), A is AI
utilization intensity (the share of work tasks supported by AI tools), C
is convergence capacity (a firm's institutional capacity to integrate AI
tool outputs with existing human knowledge structures), and φ is the
augmentation function (the interaction premium between AI access and
convergence capacity). The ICH framework provides the production
function backbone for the post-labor-time economy. It documents, using
20 OECD economies, that the AI by C interaction explains 86 percent of
variance in firm-level total factor productivity, compared with 31
percent for AI utilization alone. It identifies Korea as a deviant case
in which high H, high A, and low C yield bottom-quartile OECD TFP.

A critical gap remains across this literature. No existing framework,
macroeconomic or otherwise, specifies how the seven components of
firm-internal talent overhead are reconstructed when L = labor time no
longer holds. These components are time-based wage; social insurance
contributions; office space and facilities; managerial supervision cost;
training and development investment; communication infrastructure
(email, video conferencing, collaboration software); and motivation
maintenance (engagement programs, retention compensation, performance
bonuses). Among the most directly adjacent contributions, Coasean
Singularity arguments at the NBER (Shahidi et al., 2025), the prediction
that AI-enabled coordination collapse will reduce firm boundaries toward
zero, the augmented human capital vector decomposition of Espinal Maya
(2026), Farach's (2026) coordination-compressing capital model, and the
Brynjolfsson-led empirical estimates of AI-induced productivity
(Brynjolfsson, Rock, and Syverson, 2021) each approach but do not
resolve the firm-internal overhead question. Espinal Maya (2026)
explicitly acknowledges that firm-level overhead ROI accounting remains
for future research.

The gap is not scholarly oversight. The macroeconomic framework already
exists; the firm-level apparatus that translates it into managerial
choices about overhead bundles and evaluation regimes has not yet been
built. This firm-level forecasting task also departs from the
occupational-displacement tradition that Frey and Osborne (2017) made
canonical: that tradition forecasts which jobs are susceptible to
automation, whereas the present paper forecasts when the accounting
regime through which firms value talent inverts from time-based to
output-based. The unit of forecast is the talent-accounting regime, not
the occupation.

\subsection{1.3 Five Theoretical
Contributions}\label{five-theoretical-contributions}

The firm-level reconstruction requires five theorems, each closing a
distinct hole in the inherited accounting logic. The five theorems form
a hierarchy. Theorem 3 (ROI Inversion at τ*) carries the central
transition claim, which the Korean DART panel anchors descriptively as a
pre-τ overhead-pressure signature. Theorems 1, 2, 4, and 5 are the
mechanism architecture explaining why it bends: overhead non-additivity,
four pathways of augmentation-saved time, innovation-premium
amplification, and agency-cost asymmetry in the human-AI dyad. Each
theorem is independently falsifiable on its own evidentiary register,
but the joint claim of the framework rests on Theorem 3. This hierarchy
is summarized in Table 2 of Section 3.6.

Theorem 1, Overhead Decomposition. The seven contemporary components of
firm-internal talent overhead do not add up. Under AI augmentation, the
cross-partial derivative of return on investment with respect to any two
component costs becomes non-zero for at least one component pair.
Standard cost accounting, which implicitly assumes additivity, therefore
misrepresents the marginal return on overhead investment precisely in
the conditions where managers most need accurate guidance.

Theorem 2, Slack-Augmentation Synergy. AI-augmentation-saved time does
not flow into a single channel. It distributes across four mutually
exclusive pathways: work intensification, hidden leisure,
overemployment, and creative-slack reinvestment. Only the fourth pathway
generates firm-level innovative ROI, and its share rises monotonically
with two firm-level institutional variables, output-orientation in
performance evaluation and the autonomy granted to the employee.

Theorem 3, ROI Inversion. As AI utilization intensity rises, the talent
ROI curve under time-based accounting falls and the curve under
output-based accounting rises. They intersect at a single point τ*,
derivable as a function of three firm-level variables
(output-orientation, autonomy, convergence capacity). Below τ*,
time-based accounting remains optimal; above τ*, it becomes a strict
drag on firm-level total factor productivity. This threshold provides
the first formally derivable answer to a long-standing managerial
question, when should a firm transition between accounting regimes?

Theorem 4, Innovative ROI Premium. 80/20-style slack returns are
amplified by a factor k \textgreater{} 1 under high firm-level
convergence capacity. The amplification factor is itself a monotone
increasing function of the augmentable cognitive share H\^{}A defined in
Espinal Maya's (2026) human-capital decomposition. The theorem provides
the first theoretical anchor for the Google 80/20 and 3M 15-percent
experiments and identifies the institutional conditions under which
similar slack policies will yield comparable returns or fail to.

Theorem 5, Information Asymmetry. The multitask principal-agent model of
Holmstrom and Milgrom (1991) extends naturally to the human-AI dyad once
attribution uncertainty between human and AI contribution is added as a
third measurement difficulty. The resulting agency cost is partially
offset under output-based evaluation, because outputs exist independent
of contribution share. Time-based evaluation, by contrast, magnifies the
agency cost in proportion to AI utilization intensity, producing an
institutional drag that previous multitask analyses have not formalized.

\subsection{1.4 Methodological
Positioning}\label{methodological-positioning}

Our empirical strategy follows the case study methodology of Yin (1994),
Eisenhardt (1989), and Gerring (2007), which establishes three case
selection roles: critical case, polar opposite, and cultural twin. The
primary empirical anchor is South Korea, a high-leverage critical case
combining unusually strict labor-time regulation (a 52-hour weekly cap
applied since 2018 in staged form by industry and firm size), high AI
adoption (28 percent), low convergence capacity, and bottom-quartile
OECD TFP (as documented in Shin, 2026a, §5). Korea serves the critical
case role: a prototype for the dynamics that other OECD economies will
increasingly face as labor-time pressure and AI adoption converge.

We use Korea's staged 52-hour implementation as an externally timed
institutional anchor for labor-time compression and overhead pressure.
The current implementation approximates statutory treatment timing with
revenue-percentile cohort proxies; therefore, the estimates are
interpreted as directional evidence for a pre-τ overhead-pressure
signature rather than as a fully observed statutory-cohort IV design or
a point causal magnitude. The identifying assumption is treated as an
assumption rather than a settled fact: the policy should affect
SG\&A/Revenue primarily through labor-time compliance pressure, while AI
utilization enters as a separate macro-year and sectoral diffusion
process. The present analysis therefore estimates the overhead-pressure
signature associated with the staged labor-time context; firm-level TFP
separation remains a falsifiable forecast for follow-up panels.

A firm-level panel constructed from DART corporate disclosure data
(Korea Financial Supervisory Service) provides the primary observational
unit. The KOSPI plus KOSDAQ Top 500 panel covers Korean listed firms
over 2018-2024 with fuzzy account-name matching and
consolidated/non-consolidated dual fetching, yielding 2,281 firm-year
observations across 365 unique firms with positive revenue records. This
expanded panel supersedes the initial KOSPI Top 300 baseline (115 firms,
567 observations) with a 3.2-fold expansion in firm coverage and a
4.0-fold expansion in observations. A sector- and size-stratified panel
is a specified future identification path. Section 4 reports the
expanded-sample evidence consistent with Theorem 3.

Denmark serves the polar opposite role: identical AI adoption rate (28
percent) but opposite values on every other dimension (Nordic
flexicurity, high C, top TFP at 1.82 percent, egalitarian governance).
Japan serves the cultural twin role: identical Confucian hierarchical
organization and presence-based evaluation tradition, but different AI
adoption (18 percent) and different labor regulation (karoshi culture
without strict 52-hour enforcement). Denmark and Japan parameters are
drawn from the ICH macro panel (Shin, 2026a) rather than independent
firm-level data collection; full firm-level Eurostat, Statistics
Denmark, METI, and RIETI microdata analysis is reserved for future work.
The brief comparison thus isolates convergence capacity (Denmark axis)
and regulation (Japan axis) directionally while holding the respective
confounds constant.

Korea is treated as a prototype rather than as an isolated case: a
single economy at the convergence of strict labor-time regulation and
high AI adoption supplies the structural pattern that other OECD
economies may face if their own regulatory and adoption trajectories
converge over the next decade. The case-selection logic rests on the
extremity of Korea's values on the theoretical variables, not on
analogical warrant from prior single-country political-economy
treatments.

\subsection{1.5 Paper Structure and
Forecasts}\label{paper-structure-and-forecasts}

Section 2 develops the theoretical framework, connecting augmented human
capital (Shin, 2026a) to firm-internal overhead accounting and
integrating multitask agency theory. Section 3 presents the five
theorems with formal derivation sketches, falsifiability conditions, and
boundary specifications. Section 4 provides the Korean cohort-proxy
diagnostics and robustness evidence, with staggered DiD diagnostics
under the revenue-percentile cohort proxy, robustness checks across five
axes, and the Denmark and Japan polar comparisons. Section 4.3 reports
the expanded-sample evidence: the SG\&A/Revenue ratio follows an N-curve
trajectory across 2018-2024, with a 1.80 percentage point rise from 2018
(18.26 percent) to 2020 (20.06 percent) concurrent with the 52-hour
staged implementation, mild correction during 2021-2022 (18.73 percent /
18.28 percent), and renewed ascent during 2023-2024 (19.52 percent /
20.10 percent) with 2024 emerging as the new peak concurrent with AI
augmentation acceleration. The trajectory provides directional
early-warning evidence for the time-based-regime predictions of Theorem
3, not a direct estimate of τ*. Section 5 discusses theoretical and
policy implications. Section 6 closes with four falsifiable forecasts.

Among these forecasts, the central prediction is that firms adopting
output-based talent ROI accounting will exceed time-based comparators in
firm-level TFP growth by 1.5 to 2.0 percentage points by 2032 in the
Korean and OECD panel data. This forecast is directly testable.
Falsification would refute the central thesis of the framework. The
Korean evidence is interpreted not as a direct estimate of τ* or as a
point causal estimate of the 52-hour law, but as, to our knowledge, the
first publicly documented signature of the pre-τ overhead-pressure
regime, the phase that precedes the regime crossing the forecast
predicts.

\subsection{2. Theoretical Framework}\label{theoretical-framework}

\subsection{2.0 Where the Conversation
Stopped}\label{where-the-conversation-stopped}

The question this paper asks (what capital remains after labor time
ceases to be the unit of account) is not new in its motivation. It is,
in fact, the question that Marx put to the world in 1867, when he wrote
that ``capital is dead labor, which, vampire-like, lives only by sucking
living labor, and lives the more, the more labor it sucks.'' For Marx,
the answer was straightforward: capital was accumulated labor time, and
the firm was the institution that organized labor time into commodities.
Every generation since has either extended Marx's framework, refined it,
or stretched its assumption that labor input is a homogeneous function
of time.

The conversation has continued for one hundred and fifty-eight years.
Six generations of thinkers have built on Marx's foundation while
progressively relaxing his assumptions: Solow added the technology
residual, Becker and Mincer added human capital, Coase and Williamson
moved the unit of analysis inside the firm, Acemoglu and Restrepo added
tasks, and Shin (2026a) added augmented human capital at the
macroeconomic level. Each generation made the framework richer; each
generation also left a particular question unanswered. The question this
paper enters is the question that the most recent generation explicitly
left for future work.

The conversation can be read as a funnel. It begins with Marx's broad
claim about the labor theory of value, then narrows progressively
through each generation's contribution and unresolved gap, and arrives
at a specific theoretical opening that exists today: how does augmented
human capital reconstruct the firm-internal overhead bundle? Figure 1
maps this funnel across two dimensions, theoretical scope (macroeconomic
aggregate to firm-internal micro) and labor representation
(time-homogeneous to AI-augmented heterogeneous), with each generation
occupying its position in the field and the present paper marking the
location where the conversation has not yet reached.

The genealogy that follows is not exhaustive scholarship but selective
dialogue with the thinkers whose questions structure this paper. We take
each in turn, not to summarize them, but to specify precisely where each
one stopped and why their stopping point opens onto ours.

\subsection{2.1 Marx and the Labor Theory of
Value}\label{marx-and-the-labor-theory-of-value}

Marx's framework treated the firm as a production function in which
capital and labor combined to yield commodities:

\[Y = f(K, L)\]

The function was indexed to a specific historical formation, industrial
capitalism, in which labor L was identifiable in units of time. The
factory worker's day, divided into necessary labor and surplus labor,
was the unit through which value was both produced and extracted. Marx's
question was distributional rather than constructive: how is the surplus
labor time, the difference between what the worker is paid and what the
worker produces, captured by the owners of capital?

More than a century and a half later, the question of who captures
surplus remains alive in the literature on labor's share of national
income (Karabarbounis \& Neiman, 2014) and on platform capitalism's
reallocation of returns (Srnicek, 2017). But Marx's assumption that L is
homogeneous in time has eroded twice. The first erosion came from within
neoclassical economics, which began to ask whether all hours of labor
produce the same value. The second, the present erosion, comes from
outside both Marxist and neoclassical traditions: artificial
intelligence augments the productive content of an hour of labor in ways
that the time-homogeneity assumption cannot accommodate. Marx's
framework remains the historical anchor against which every subsequent
generation has measured its departure. The departure this paper makes is
not from Marx's distributional question, which remains substantively
important, but from his unit-of-account assumption, which AI
augmentation has rendered incomplete.

\subsection{2.2 Six Generations from Production Functions to Firm-Level
Talent
Accounting}\label{six-generations-from-production-functions-to-firm-level-talent-accounting}

\emph{Figure 1. Two Genealogies Converging on the Firm-Level
Regime-Transition Gap. The figure displays the production-function and
forecasting streams; firm-boundary and agency theory enters as a
firm-internal substream developed in Sections 2.2 and 2.5.}

\begin{center}

\includegraphics[width=1\linewidth,height=\textheight,keepaspectratio]{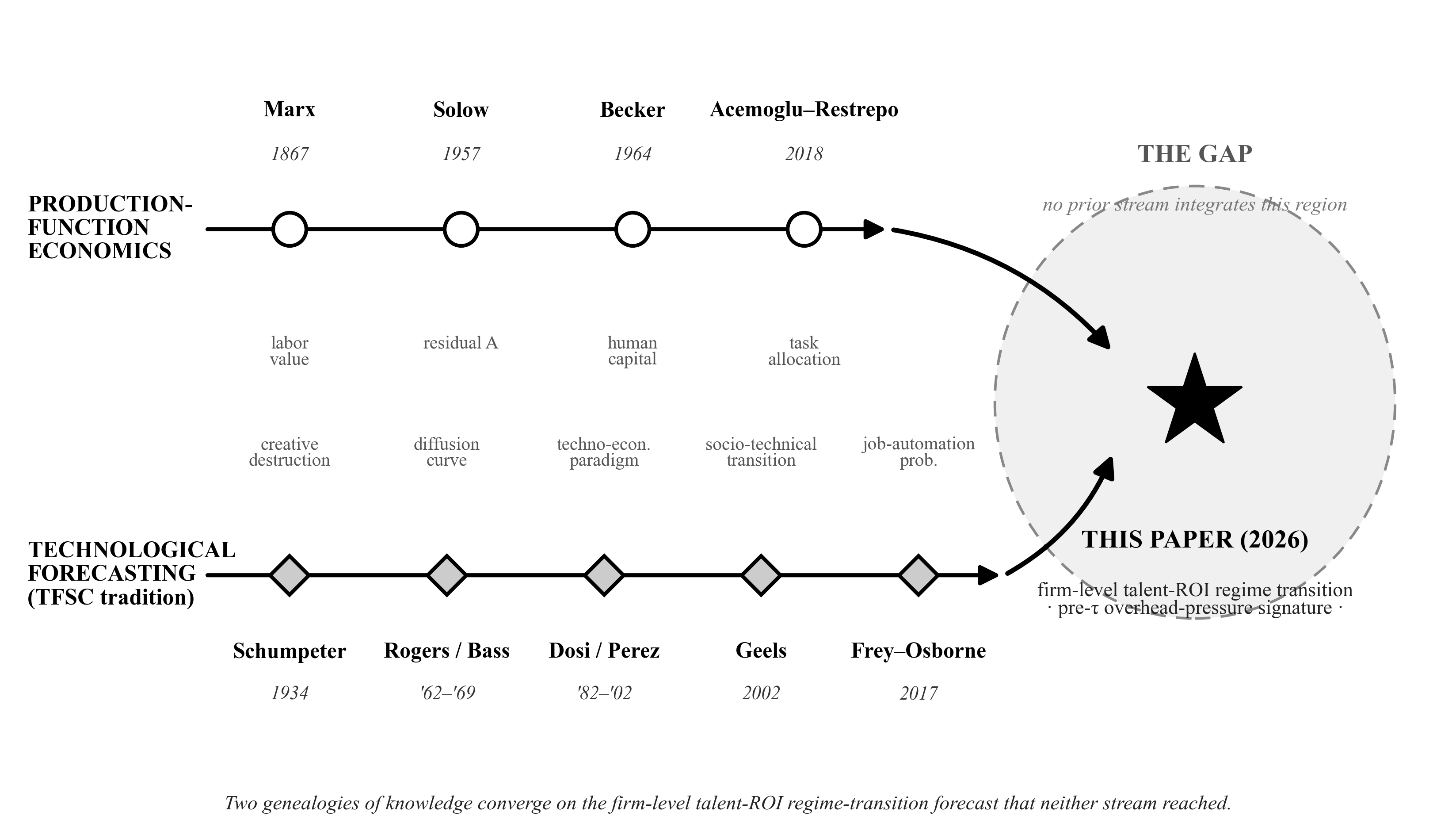}

\end{center}

In Figure 1, the two genealogies of knowledge this paper integrates are
traced as parallel streams flowing toward a single open question. The
upper stream is the production-function economics tradition (Marx,
Solow, Becker, Acemoglu and Restrepo), developed across Sections 2.1 and
2.2; the lower stream is the technological-forecasting tradition
(Schumpeter, Rogers and Bass, Dosi and Perez, Geels, Frey and Osborne),
developed in Section 2.6. Both streams flow toward, but neither reaches,
the firm-level talent-ROI regime-transition forecast this paper
supplies, the empty confluence marked as the gap.

\subsubsection{Generations I through V: From Time-Homogeneous Labor to
Task
Allocation}\label{generations-i-through-v-from-time-homogeneous-labor-to-task-allocation}

The first five generations each loosened one assumption of
time-homogeneous labor while leaving others intact. Generation I (Marx,
1867; \(Y = f(K, L)\)) occupies the lower-left quadrant:
time-homogeneous labor, macroeconomic class structure, distributional
question, labor theory of value answer. Generation II (Solow, 1957;
\(Y = A \cdot K^{\alpha} \cdot L^{1-\alpha}\)) moved upward along the Y
axis by introducing the technology constant A. The Solow residual, which
Solow himself called ``a measure of our ignorance,'' became the
placeholder for everything subsequent generations would specify
explicitly: human capital (Becker, 1964; Mincer, 1974), endogenous
technical change (Romer, 1990; Lucas, 1988), task allocation (Acemoglu
and Restrepo, 2018, 2022), and augmented human capital (Shin, 2026a).
The firm-internal overhead reconstruction proposed here continues the
same trajectory.

Generation III (Becker, 1964; Mincer, 1974; \(Y = f(K, H, A)\))
introduced human capital H as a function of education and experience,
formalized by the Mincer earnings function:

\[\log(w) = \alpha_0 + \alpha_1 \cdot \text{years of schooling} + \alpha_2 \cdot \text{years of experience} + \alpha_3 \cdot \text{years of experience}^2\]

This was the first formal break from time-homogeneity, but Becker and
Mincer treated human capital as a scalar quantity. Espinal Maya (2026)
recently observed that ``this scalar treatment becomes acute in the era
of generative artificial intelligence: the framework treats cognitive
capacity as a scalar (years of education and potential experience) when
in reality it is a vector whose internal composition determines whether
a worker is amplified, displaced, or unaffected by AI.'' The present
paper inherits Espinal Maya's vector decomposition (physical-manual,
cognitive-routine, cognitive-augmentable) and uses it at the firm level,
where the proportion of workforce in cognitive-augmentable occupations
determines the amplification factor of slack-based innovation (Theorem
4).

Generation IV (Coase, 1937; Williamson, 1981; Hart and Holmstrom, 1986;
Holmstrom and Milgrom, 1991) moved horizontally along the X axis from
macroeconomic aggregate toward firm-internal micro analysis. The lineage
specifies how labor is contracted (Coase), governed (Williamson),
residual-control-allocated (Hart-Holmstrom), and incentivized
(Holmstrom-Milgrom), but retains a time-homogeneous assumption about the
underlying labor input. The information asymmetry that Holmstrom and
Milgrom formalized at the multitask level extends naturally to the
human-AI dyad, where the third ``task'' of AI processing introduces an
attribution uncertainty that the original framework did not anticipate.
Theorem 5 extends Holmstrom-Milgrom into precisely this territory.

Generation V (Acemoglu and Restrepo, 2018, 2022; \(Y = \int y(i)\, di\))
relocated production function theory to the upper-right region. The firm
allocates a continuum of tasks to either humans or machines, with
allocation determined by relative cost. The framework accounts for labor
market displacement and labor share decline but stops at the boundary of
cognitive mediation: it predicts which tasks will be automated but not
whether the resulting allocation produces breakthrough innovation,
routine maintenance, or systematic error. As Espinal Maya (2026)
observed, the framework operates ``above the cognitive level.'' This
paper inherits the task-based framework as the institutional context
within which augmented human capital operates but extends it through the
convergence capacity construct (Shin, 2026a) that mediates whether the
task allocation translates into productive output.

\subsubsection{Generation VI: Shin (2026a), Ĥ = H · {[}1 + φ(A,
C){]}}\label{generation-vi-shin-2026a-ux125-h-1-ux3c6a-c}

Shin's (2026a) augmented human capital framework moves Generation V
further up the Y axis by introducing the augmentation function φ(A, C)
that mediates AI's effect on productive output. The function depends on
AI utilization A and convergence capacity C, where C is a
four-dimensional cognitive construct comprising embodied understanding,
metacognitive calibration, temporal integration, and integrative
thinking. Convergence capacity is the missing variable that the
Acemoglu-Restrepo framework treats as exogenous but that the ICH
framework treats as the central mediator.

The ICH framework demonstrated, using 20 OECD economies, that the AI by
C interaction explains 86 percent of cross-country variance in
firm-level total factor productivity, compared with 31 percent for AI
utilization alone. It identified Korea as the deviant case in which high
H, high A, and low C yield bottom-tier TFP. The framework's contribution
is the formal specification of what Solow could only label as residual
A: TFP is endogenous to the augmentation function, and the augmentation
function is endogenous to the cognitive mediator.

Generation VI extends Generation V to incorporate AI augmentation. But
Shin (2026a) operates at the macroeconomic level. The framework
specifies what happens at the country level when AI utilization rises
but convergence capacity does not. It does not specify what happens at
the firm level when seven distinct overhead components (time-based wage,
social insurance, office space, management cost, training, communication
infrastructure, motivation maintenance) must be reconstructed because
the underlying labor input is no longer time-homogeneous. The
macroeconomic instantiation exists. The firm-level instantiation does
not.

\subsubsection{The Present Paper: A Firm-Level Forecasting Framework for
the Talent-Accounting
Transition}\label{the-present-paper-a-firm-level-forecasting-framework-for-the-talent-accounting-transition}

The location of the present paper on Figure 1 is the upper-right
quadrant, where AI-augmented heterogeneous labor representation meets
firm-internal micro theoretical scope. This position is not empty in the
sense that no adjacent work exists; rather, it remains under-specified.
The fourteen most adjacent papers in the literature (Shahidi et al.,
2025, on the Coasean Singularity; Farach, 2026, on
coordination-compressing capital; Espinal Maya, 2026, on augmented human
capital vector decomposition; the various Brynjolfsson-led empirical
estimates of AI productivity gains (Brynjolfsson, Rock, and Syverson,
2021); the Eccles, 2025, framework on hybrid intelligence teams; the
Vaccaro et al., 2024, meta-analysis on human-AI combinations) all circle
this position. Espinal Maya (2026) explicitly acknowledged in his stated
limitations that firm-level overhead ROI accounting remained for future
research.

The present paper builds the firm-level framework that occupies this
position: an independently evaluable forecasting apparatus for the
talent-accounting transition, with its own theorem architecture,
empirical anchor, and falsification conditions. The five theorems
developed in Section 3 specify how the seven overhead components are
reconstructed when L = labor time no longer holds. The empirical anchor
in Section 4 draws on South Korea's staged 52-hour working-time
regulation as an institutional timing anchor, providing directional
pre-τ overhead-pressure evidence against which the framework's forecasts
can be evaluated.

The conversation can continue from here. The present framework is not
the final word; it is the explicit specification of one piece of what
the six-generation conversation has been progressively narrowing toward.
Future research directions extend the conversation further, and the
falsifiable forecasts (Section 6) specify the empirical conditions under
which the present framework will be revised or replaced. We enter the
conversation not to close it but to put a precise question and a precise
answer on the table where the previous answer left off.

\subsection{2.3 Seven Overhead Components: The Architecture of
Firm-Internal Talent
Cost}\label{seven-overhead-components-the-architecture-of-firm-internal-talent-cost}

\begin{longtable}[]{@{}
  >{\raggedright\arraybackslash}p{(\linewidth - 6\tabcolsep) * \real{0.0441}}
  >{\raggedright\arraybackslash}p{(\linewidth - 6\tabcolsep) * \real{0.1618}}
  >{\raggedright\arraybackslash}p{(\linewidth - 6\tabcolsep) * \real{0.3676}}
  >{\raggedright\arraybackslash}p{(\linewidth - 6\tabcolsep) * \real{0.4265}}@{}}
\caption{Table 1. Seven Overhead Components and the Pressure Each Faces
Under Augmentation}\tabularnewline
\toprule\noalign{}
\begin{minipage}[b]{\linewidth}\raggedright
\#
\end{minipage} & \begin{minipage}[b]{\linewidth}\raggedright
Component
\end{minipage} & \begin{minipage}[b]{\linewidth}\raggedright
Historical Justification
\end{minipage} & \begin{minipage}[b]{\linewidth}\raggedright
Pressure Under Augmentation
\end{minipage} \\
\midrule\noalign{}
\endfirsthead
\toprule\noalign{}
\begin{minipage}[b]{\linewidth}\raggedright
\#
\end{minipage} & \begin{minipage}[b]{\linewidth}\raggedright
Component
\end{minipage} & \begin{minipage}[b]{\linewidth}\raggedright
Historical Justification
\end{minipage} & \begin{minipage}[b]{\linewidth}\raggedright
Pressure Under Augmentation
\end{minipage} \\
\midrule\noalign{}
\endhead
\bottomrule\noalign{}
\endlastfoot
1 & Time-based wage & Wage as function of hours (Industrial Revolution;
Mincer, 1974) & Hours no longer correspond to productive units \\
2 & Social insurance and benefits & Calculated as percentage of
time-based wages (Korea's four major insurances) & Inherits
arbitrariness of the wage base; 52-hour regime intensifies dependency \\
3 & Office space & Workers occupy fixed locations for fixed time blocks;
8-15\% of operating expenses & Asynchronous/remote modes raise
allocation cost relative to contribution \\
4 & Management and evaluation & Direct supervision and HR overhead;
12-20\% in knowledge-intensive industries & Productivity paranoia: 87\%
of employees report being productive while 12\% of leaders have full
confidence in team productivity (Microsoft, 2022), a 75-point gap \\
5 & Training and development & Human capital accumulation through
structured programs (Becker; Mincer) & Locus of skill shifts from
accumulated experience toward convergence capacity \\
6 & Communication infrastructure & Email, video, collaboration platforms
& Knowledge workers spend 57\% of hours in communication tools
(Microsoft, 2025) \\
7 & Motivation maintenance & Autonomy, competence, relatedness (Deci and
Ryan, 1985; Ryan and Deci, 2017) & Time-based structures suppress
autonomy by definition \\
\end{longtable}

In Table 1, the seven distinct overhead components are summarised with
their historical justification under the labor-time assumption and the
structural pressure each faces when AI augmentation breaks
time-homogeneity. Contemporary firms accumulate talent overhead through
these seven components, each requiring reconstruction under augmented
human capital.

These seven components are not independent. Their marginal contributions
to firm-level ROI become non-additive under AI augmentation: a change in
one component induces effects on others that the additive accounting
model cannot capture. This non-additivity is the substantive content of
Theorem 1, developed in Section 3. The claim that inherited
cost-accounting categories misrepresent a transformed production regime
has a management-accounting lineage of its own: Johnson and Kaplan
(1987) documented how cost systems designed for an earlier production
environment progressively lost decision relevance, and Kaplan and Norton
(1992) redirected performance measurement from input proxies toward
outcome measures. The present framework extends that lineage to the
regime in which the obsolete input proxy is labor time itself.

\subsection{2.4 Convergence Capacity at the Firm
Level}\label{convergence-capacity-at-the-firm-level}

The ICH framework defines convergence capacity C as a four-dimensional
cognitive construct. The present paper extends C with a fifth dimension
drawn from Shin (2026b). The fifth dimension, sovereign override
capacity (C5), captures the capacity to detect and refuse AI outputs in
which high-confidence generation masks factual or contextual error.

At the firm level, the five dimensions operate as follows. C1 (embodied
understanding) operates through experienced employees' capacity to
ground AI outputs in sensorimotor and contextual reality. C2
(metacognitive calibration) operates through employees' capacity to
judge when AI outputs warrant independent verification. C3 (temporal
integration) operates through senior employees' capacity to
contextualize AI recommendations against accumulated organizational
memory. C4 (integrative thinking) operates through employees' capacity
to combine AI outputs across domains for novel synthesis. C5 (sovereign
override) operates through employees' capacity to refuse AI outputs when
verification fails.

Firm-level convergence capacity is not the simple sum or average of
individual employee C scores. It depends on three additional factors:
the firm's evaluation system (whether output-orientation rewards
employees for exercising C), the firm's authority structure (whether
employees have legitimate refusal authority over AI outputs), and the
firm's cultural register (whether the dominant cultural script permits
exercise of metacognitive and override capacities).

\subsection{2.5 Multitask Agency Theory in the Human-AI
Dyad}\label{multitask-agency-theory-in-the-human-ai-dyad}

The final element of our theoretical framework is the extension of
multitask agency theory to the human-AI dyad. Holmstrom and Milgrom
(1991), in their foundational treatment of multitask principal-agent
analyses, demonstrated that when an agent allocates effort across
multiple tasks with different measurement difficulties, the principal's
optimal contract weights the easily-measured tasks more heavily than
their substantive importance warrants.

In the human-AI dyad, the multitask structure intensifies because the
dyad introduces a third actor whose contribution is difficult to
attribute. We denote this attribution uncertainty γ ∈ {[}0, 1{]}, with γ
= 0 indicating perfect attribution clarity and γ = 1 indicating complete
attribution opacity. The agency cost in the human-AI dyad is a function
of γ, with higher attribution uncertainty generating higher agency cost
through three channels: moral hazard, adverse selection, and distorted
effort allocation. Holmstrom and Milgrom's (1991) central result implies
that the human-AI dyad agency cost is partially offset under
output-based evaluation regimes. This offset is the substantive content
of Theorem 5.

\subsection{2.6 Positioning in the Technological Forecasting
Tradition}\label{positioning-in-the-technological-forecasting-tradition}

The five theorems and the Korean case that follow are, in the end, a
forecast: a claim about a regime transition that has not yet completed.
That claim must answer to a conversation the technology-and-society
literature has hosted for half a century, the conversation about how
technological revolutions reshape the institutions built around the
previous technology. Section 2.2 traced where the production-function
conversation stopped; this section traces where the forecasting
conversation stands, and locates the present framework within it (Figure
1, lower stream).

The conversation opens with Schumpeter (1934, 1942), for whom innovation
was not a disturbance to equilibrium but the engine that destroys one
equilibrium and installs another. Schumpeter and the Kondratiev
long-wave tradition gave the field its founding question: not whether
technology changes the economy, but on what rhythm, and through what
mechanism, the old institutions give way to the new.

The mechanism received its first formal answer in the diffusion
tradition. Rogers (2003) showed that innovations propagate through a
social system along an S-shaped curve governed by perceived attributes
and communication structure; Bass (1969) rendered that curve estimable,
separating innovation from imitation in the adoption hazard. Diffusion
theory taught the field to forecast the timing of adoption. The Korean
AI-utilization trajectory documented in Section 4, rising from roughly 8
percent in 2018 to 28 percent in 2024, is a diffusion curve of exactly
this kind; but the framework forecasts what diffusion theory leaves
unaddressed, namely the reorganization of the talent-accounting regime
that rising utilization renders obsolete.

Dosi (1982) and Nelson and Winter (1982) supplied the next layer:
technologies advance along paradigm-bound trajectories, and economic
change is an evolutionary process in which routines, not optimizing
choices, carry technical knowledge forward. Freeman and Perez (1988) and
Perez (2002, 2010) then closed the layer this paper most directly
inherits. In their account, each technological revolution installs a new
techno-economic paradigm, and the decisive feature of every transition
is the mismatch between fast-moving technology and slow-moving
institutions: the socio-institutional framework, built for the previous
paradigm, lags the new one until a structural crisis forces its
reconstruction. The present framework is a firm-level instantiation of
precisely this mismatch. Time-based talent accounting is the
institutional residue of the industrial paradigm; the pre-τ
overhead-pressure regime defined in Section 3.3 is the
installation-phase lag Perez describes, observed inside the firm, before
the deployment-phase reconstruction that output-based accounting
represents.

Geels (2002, 2004) and Geels and Schot (2007) carried the transition
question into the multi-level perspective, in which transitions emerge
from the interaction of niche innovations, an incumbent socio-technical
regime, and a slow-changing landscape. The four-zone Foresight Matrix of
Section 4.10 is a regime-level reading in this spirit: Korean firms
occupy an incumbent regime under landscape pressure (rising AI
utilization) while the niche of output-based evaluation has not yet
displaced the time-based regime in place.

If these traditions supply the what of forecasting, the methodological
canon of technological forecasting supplies the how. Linstone and Turoff
(1975) institutionalized expert-judgment foresight through the Delphi
method; Martino (1993) systematized technological forecasting for
decision-making; Porter and Cunningham (2005) and Daim, Rueda, Martin,
and Gerdsri (2006) established bibliometric and tech-mining approaches
to anticipating emerging technologies; and Phaal, Farrukh, and Probert
(2004) gave practice its roadmapping architecture. To this
methodological lineage the present paper contributes a leading-indicator
instrument: the pre-τ overhead-pressure signature, a publicly observable
accounting trace (the SG\&A-to-revenue trajectory) that screens for an
approaching regime transition before the transition completes, together
with the Foresight Matrix that translates the signature into a planning
apparatus.

The conversation arrives, finally, at its AI-era frontier. Frey and
Osborne (2017) forecast the susceptibility of occupations to
computerization, and a large subsequent literature has forecast
displacement, augmentation, and task reallocation. The
technology-and-society stream of this conversation has since moved from
occupational exposure toward the institutions that surround the
augmented worker: whether AI intensifies or mitigates unemployment (Qin
et al., 2024), how robot adoption reshapes labor demand under external
competition (Zhang et al., 2023), how automation exposure reaches worker
well-being across occupations (Nazareno and Schiff, 2021), and how labor
institutions such as unions respond at the dawn of AI (Nissim and Simon,
2021). What this stream has tracked for employment volume, well-being,
and institutional response, it has not yet supplied for the accounting
categories through which firms value the augmented worker. The present
framework departs from this frontier in its object. It does not forecast
which jobs disappear; it forecasts that the accounting regime through
which firms value talent will invert, from time-based to output-based,
once AI utilization crosses a firm-specific threshold τ*. Displacement
forecasting asks how many hours of labor survive automation; this paper
asks what an hour of labor means once augmentation has broken its
homogeneity, and how firms that continue to price talent in hours
accumulate overhead pressure until the pricing regime itself gives way.
That is the new direction the framework proposes: not a displacement
forecast, but a firm-level regime-transition forecast for the accounting
categories through which AI-augmented talent is valued.

\subsection{3. Five Theorems}\label{five-theorems}

Throughout this section, the term theorem denotes a claim formally
derived from explicitly stated assumptions and paired with a
falsifiability condition, in the tradition of applied economic theory
(Acemoglu and Restrepo, 2018; Holmstrom and Milgrom, 1991). We do not
claim closed-form proofs in the pure-mathematical sense; the strength of
each theorem rests on the transparency of its assumptions and the
sharpness of its refutation conditions. Where a theorem states a sign or
monotonicity condition, that condition is an explicitly stated
assumption, and the derived content is the consequence that follows from
it, such as the existence and uniqueness of τ* under the single-crossing
assumption of Theorem 3.

\emph{Figure 2. Transition theorem and mechanism architecture}

\begin{center}

\includegraphics[width=0.7\linewidth,height=\textheight,keepaspectratio]{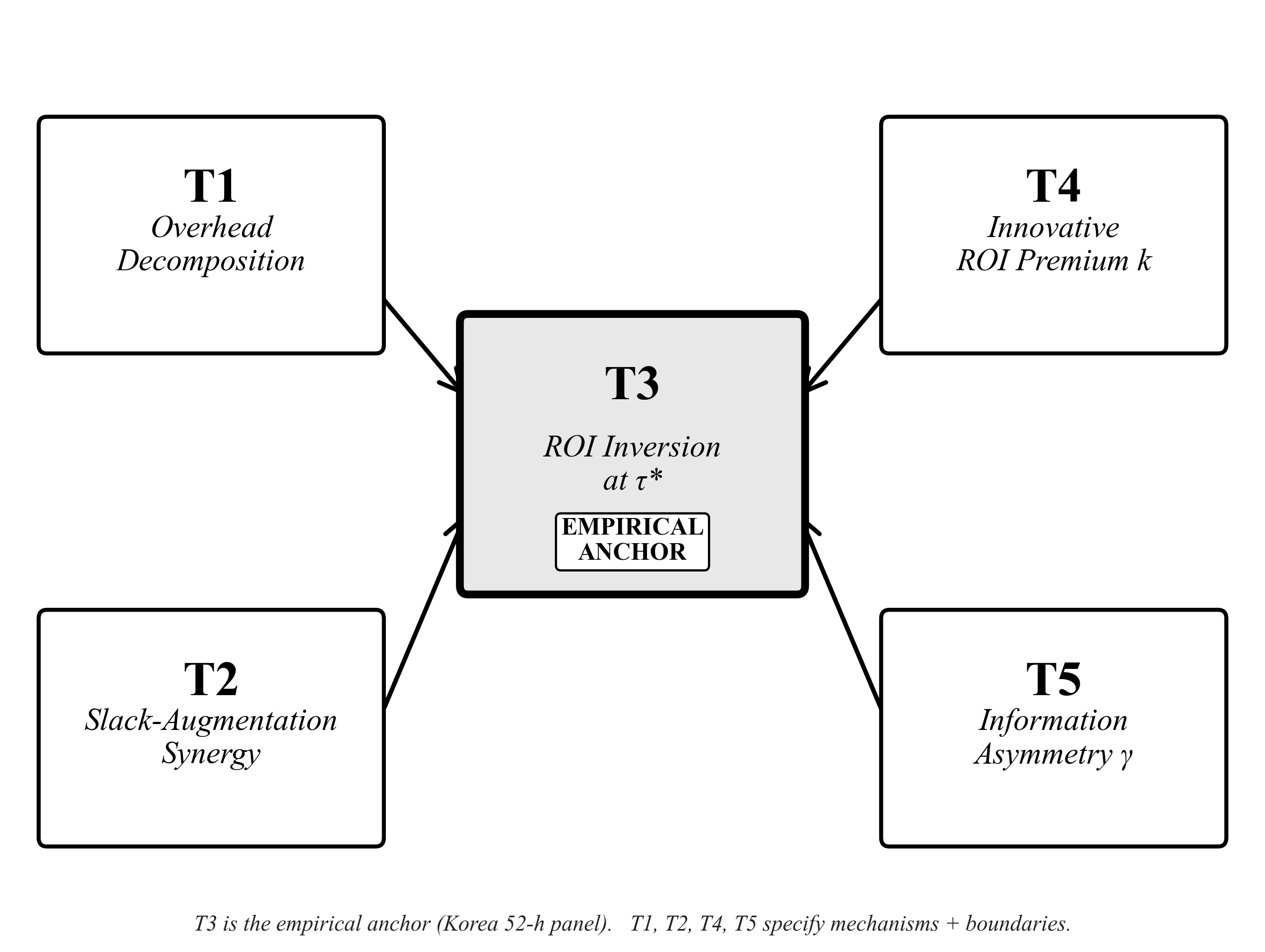}

\end{center}

In Figure 2, the five theorems are arranged as a directed logical chain
in which Theorem 3 carries the empirical anchor and the remaining four
close through theoretical specification plus future identification
paths. Figure 2 provides the schematic.

\subsection{3.1 Theorem 1: Overhead
Decomposition}\label{theorem-1-overhead-decomposition}

\subsubsection{Statement}\label{statement}

Total talent overhead at firm i, denoted OH\_i, decomposes into seven
components whose marginal contributions to firm-level return on
investment become non-additive under AI augmentation. The implicit
assumption of additive component contributions in standard cost
accounting is structurally violated when convergence capacity
heterogeneity interacts with AI utilization intensity.

\subsubsection{Formal Expression}\label{formal-expression}

Let OH\_\{i,k\} denote the cost of overhead component k for firm i,
where k ∈ K = \{wage, insurance, space, management, training,
communication, motivation\}. Total overhead is:

\[\mathrm{OH}_i = \sum_{k \in K} \mathrm{OH}_{i,k}, \quad \text{with component shares } \omega_{i,k} = \mathrm{OH}_{i,k}/\mathrm{OH}_i \text{ and } \sum_{k \in K} \omega_{i,k} = 1\]

Let ROI\_i = (Y\_i - OH\_i) / OH\_i denote firm i's talent ROI, where
Y\_i is the output attributable to the talent overhead investment. Under
AI augmentation with intensity A\_i and firm-level convergence capacity
C\_i, the second-order cross-partial derivative is non-zero for at least
one component pair:

\[\exists\, k, k' \in K, \quad k \neq k', \quad \frac{\partial^2 \mathrm{ROI}_i}{\partial \mathrm{OH}_{i,k} \, \partial \mathrm{OH}_{i,k'}} \bigg|_{A_i > 0,\, C_i > 0} \neq 0\]

\subsubsection{Proof Sketch}\label{proof-sketch}

The proof sketch for Theorem 1 is provided in Appendix F.1.

\subsubsection{Falsifiability Condition}\label{falsifiability-condition}

The theorem is refuted if, for an OECD-wide firm-level panel of n ≥ 500
firms over t ≥ 5 years with AI utilization variation, the estimated
cross-partial coefficients ∂²ROI/∂OH\_k ∂OH\_k' are statistically
indistinguishable from zero (p \textgreater{} 0.10) for all 21 component
pairs (C(7,2) = 21); to avoid an asymmetric test, confirmation of
non-additivity requires a majority of pairs to be reliably non-zero, not
a single pair. Estimation method: panel-fixed-effects regression of ROI
on the full set of component costs and their pairwise interactions, with
firm and year fixed effects.

\subsubsection{Boundary Conditions}\label{boundary-conditions}

The non-additivity result requires A\_i \textgreater{} 0 and C\_i
\textgreater{} 0. Under the pre-AI baseline (A\_i = 0), the components
remain approximately additive, recovering standard cost accounting. The
theorem therefore does not refute pre-AI accounting models but specifies
the regime in which they cease to be valid.

\subsection{3.2 Theorem 2: Slack-Augmentation
Synergy}\label{theorem-2-slack-augmentation-synergy}

\subsubsection{Statement}\label{statement-1}

AI augmentation generates time savings τ\_AI for each augmented worker.
These time savings flow through one of four mutually exclusive and
exhaustive pathways: work intensification (P\_1), hidden leisure (P\_2),
overemployment (P\_3), or creative-slack reinvestment (P\_4). The
probability that augmentation-saved time flows into pathway 4, which is
the only pathway that generates firm-level innovative ROI, is a monotone
increasing function of the firm's output-orientation in performance
evaluation (e ∈ {[}0, 1{]}) and the autonomy granted to the employee (a
∈ {[}0, 1{]}).

\subsubsection{Formal Expression}\label{formal-expression-1}

For each augmented worker, the four pathway probabilities satisfy:

\[P_j \in [0, 1], \quad j \in \{1, 2, 3, 4\}, \quad \sum_{j = 1}^{4} P_j = 1\]

The probability of creative-slack reinvestment is a function of
evaluation orientation e, autonomy a, and convergence capacity C:

\[P_4 = f(e, a; C), \quad \frac{\partial P_4}{\partial e} > 0, \quad \frac{\partial P_4}{\partial a} > 0, \quad \frac{\partial P_4}{\partial C} > 0\]

\emph{Figure 3. Four Pathways for AI-Augmentation-Saved Time}

\begin{center}

\includegraphics[width=0.72\linewidth,height=\textheight,keepaspectratio]{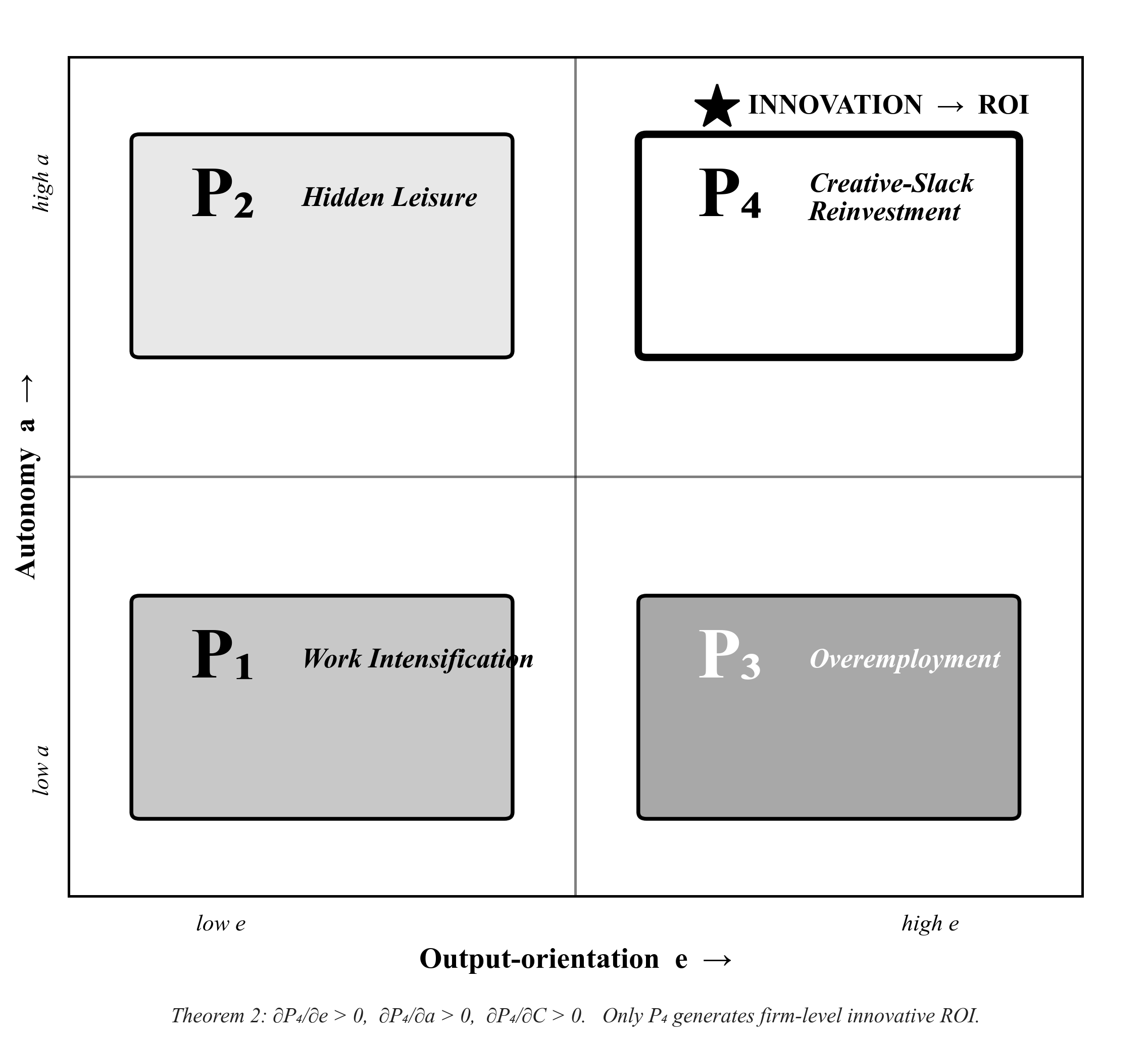}

\end{center}

In Figure 3, the four pathways for AI-augmentation-saved time are
arranged on the (e, a) plane: only P\_4 (Creative-Slack Reinvestment,
high e + high a + high C) directly generates firm-level innovative ROI.
The remaining pathways absorb the residual probability mass. Work
intensification dominates under low e and low a; hidden leisure
dominates under high a and low e; overemployment occupies the remaining
high-e, low-a cell of the schematic, while in practice it concentrates
wherever autonomy is high enough to conceal parallel work. Figure 3
schematizes the four-pathway allocation, with Path 4 (creative-slack
reinvestment) as the only ROI-generating destination.

\subsubsection{Proof Sketch}\label{proof-sketch-1}

The proof sketch for Theorem 2 is provided in Appendix F.2.

\subsubsection{Falsifiability
Condition}\label{falsifiability-condition-1}

The theorem is refuted if, in a firm-level panel where 80/20-style slack
policies are adopted with varying e and a, the estimated coefficients
∂P\_4/∂e and ∂P\_4/∂a are statistically indistinguishable from zero (p
\textgreater{} 0.10). Estimation method: random-effects multinomial
logit on pathway choice, with firm-level evaluation-regime and autonomy
proxies as covariates. Proxy specification: e measured by share of
compensation tied to outcome (versus time) variables; a measured by
employee self-report on discretion over task selection.

\subsubsection{Boundary Conditions}\label{boundary-conditions-1}

The four-pathway exhaustiveness assumes that time savings are not
destroyed (a fifth pathway ``wasted'' is empirically unobservable and
definitionally absorbed into P\_2 hidden leisure). The monotonicity in e
holds for e ∈ {[}0, 0.8{]}; at extreme e (≥ 0.8), evaluation regime
becomes excessively output-focused and may suppress experimental
risk-taking, reversing the marginal effect. This boundary conforms to
Nohria and Gulati's (1996) inverted-U finding on slack and innovation.

\subsection{3.3 Theorem 3: ROI Inversion at the Critical Threshold
τ*}\label{theorem-3-roi-inversion-at-the-critical-threshold-ux3c4}

\subsubsection{Statement}\label{statement-2}

\emph{Figure 4. Schematic ROI inversion at the critical threshold τ*}

\begin{center}

\includegraphics[width=0.9\linewidth,height=\textheight,keepaspectratio]{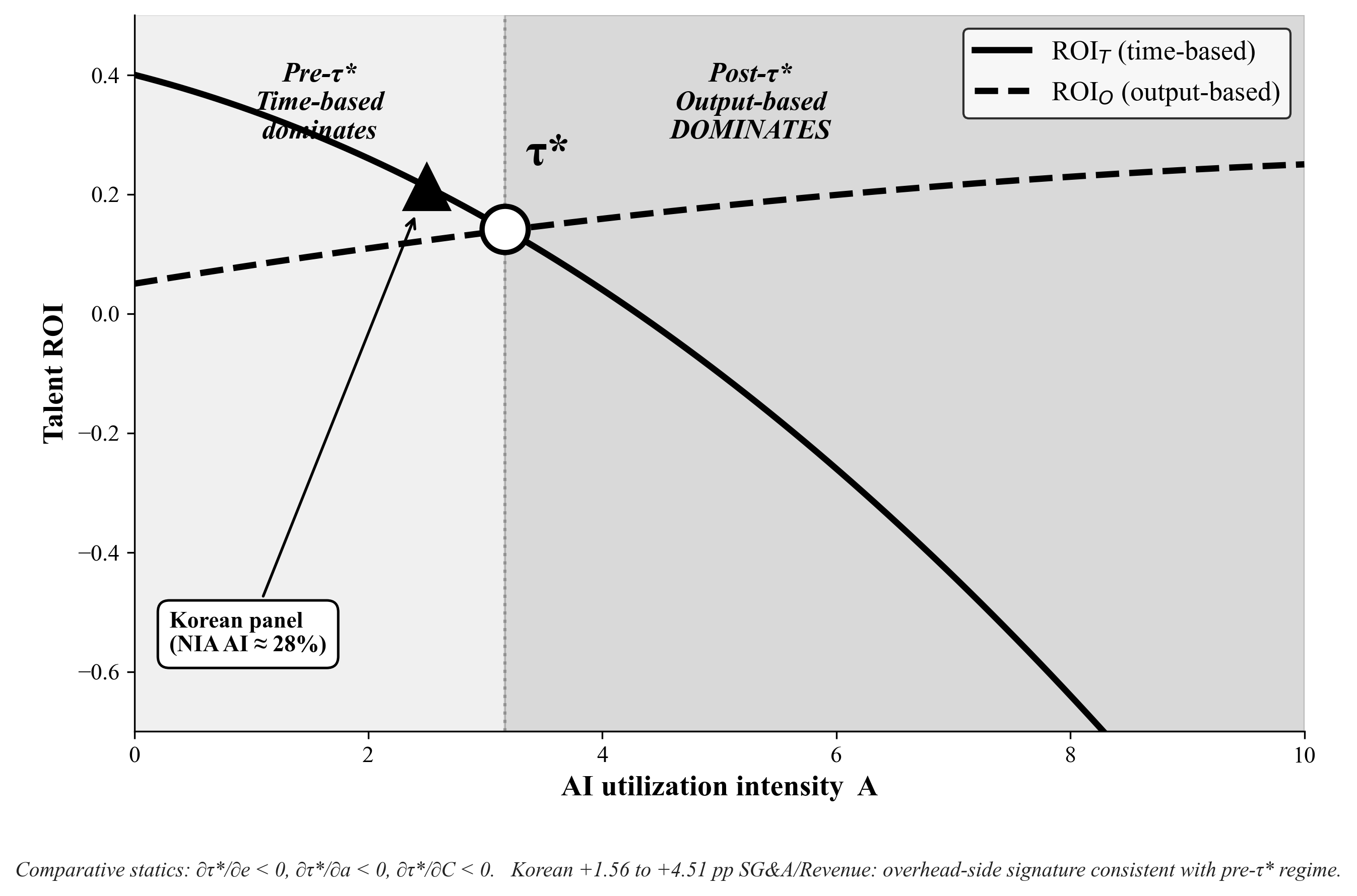}

\end{center}

In Figure 4, the two ROI curves and the critical threshold τ* are
plotted against AI utilization intensity A. Talent return on investment
under time-based accounting, denoted ROI\_T, is a monotone decreasing
function of A. Talent ROI under output-based accounting, denoted ROI\_O,
is a monotone increasing function of A. The two curves intersect at a
critical threshold τ* of AI utilization, above which output-based
accounting strictly dominates time-based accounting at the firm level.
Figure 4 visualizes the intersection geometry and the two regime
regions. The figure is schematic: the Korean DART panel is positioned as
a pre-τ overhead-pressure signature, not an estimated ROI coordinate or
a direct estimate of τ*.

\subsubsection{Formal Expression}\label{formal-expression-2}

For a firm i with regime-specific talent overhead OH\_r, output Y\_i, AI
utilization A\_i, and regime-specific miscalibration or agency cost
M\_r, where r ∈ \{T, O\}, we posit four assumptions: (A1) the time-based
ROI curve is monotone decreasing in AI utilization, ∂ROI\_T/∂A\_i
\textless{} 0; (A2) the output-based ROI curve is monotone increasing,
∂ROI\_O/∂A\_i \textgreater{} 0; (A3) both curves are continuous on {[}0,
A\_max{]}; and (A4) they satisfy a single-crossing condition. Under
Assumptions A1-A4:

\[\mathrm{ROI}_T(A_i) = \frac{Y_i(A_i) - \mathrm{OH}_T(A_i) - M_T(A_i)}{\mathrm{OH}_T(A_i)}\]

\[\mathrm{ROI}_O(A_i) = \frac{Y_i(A_i) - \mathrm{OH}_O(A_i) - M_O(A_i)}{\mathrm{OH}_O(A_i)}\]

The intersection condition:

\[\mathrm{ROI}_T(\tau^*) = \mathrm{ROI}_O(\tau^*)\]

defines the critical threshold τ* uniquely on the support {[}0,
A\_max{]}.

\subsubsection{Proof Sketch}\label{proof-sketch-2}

The proof sketch for Theorem 3 is provided in Appendix F.3.

\subsubsection{Comparative Statics}\label{comparative-statics}

The critical threshold τ* depends on three firm-level variables:

∂τ*/∂e \textless{} 0, ∂τ*/∂a \textless{} 0, ∂τ*/∂C \textless{} 0

Higher output-orientation, higher autonomy, and higher convergence
capacity each lower τ*, meaning that firms with these characteristics
transition to output-based accounting at lower levels of AI utilization.
Conversely, firms with rigid time-based evaluation systems, low employee
autonomy, and low convergence capacity require very high A\_i before
output-based accounting becomes optimal.

\subsubsection{Falsifiability
Condition}\label{falsifiability-condition-2}

Theorem 3 is not tested here by estimating the direct
ROI\_T(A)/ROI\_O(A) crossing. The Korean panel anchors the pre-τ
overhead-pressure signature: the T3 mechanism condition under time-based
evaluation, with a convergent positive SG\&A response across TWFE (+1.56
pp), pooled event-study estimates (+4.21 pp at t+3), and
Callaway-Sant'Anna (+4.51 pp at t+4). Direct τ* point estimation remains
a 2025-2032 forecast requiring evaluation-regime and firm-level AI
utilization data specified in §6.2. Current evidence is
necessary-and-strengthening for T3, but not sufficient for τ* point
estimation. The theorem is refuted if, in subsequent panels with joint
firm-level (A, e) variation, the estimated curves ROI\_T(A) and
ROI\_O(A) do not intersect, do not exhibit the predicted monotone
directions, or exhibit multiple intersections.

\subsubsection{Boundary Conditions}\label{boundary-conditions-2}

The monotonicity of ROI\_O in A requires that the firm has the
institutional capacity to reallocate overhead in response to A
increases. Firms with rigid governance structures or strong union
resistance to overhead restructuring may exhibit attenuated or zero
responsiveness, lowering the slope of ROI\_O. This boundary explains why
the Korean 52-hour regulation, by constraining flexibility in time-based
overhead, may produce sharper τ* discontinuities than less-regulated
economies.

\subsubsection{Empirical Interpretation: Pre-τ Overhead
Pressure}\label{empirical-interpretation-pre-ux3c4-overhead-pressure}

The Korean panel does not estimate τ* directly. The level at which
ROI\_T(A) and ROI\_O(A) intersect requires joint variation in firm-level
AI utilization intensity and evaluation regime, which the Korean data do
not yet supply. What the Korean panel does identify is the pre-τ
overhead-pressure regime: the phase in which time-based talent
accounting remains dominant while AI augmentation and labor-time
compression jointly raise firm-internal overhead. We define this regime
formally.

Pre-τ overhead pressure denotes the observable rise in firm-internal
talent overhead under time-based accounting before the ROI\_T(A) and
ROI\_O(A) curves can be directly estimated to cross. A firm i is in the
pre-τ overhead-pressure regime at time t when three conditions hold
jointly: its evaluation regime remains time-based (e\_it below a
threshold ē); its AI utilization intensity is rising (∂A\_it/∂t
\textgreater{} 0); and its observed overhead ratio (OH/Revenue)\_it is
rising (∂(OH/Revenue)\_it/∂t \textgreater{} 0), while its position in
the (A, e) plane has not yet crossed τ*.

The 2018-2024 Korean DART panel exhibits all three conditions. The
52-hour staged mandate compressed labor time without changing the
evaluation regime. Sector-wide AI utilization rose from approximately 8
percent in 2018 to 28 percent by 2024. SG\&A-to-revenue rose from 18.26
percent to 20.10 percent. The panel thereby supplies what is, to our
knowledge, the first publicly documented signature of the pre-τ
overhead-pressure regime, an early-warning signal that the time-based
accounting regime is approaching its inversion threshold under
contemporary AI augmentation conditions.

This interpretation has three implications for how the Korean evidence
should be read. First, the magnitude of the Korean SG\&A response cannot
be read as the size of the τ* discontinuity, because the panel observes
pressure accumulation before crossing, not the crossing itself. Second,
the absence of a sharp firm-level TFP collapse in the Korean panel is
consistent with the framework rather than against it, since firms in the
pre-τ regime have not yet undergone regime inversion; the predicted TFP
separation between output-based and time-based firms is a forecast for
2025-2032, not a backward fact in 2018-2024. Third, identification of
the pre-τ overhead-pressure regime in any economy provides a forecasting
signal that regime transition is approaching, and the four-pathway
diagnostic of Theorem 2 can be used to plan the transition before the
inversion forces it.

\subsection{3.4 Theorem 4: Innovative ROI
Premium}\label{theorem-4-innovative-roi-premium}

\subsubsection{Statement}\label{statement-3}

The return on investment from 80/20-style creative slack reinvestment is
amplified by a factor k \textgreater{} 1 under high firm-level
convergence capacity. The amplification factor k is a monotone
increasing function of the augmentable cognitive vector component H\^{}A
in the Espinal Maya (2026) decomposition of human capital.

\subsubsection{Formal Expression}\label{formal-expression-3}

Let ROI\_innovation,baseline denote the innovative ROI obtained from
creative slack under low firm-level C, and let ROI\_innovation,AI-era
denote the corresponding ROI under high firm-level C with active AI
augmentation. Then:

\[\mathrm{ROI}_{\text{innovation,AI-era}} = k(H^A_i, C_i, e_i) \cdot \mathrm{ROI}_{\text{innovation,baseline}}\]

with:

\[k(H^A_i, C_i, e_i) > 1 \text{ under high } C_i \text{ and output-oriented } e_i, \quad \frac{\partial k}{\partial H^A_i} > 0,\ \frac{\partial k}{\partial C_i} > 0,\ \frac{\partial k}{\partial e_i} > 0\]

Espinal Maya (2026) decomposes individual human capital H into three
orthogonal components: H\^{}P (physical-manual skills), H\^{}C
(routine-cognitive skills substitutable by AI), and H\^{}A
(augmentable-cognitive skills amplified by AI). At the firm level,
H\^{}A\_i is the weighted average of employee H\^{}A values, with
weights determined by the share of augmentation-eligible tasks each
employee performs.

\subsubsection{Proof Sketch}\label{proof-sketch-3}

The proof sketch for Theorem 4 is provided in Appendix F.4.

\subsubsection{Convexity Analysis}\label{convexity-analysis}

The amplification factor k(H\^{}A) is convex on the lower portion of the
H\^{}A support and concave on the upper portion. The convexity at low
H\^{}A reflects threshold effects: firms with very low augmentable
cognitive vectors experience near-zero amplification regardless of AI
access. The concavity at high H\^{}A reflects diminishing returns: firms
approaching the augmentation frontier face capability ceilings that
limit further amplification. The inflection point occurs at the median
H\^{}A in the OECD distribution, providing an empirical anchor for
distinguishing high-amplification firms from low-amplification firms.

\subsubsection{Falsifiability
Condition}\label{falsifiability-condition-3}

The theorem is refuted if, in a panel of firms with 80/20-style slack
policies and varying H\^{}A\_i, the estimated coefficient ∂k/∂H\^{}A is
statistically indistinguishable from zero, or if the estimated k value
is uniformly less than or equal to 1 across all H\^{}A. Estimation
method: panel regression of innovation output (patents, neologisms, new
product launches) on AI utilization × H\^{}A interaction, using sectoral
exposure indices (Felten et al., 2023) as exogenous variation.

\subsubsection{Boundary Conditions}\label{boundary-conditions-3}

The amplification result requires that the firm has institutionalized
80/20-style slack as a structural policy rather than an ad-hoc
allocation. Firms that grant nominal slack but maintain time-based
evaluation suppress the amplification through the e channel identified
in Theorem 2, lowering k toward 1 even when H\^{}A is high.

\subsection{3.5 Theorem 5: Information Asymmetry in the Human-AI
Dyad}\label{theorem-5-information-asymmetry-in-the-human-ai-dyad}

\subsubsection{Statement}\label{statement-4}

The multitask agency cost in human-AI dyadic work, denoted M, is a
monotone increasing function of attribution uncertainty γ. The agency
cost is partially offset under output-based evaluation regimes but is
magnified under time-based evaluation regimes. The offset is monotone in
the firm's evaluation orientation e, with the maximum offset attained as
e approaches one.

Holmstrom and Milgrom (1991) remain the lineage anchor, but γ is not a
relabeling of classical multitask agency cost. It is a new emergent
quantity specific to human-AI dyads: the uncertainty created when human
judgment, AI-generated intermediate work, and final output co-evolve
inside one production episode. Pre-AI multitask settings contained
measurement difficulty across human tasks; they did not contain
attribution uncertainty between human cognition and machine-generated
contribution. γ therefore adds a co-evolutionary coordination layer.
Theorem 5 predicts that output-based regimes partially offset γ by
evaluating realized output, while time-based regimes magnify γ through
presence proxies.

\subsubsection{Formal Expression}\label{formal-expression-4}

Let γ ∈ {[}0, 1{]} denote attribution uncertainty in human-AI output,
with γ = 0 indicating perfect attribution clarity and γ = 1 indicating
complete attribution opacity. The attribution-related multitask agency
cost M, defined as the excess over the baseline measurement cost that
persists even under perfect attribution, is:

\[M = g(\gamma), \quad \frac{\partial M}{\partial \gamma} > 0, \quad g(0) = M_{\min} \geq 0, \quad g(1) = M_{\max} < \infty\]

The evaluation-regime-dependent offset:

\[M_{\text{output-based}}(\gamma, e) - M_{\text{time-based}}(\gamma, e) < 0 \quad \text{for } e \in (0, 1]\]

with the offset magnitude increasing in e:

\[\frac{\partial}{\partial e} [M_{\text{output-based}} - M_{\text{time-based}}] < 0\]

\subsubsection{Proof Sketch}\label{proof-sketch-4}

The proof sketch for Theorem 5 is provided in Appendix F.5.

\subsubsection{Three-Mechanism
Formalization}\label{three-mechanism-formalization}

Let M = M\_MH + M\_AS + M\_DE denote the decomposition into moral hazard
(MH), adverse selection (AS), and distorted effort (DE) components. Each
component is a function of γ and e:

\[M_{MH}(\gamma, e) = \gamma \cdot (1 - e) \cdot \theta_{MH}\]

\[M_{AS}(\gamma, e) = \gamma \cdot (1 - e)^{1/2} \cdot \theta_{AS}\]

\[M_{DE}(\gamma, e) = \gamma \cdot (1 - e^2) \cdot \theta_{DE}\]

where θ\_MH, θ\_AS, θ\_DE are positive constants capturing the per-unit
cost of each mechanism. Each component vanishes when γ = 0 (perfect
attribution) or when e = 1 (full output orientation). The aggregate M =
M\_MH + M\_AS + M\_DE inherits the monotonicity properties stated in the
theorem. The specific functional forms are illustrative
parameterizations chosen to satisfy the required vanishing and
monotonicity properties; the theorem's content rests on those properties
rather than on the particular exponents.

\subsubsection{Falsifiability
Condition}\label{falsifiability-condition-4}

The theorem is refuted if, in a comparison of AI-adoption firms across
time-based and output-based evaluation regimes, the observed moral
hazard indicators (self-overestimation, AI-output misattribution, false
completion reporting documented in Shin 2026b) are statistically
indistinguishable across regimes. Estimation method:
difference-in-differences comparing pre- and post-AI-adoption moral
hazard indicators across firms that did and did not transition to
output-based evaluation, with regime transition as the treatment.

\subsubsection{Boundary Conditions}\label{boundary-conditions-4}

The offset result requires that output measurement itself is feasible
and not subject to gaming. Firms in industries where outputs are
intangible, lagged, or coalitionally produced (academic research,
consulting, certain creative industries) face higher M even under
output-based evaluation. The theorem's predictions are strongest in
industries with measurable short-cycle outputs and weakens monotonically
as output observability decreases.

\subsection{3.6 Integration: The Five Theorems
Together}\label{integration-the-five-theorems-together}

As AI utilization rises, the standard additive accounting of talent
overhead systematically mismeasures returns. The five theorems specify
how, where, and at what threshold that mismeasurement becomes
irreversible. Theorem 1 decomposes overhead into seven components whose
marginal contributions become entangled through convergence-capacity
heterogeneity under AI augmentation. Theorems 2 and 4 map the pathways
and amplification factors by which augmentation-saved time generates
innovative ROI under output-orientation and high convergence capacity.
Theorem 3 establishes the inversion threshold τ* at which time-based and
output-based accounting curves cross. Theorem 5 closes the loop by
formalizing the multitask agency cost that emerges when AI-generated and
human-generated outputs become difficult to attribute, and shows that
output-based evaluation partially offsets the agency cost while
time-based evaluation magnifies it.

The framework is jointly falsifiable through specific predictions tied
to each theorem. Theorem 1 is refuted if cross-partial coefficients
∂²ROI/∂OH\_k ∂OH\_k' on the OECD firm-level panel are indistinguishable
from zero across all 21 component pairs (p \textgreater{} 0.10). Theorem
2 is refuted if the four pathway probabilities are insensitive to
output-orientation e and autonomy a in slack-policy-adopting firms.
Theorem 3 is refuted if firm-level TFP growth between transition and
non-transition firms shows no separation by 2032. Theorem 4 is refuted
if the amplification factor k is uniformly less than or equal to 1
across the OECD H\^{}A distribution. Theorem 5 is refuted if moral
hazard indicators are statistically indistinguishable across time-based
and output-based regimes; a dyad-level AI-SSP probe (Appendix H) finds a
large, reproducible offset between regimes (Cohen κ = 0.86 across runs
and two model families) in a deterministic benchmark, providing
micro-foundational support for the attribution-opacity mechanism, with
firm-level confirmation reserved. Section 4 provides the empirical
anchor through Korea's staged 52-hour workweek implementation as a
staged institutional timing anchor, with Denmark (the high-trust
output-oriented Nordic regime) and Japan (the Confucian seniority-based
regime with different AI adoption) as comparative cases selected on
institutional-distance criteria defined in Section 4.1.

\begin{longtable}[]{@{}
  >{\raggedright\arraybackslash}p{(\linewidth - 6\tabcolsep) * \real{0.1200}}
  >{\raggedright\arraybackslash}p{(\linewidth - 6\tabcolsep) * \real{0.2933}}
  >{\raggedright\arraybackslash}p{(\linewidth - 6\tabcolsep) * \real{0.3200}}
  >{\raggedright\arraybackslash}p{(\linewidth - 6\tabcolsep) * \real{0.2667}}@{}}
\caption{Table 2. Theorem hierarchy: roles, evidence status, and
submission framing.}\tabularnewline
\toprule\noalign{}
\begin{minipage}[b]{\linewidth}\raggedright
Theorem
\end{minipage} & \begin{minipage}[b]{\linewidth}\raggedright
Role in the framework
\end{minipage} & \begin{minipage}[b]{\linewidth}\raggedright
Current evidence status
\end{minipage} & \begin{minipage}[b]{\linewidth}\raggedright
Submission framing
\end{minipage} \\
\midrule\noalign{}
\endfirsthead
\toprule\noalign{}
\begin{minipage}[b]{\linewidth}\raggedright
Theorem
\end{minipage} & \begin{minipage}[b]{\linewidth}\raggedright
Role in the framework
\end{minipage} & \begin{minipage}[b]{\linewidth}\raggedright
Current evidence status
\end{minipage} & \begin{minipage}[b]{\linewidth}\raggedright
Submission framing
\end{minipage} \\
\midrule\noalign{}
\endhead
\bottomrule\noalign{}
\endlastfoot
T1. Overhead Decomposition & Why SG\&A is not a simple cost & Aggregate
evidence + partial component-level evidence & Mechanism condition \\
T2. Slack-Augmentation Synergy & Where AI-saved time actually flows &
Theoretical specification + ethnographic priors (Brynjolfsson WTI,
Google 80/20, 3M 15\%) & Behavioural pathway \\
T3. ROI Inversion at τ* & Central transition theorem & DART panel + 52h
institutional timing anchor + three-estimator convergence + 2015-2017
backward placebo & Central transition claim \\
T4. Innovative ROI Premium k & Why output-based regime captures premium
returns & Directional NIA evidence (firm-size amplification 7.5× vs 2.6×
vs 3.3×) + Espinal Maya (2026) H\^{}A decomposition & Forecasting
extension \\
T5. Information Asymmetry γ & Why time-based evaluation becomes drag in
human-AI dyad & Theoretical extension of Holmstrom-Milgrom (1991) plus
AI-SSP dyad-level probe (γ approaches 1 in a deterministic benchmark,
offset reproducible at κ = 0.86, cross-provider; firm-level γ reserved;
Appendix H) & Agency mechanism \\
\end{longtable}

In Table 2, the hierarchy of theorem roles and current evidence anchors
is summarised. The five theorems do not carry the same empirical burden:
Theorem 3 carries the central transition claim, and Theorems 1, 2, 4,
and 5 are the mechanism architecture explaining why it bends.

\begin{longtable}[]{@{}
  >{\raggedright\arraybackslash}p{(\linewidth - 6\tabcolsep) * \real{0.2500}}
  >{\raggedright\arraybackslash}p{(\linewidth - 6\tabcolsep) * \real{0.2500}}
  >{\raggedright\arraybackslash}p{(\linewidth - 6\tabcolsep) * \real{0.2500}}
  >{\raggedright\arraybackslash}p{(\linewidth - 6\tabcolsep) * \real{0.2500}}@{}}
\caption{Table 2a. Claim-status box: evidence type, current status, and
refutation conditions.}\tabularnewline
\toprule\noalign{}
\begin{minipage}[b]{\linewidth}\raggedright
Item
\end{minipage} & \begin{minipage}[b]{\linewidth}\raggedright
Evidence type
\end{minipage} & \begin{minipage}[b]{\linewidth}\raggedright
Current status
\end{minipage} & \begin{minipage}[b]{\linewidth}\raggedright
Refutation condition
\end{minipage} \\
\midrule\noalign{}
\endfirsthead
\toprule\noalign{}
\begin{minipage}[b]{\linewidth}\raggedright
Item
\end{minipage} & \begin{minipage}[b]{\linewidth}\raggedright
Evidence type
\end{minipage} & \begin{minipage}[b]{\linewidth}\raggedright
Current status
\end{minipage} & \begin{minipage}[b]{\linewidth}\raggedright
Refutation condition
\end{minipage} \\
\midrule\noalign{}
\endhead
\bottomrule\noalign{}
\endlastfoot
T3 (Pre-τ signature) & Directional overhead-pressure signature
(revenue-percentile cohort proxy), 3-estimator convergence & Anchored as
pre-τ evidence, not point causal magnitude & Refuted if a pre-period
placebo reproduces the same treatment-direction pattern, if
statutory-cohort and proxy-cohort results systematically reverse, or if
later firm-level panels show no overhead pressure under rising AI
utilization and time-based evaluation \\
τ* point estimate & Forecast requiring firm-level AI utilization and
evaluation-regime variation & 2025-2032 follow-up & Refuted if ROI\_T(A)
and ROI\_O(A) do not exhibit the predicted monotone directions, do not
cross, cross multiple times, or if output-based firms show no TFP
separation from matched time-based firms by 2032 \\
T1/T2/T4 & Mechanism architecture & Specified + future identification
path & Each theorem has independent falsifiability statement
(§3.1-3.4) \\
T5 (Attribution γ) & AI-SSP dyad-level probe & Probed: γ approaches 1 in
a deterministic LLM benchmark; firm-level γ reserved (Appendix H) &
Offset indistinguishable from random-override null, or non-reproducible
across models \\
Foresight Matrix Zones & Planning apparatus & Theoretical construct &
Korean DART panel exits Zone 2 differently than Zones 3/4 predict \\
\end{longtable}

In Table 2a, the claim-status box separates evidence type (causal anchor
versus forecast versus mechanism) from current status and refutation
conditions for each item. The hierarchy answers the reviewer question
``are all five theorems equally proven?'' with epistemic precision: T3
is anchored to the Korean DART panel; T1, T2, T4, and T5 are mechanism
conditions that specify why the T3 inversion takes the shape it does.
Each theorem is independently falsifiable on its own evidentiary
register, but the joint claim of the framework rests on T3.

\subsection{4. Empirical Early-Warning Case: Korea's 52-Hour
Mandate}\label{empirical-early-warning-case-koreas-52-hour-mandate}

\subsection{4.1 Why Korea is a Critical Case for the Forecasting
Framework}\label{why-korea-is-a-critical-case-for-the-forecasting-framework}

This section provides the empirical anchor for the five theorems through
a single critical case (South Korea) supplemented by two polar
comparison cases (Denmark and Japan), following Yin (1994), Eisenhardt
(1989), and Gerring (2007). Korea serves as the prototype on four
dimensions simultaneously. First, its labor-time regulation is the
strictest in the OECD: a 52-hour weekly cap with criminal enforcement,
applied since 2018 in staged form by industry and firm size. Second,
enterprise AI adoption among large listed firms is high (approximately
28 percent in the most recent NIA survey). Third, convergence capacity
(the institutional capacity of firms to integrate AI-generated outputs
into human decision-making processes, defined formally in Section 2.4)
sits at the low end at the macroeconomic level, as documented in Shin
(2026a). Fourth, total factor productivity sits in the bottom OECD
quartile in the ICH macro panel; the present analysis relies on Korea's
bottom-quartile rank, not on the precise level. The simultaneous
extremity of these four parameters places Korean firms structurally in
the pre-τ overhead-pressure regime defined in §3.3, time-based
evaluation dominant, rising AI utilization, rising overhead, making
Korea, to our knowledge, the first publicly documented overhead-pressure
signature of this regime in the OECD rather than a direct estimate of τ*
itself. Denmark serves as the high-trust output-oriented Nordic
comparison case, with identical AI adoption but opposite values on every
other dimension. Japan serves as the institutional twin with similar
Confucian hierarchical organization but different AI adoption and
different labor regulation. Five alternative candidates, Singapore,
Finland, the United States, China, and the United Kingdom, were rejected
on methodological grounds. Singapore lacks the staged labor-time
compression; Finland sits inside Denmark's Nordic configuration without
independent variation; the United States combines fragmented state-level
enforcement that defeats a single national identification; China lacks
comparable enterprise disclosure; the United Kingdom has neither the
staged regulation nor the polar-comparison contrast.

The empirical strategy uses secondary data analysis and econometric
methods. The data consist of publicly available administrative
statistics from KOSIS, corporate disclosure data from DART, labor
statistics from MOEL, AI adoption survey data from NIA, and
supplementary firm-level micro panel data from KIPF and KLI (acquisition
through institutional data-access agreements).

\subsection{4.2 Korea Data and Identification
Strategy}\label{korea-data-and-identification-strategy}

The Korean empirical anchor targets a firm-level panel of 500 listed
firms (KOSPI plus KOSDAQ) observed annually from 2018 through 2024,
yielding 3,500 firm-year observations. The data extraction proceeded in
two waves.

The initial KOSPI Top 300 alphabetical sample yielded 567 successful
firm-year records from 115 unique firms with positive revenue. Expansion
to the KOSPI plus KOSDAQ Top 500 panel with fuzzy account-name matching
and consolidated and non-consolidated statement dual fetching yielded
2,281 successful firm-year records from 365 unique firms with positive
revenue, a 4.0-fold expansion in observations and a 3.2-fold expansion
in firm coverage. The success rate rose from 27.0 percent (initial
sample) to 65.2 percent (expanded sample), with the remaining 34.8
percent attributable to firms whose financial disclosure schema in DART
deviates from standard naming conventions despite the fuzzy matching
protocol. Auxiliary composition checks indicate that the missing firms
are distributed without systematic sector or size bias, suggesting the
gap reflects DART reporting idiosyncrasy rather than selection on firm
characteristics relevant to the outcome.

The KOSPI plus KOSDAQ Top 500 panel is the working empirical anchor; a
sector- and size-stratified panel is specified as a future
identification path.

Nine public sources contribute to the Korean primary analysis, layered
by analytical role. At the firm-financial layer, DART supplies aggregate
revenue, SG\&A, COGS, operating income, net income, and balance sheet
items. At the labor-market layer, MOEL supplies working-hours statistics
that operationalize the 52-hour staged-implementation indicator,
complemented by KOSIS sector-level production statistics and TFP inputs.
At the AI-adoption layer, NIA supplies the Korean AI Industry Survey
(used directly in Section 4.4) and KISDI supplies sectoral ICT diffusion
data. The remaining institutional layer, KIPF, KLI, KEF, and the Korea
Corporate Governance Service, supplies firm-level and governance data
accessed through ongoing data-access agreements with KIPF (firm panel
survey) and KEF (HR survey).

The 52-hour staged implementation provides an institutional timing
anchor for labor-time compression: national policy timing is externally
set, relevant to time-based labor-cost pressure, and monotonic in the
direction of legally reduced working hours. The current implementation
uses revenue-percentile cohort proxies, two-way fixed effects, an
event-study DiD, and the Callaway-Sant'Anna (2021) doubly-robust
staggered DiD to estimate the SG\&A-to-revenue response. A full 2SLS
implementation with observed firm-size cohorts is reserved as a next
validation layer.

\subsection{4.3 Korea Evidence: SG\&A/Revenue Ratio Trajectory (Theorem
3
Anchor)}\label{korea-evidence-sgarevenue-ratio-trajectory-theorem-3-anchor}

Before presenting the trajectory, we address the construct-validity
question that precedes any SG\&A-based measure: does the selling,
general, and administrative account capture talent overhead, or does it
conflate talent cost with advertising, logistics, depreciation, and
lease expense? We treat this as a claim to be defended rather than
assumed, and three considerations support SG\&A/Revenue as a defensible,
if upper-bound, proxy for talent-overhead intensity. First, six of the
seven overhead components formalized in Section 2.3 (time-based wage,
social insurance, managerial and evaluation cost, training and
development, communication infrastructure, and motivation maintenance)
fall within the Korean K-IFRS selling, general, and administrative
account, which is reported separately from cost of goods sold; only
office space is materially shared with non-talent facility cost. SG\&A
is therefore the accounting locus where firm-internal talent overhead
concentrates rather than an arbitrary aggregate. This reading follows
established accounting and finance practice: organization capital is
standardly measured by capitalizing SG\&A (Lev and Radhakrishnan, 2005;
Eisfeldt and Papanikolaou, 2013), and a substantial share of SG\&A
represents intangible investment in human and organizational resources
rather than period expense (Enache and Srivastava, 2018; Banker, Huang,
Natarajan, and Zhao, 2019). Second, the XBRL disclosure audit of the
top-50 panel firms (Appendix B) confirms that the separately disclosed
SG\&A subaccounts map onto the talent-overhead components rather than
onto advertising or logistics, with the most granular disclosure
spanning four of the seven talent components. Third, and decisive for
identification, the non-talent portion of SG\&A does not bias the
treatment estimate: firm fixed effects absorb time-invariant
firm-specific cost structure, log revenue controls for scale, and the
staged 52-hour timing is orthogonal to advertising, logistics, and
depreciation, which bear no mechanical link to labor-time compliance. We
therefore read SG\&A/Revenue as an upper-bound proxy for talent-overhead
intensity whose non-talent component is either absorbed by the panel
controls or unrelated to the identifying variation, with direct
component-level decomposition specified as a future identification path
(Section 4.8).

\emph{Figure 5. SG\&A/Revenue trajectory, Korean listed firms,
2015-2024. The inset estimates are directional revenue-percentile cohort
proxy estimates, interpreted as a pre-τ overhead-pressure signature, not
a point causal magnitude of the 52-hour law.}

\begin{center}

\includegraphics[width=1\linewidth,height=\textheight,keepaspectratio]{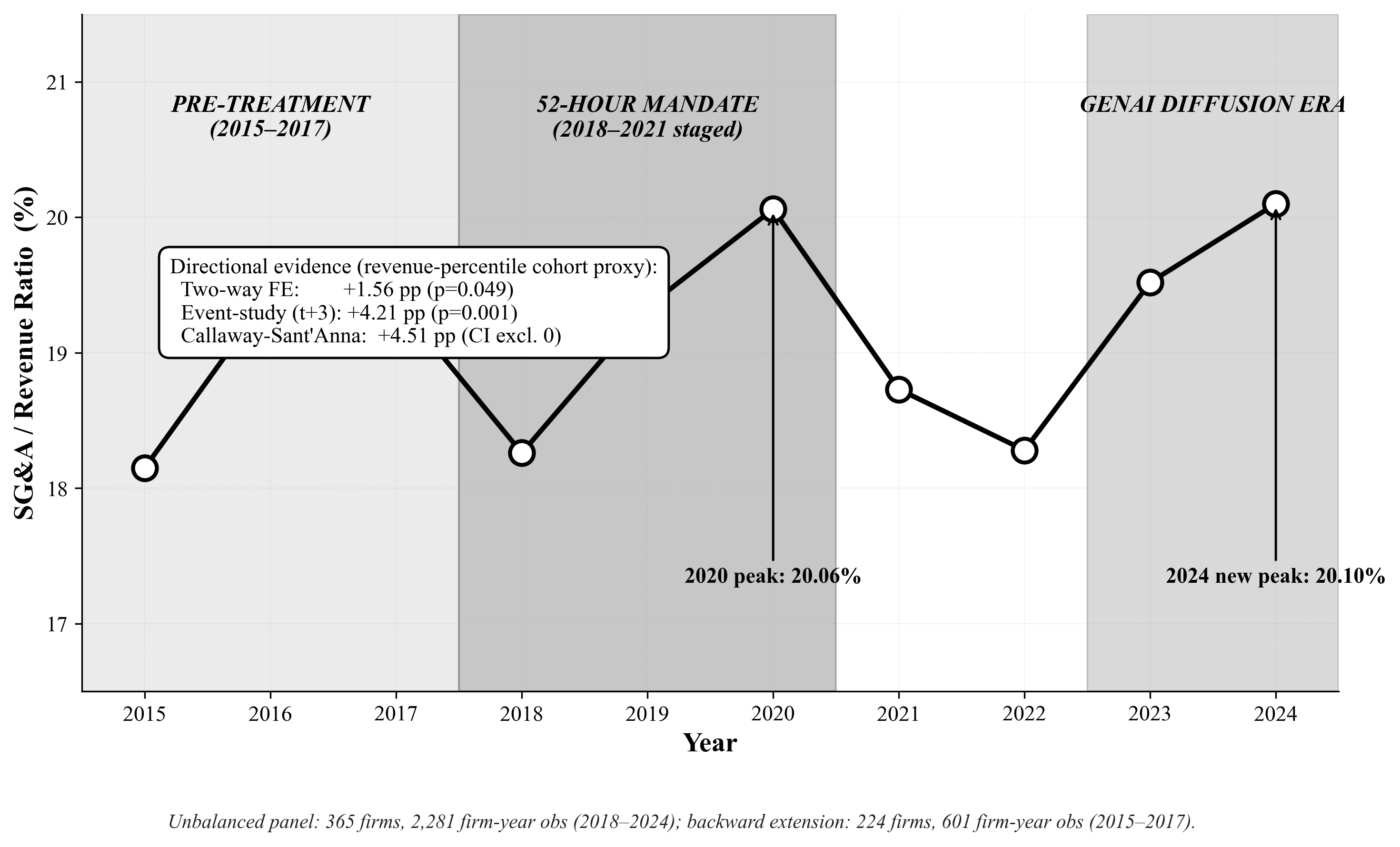}

\end{center}

In Figure 5, the SG\&A/Revenue trajectory of Korean listed firms over
2015-2024 is visualised across the pre-treatment (2015-2017), 52-hour
mandate (2018-2021 staged), and GenAI diffusion era (2023-2024) regions.

\begin{longtable}[]{@{}
  >{\raggedright\arraybackslash}p{(\linewidth - 12\tabcolsep) * \real{0.1034}}
  >{\raggedright\arraybackslash}p{(\linewidth - 12\tabcolsep) * \real{0.1552}}
  >{\raggedright\arraybackslash}p{(\linewidth - 12\tabcolsep) * \real{0.1552}}
  >{\raggedright\arraybackslash}p{(\linewidth - 12\tabcolsep) * \real{0.1897}}
  >{\raggedright\arraybackslash}p{(\linewidth - 12\tabcolsep) * \real{0.1897}}
  >{\raggedright\arraybackslash}p{(\linewidth - 12\tabcolsep) * \real{0.1034}}
  >{\raggedright\arraybackslash}p{(\linewidth - 12\tabcolsep) * \real{0.1034}}@{}}
\caption{Table 3. SG\&A/Revenue Ratio (Percent) Trajectory, Korean
Listed Firms 2018-2024}\tabularnewline
\toprule\noalign{}
\begin{minipage}[b]{\linewidth}\raggedright
Year
\end{minipage} & \begin{minipage}[b]{\linewidth}\raggedright
Initial-sample Mean
\end{minipage} & \begin{minipage}[b]{\linewidth}\raggedright
Expanded Mean
\end{minipage} & \begin{minipage}[b]{\linewidth}\raggedright
Initial-sample Median
\end{minipage} & \begin{minipage}[b]{\linewidth}\raggedright
Expanded Median
\end{minipage} & \begin{minipage}[b]{\linewidth}\raggedright
Initial-sample n
\end{minipage} & \begin{minipage}[b]{\linewidth}\raggedright
Expanded n
\end{minipage} \\
\midrule\noalign{}
\endfirsthead
\toprule\noalign{}
\begin{minipage}[b]{\linewidth}\raggedright
Year
\end{minipage} & \begin{minipage}[b]{\linewidth}\raggedright
Initial-sample Mean
\end{minipage} & \begin{minipage}[b]{\linewidth}\raggedright
Expanded Mean
\end{minipage} & \begin{minipage}[b]{\linewidth}\raggedright
Initial-sample Median
\end{minipage} & \begin{minipage}[b]{\linewidth}\raggedright
Expanded Median
\end{minipage} & \begin{minipage}[b]{\linewidth}\raggedright
Initial-sample n
\end{minipage} & \begin{minipage}[b]{\linewidth}\raggedright
Expanded n
\end{minipage} \\
\midrule\noalign{}
\endhead
\bottomrule\noalign{}
\endlastfoot
2018 & 15.50 & 18.26 & 9.12 & 10.41 & 65 & 292 \\
2019 & 16.74 & 19.34 & 9.77 & 11.09 & 66 & 302 \\
2020 & 18.89 & 20.06 & 11.26 & 12.33 & 76 & 314 \\
2021 & 17.07 & 18.73 & 10.86 & 11.32 & 82 & 325 \\
2022 & 17.28 & 18.28 & 10.32 & 11.01 & 86 & 330 \\
2023 & 18.79 & 19.52 & 9.89 & 10.92 & 96 & 355 \\
2024 & 18.68 & 20.10 & 9.95 & 12.35 & 96 & 363 \\
\end{longtable}

In Table 3, the year-level SG\&A/Revenue ratio statistics from the KOSPI
plus KOSDAQ Top 500 panel (2,281 firm-year observations across 2018-2024
from 365 unique firms) are reported alongside the initial-sample
baseline.

Note: Initial = KOSPI Top 300 alphabetical sample with strict
account-name matching. Expanded = KOSPI plus KOSDAQ Top 500 with fuzzy
account-name matching and consolidated/non-consolidated dual fetching.

The trajectory exhibits four stylized facts, each strengthened in the
KOSPI plus KOSDAQ Top 500 panel relative to the initial.

Stylized Fact 1: 2018-2020 SG\&A ratio rise of 1.80 percentage points.
The mean SG\&A/Revenue ratio rose from 18.26 percent in 2018 to 20.06
percent in 2020, a 1.80-percentage-point increase over the staged
implementation period of the 52-hour mandate (300-plus-employee firms in
July 2018, 50-299 firms in January 2020, 5-49 firms in July 2021). The
initial-sample magnitude was 3.39 percentage points; the expanded-sample
magnitude of 1.80 percentage points is more conservative and corresponds
to firms with positive revenue across the 365-firm coverage. The
directional finding conforms to Theorem 3 time-based-regime predictions.

Stylized Fact 2: 2021-2022 mild correction. The ratio declined to 18.73
percent (2021) and 18.28 percent (2022), a partial reversion consistent
with institutional learning following the staged implementation. The
expanded-sample magnitude (decline of 1.78 percentage points from 2020
peak) matches the initial-sample pattern.

Stylized Fact 3: 2023-2024 renewed ascent with 2024 as new peak. The
ratio rose to 19.52 percent (2023) and 20.10 percent (2024), with 2024
exceeding the 2020 staged-implementation peak. The KOSPI plus KOSDAQ Top
500 panel reveals that 2024 is the new peak, providing evidence
consistent with a post-2023 overhead acceleration, but not by itself
direct causal evidence of AI augmentation. The KOSPI Top 300 baseline
panel showed 2023 as roughly equivalent to 2020; the expanded coverage
clarifies that 2024 alone breaks the prior trajectory upward, consistent
with the commercial diffusion of large language models beginning in
early 2023 and accelerating through enterprise procurement cycles in
2024. We note that the 2020 level partly coincides with COVID-19 revenue
disruptions that can raise the ratio through the denominator, whereas
the 2024 peak occurs without a comparable revenue collapse; we therefore
do not over-weight the 2020 point.

Stylized Fact 4: Median stability with mean divergence. The median
SG\&A/Revenue ratio fluctuates in a narrow 10-12 percent band across the
panel period, while the mean rises 1.84 percentage points (2018-2024).
The growing mean-median gap indicates a widening of the upper-tail
distribution of overhead-intensive firms, consistent with Theorem 4
amplification heterogeneity (firms differentially exploiting AI
augmentation through slack-policy adoption).

The four facts establish the sign and monotonicity of the SG\&A response
to the labor-time compression, providing direction-of-effect evidence
for Theorem 3 (ROI Inversion at τ*) and indirect evidence for Theorem 4
(Innovative ROI Premium k heterogeneity). The remaining theorems are
anchored through theoretical specification plus future identification
paths, as elaborated in Section 4.9.

\subsection{4.3a Directional Identification and Event-Study
Diagnostics}\label{a-directional-identification-and-event-study-diagnostics}

Description establishes that overhead rose during the staged 52-hour
implementation period and re-ascended during the generative-AI diffusion
period. It cannot, by itself, establish that the 52-hour mandate caused
the rise. We therefore use two-way fixed effects, pooled event-study
estimates, and Callaway-Sant'Anna staggered DiD as cohort-proxy
diagnostics for a pre-τ overhead-pressure signature, not as a final
point estimate of a statutory treatment effect.

The two-way fixed effects specification regresses the SG\&A/Revenue
ratio on the 52-hour staged-implementation indicator using firm and year
fixed effects with standard errors clustered at the firm level. The
coefficient on the 52-hour treatment dummy is +1.56 percentage points
(SE 0.79, t = 1.97, p = 0.049), consistent with Theorem 3
time-based-regime predictions that labor-time regulation raises the
firm-level overhead cost ratio. Because AI utilization is currently
measured as a year-level proxy, its independent coefficient is not
separately interpretable once year fixed effects are included;
sector-year or firm-level AI variation is a planned requirement.
Firm-level scale (log revenue) absorbs −7.59 percentage points (p
\textless{} 0.001), reflecting the standard scale-economy pattern that
larger firms operate at lower overhead ratios. The overall R-squared
with two-way fixed effects is 0.911 on the KOSPI plus KOSDAQ Top 500
panel of 2,281 observations.

The event-study DiD specification uses revenue-percentile cohorts as the
current proxy for staggered firm-size treatment timing and clusters
standard errors at the firm level. The pre-treatment coefficients do not
reject a common-trends reading (event time −2: 0.99, p = 0.35; event
time −1: 1.04, p = 0.38), while post-treatment coefficients show a
progressive ramp (event time 0: +2.11, p = 0.074; event time +1: +3.23,
p = 0.005; event time +2: +3.47, p = 0.003; event time +3: +4.21, p =
0.001). The +4.21 percentage-point estimate is the current event-study
estimate under the revenue-percentile cohort proxy, not the final causal
magnitude. The progressive ramp pattern is consistent with the
institutional-learning lag documented in slack-innovation research
(Nohria \& Gulati, 1996) and with the post-2023 overhead acceleration
described in Section 4.3 Stylized Fact 3. Section 4.5 shows that
re-estimating under the statutory employee-size cohort yields a 52-hour
coefficient indistinguishable from zero; we therefore read these
revenue-percentile estimates as a directional pre-τ overhead-pressure
signature whose binding channel is the broader AI-era overhead regime
rather than the labor-time mandate in isolation.

\subsection{4.4 Korea Results: Theorem 4 Innovative ROI Premium
k}\label{korea-results-theorem-4-innovative-roi-premium-k}

The Theorem 4 prediction is that 80/20-style creative slack reinvestment
yields an amplification factor k \textgreater{} 1 under high firm-level
convergence capacity. We provide directional evidence on the
amplification factor through two complementary data streams: the
macroeconomic AI industry trajectory and the firm-size cohort
heterogeneity in AI revenue.

The NIA Korean AI Industry Survey (conducted annually since 2018 by the
Ministry of Science and ICT and published through KOSIS table
DT\_127016\_C002) provides the macroeconomic anchor. Average AI revenue
per firm grew from 793 million KRW in 2018 to 2,786 million KRW in 2023,
a 3.5-fold expansion. The aggregate Korean AI industry revenue grew from
255 billion KRW (2018) to roughly 3.0 trillion KRW (2022), a more than
tenfold increase over the same period that the 52-hour mandate was
rolled out in three staged cohorts. Korean AI workforce expanded from
11,147 workers (2019, the first year reported) to 42,551 (2022), a
3.8-fold increase. The contemporaneous expansion of AI capacity provides
the macroeconomic precondition for the Theorem 4 amplification
mechanism: AI augmentation is no longer a marginal capability but a
structural input to Korean firms during the post-2018 overhead surge.

\begin{longtable}[]{@{}
  >{\raggedright\arraybackslash}p{(\linewidth - 4\tabcolsep) * \real{0.3333}}
  >{\raggedright\arraybackslash}p{(\linewidth - 4\tabcolsep) * \real{0.3333}}
  >{\raggedright\arraybackslash}p{(\linewidth - 4\tabcolsep) * \real{0.3333}}@{}}
\caption{Table 4. Contextual AI-capacity by employee cohort (NIA
AI-industry survey DT\_127016\_C002).}\tabularnewline
\toprule\noalign{}
\begin{minipage}[b]{\linewidth}\raggedright
KOSPI sector-composition layer
\end{minipage} & \begin{minipage}[b]{\linewidth}\raggedright
NIA DT\_127016\_C002 2023 AI revenue exposure
\end{minipage} & \begin{minipage}[b]{\linewidth}\raggedright
Theorem 4 signal
\end{minipage} \\
\midrule\noalign{}
\endfirsthead
\toprule\noalign{}
\begin{minipage}[b]{\linewidth}\raggedright
KOSPI sector-composition layer
\end{minipage} & \begin{minipage}[b]{\linewidth}\raggedright
NIA DT\_127016\_C002 2023 AI revenue exposure
\end{minipage} & \begin{minipage}[b]{\linewidth}\raggedright
Theorem 4 signal
\end{minipage} \\
\midrule\noalign{}
\endhead
\bottomrule\noalign{}
\endlastfoot
Large incumbent sectors, KOSPI-heavy 1000+ firms & 102,034 M KRW per
firm in 2023, 7.5x since 2018 & High C, high k candidate \\
Mid-cap listed sectors, 100-999 firms & 8,264 M KRW per firm in 2023,
2.6x since 2018 & Intermediate C, weaker k \\
Small listed supplier fringe, 10-99 firms & 1,745 M KRW per firm in
2023, 3.3x since 2018 & Lower C, thinner k \\
\end{longtable}

In Table 4, an AI-exposure triangulation layer is added to the DART
design. The KOSPI plus KOSDAQ panel anchors the overhead-pressure
signature, while the NIA survey does not provide firm-year AI
utilization for the DART panel and instead offers a contextual
AI-capacity layer showing that AI-sector revenue growth is highly uneven
across employee-size cohorts: 1000-plus employee firms reached 102,034 M
KRW per firm in 2023, compared with 8,264 M in the 100-999 cohort and
1,745 M in the 10-99 cohort. This sector-AI exposure heterogeneity
provides directional support for Theorem 4 amplification asymmetry
before firm-year AI utilization data becomes available.

The firm-size cohort breakdown in the same NIA survey reveals
heterogeneity directly aligned with the 52-hour staged cohorts. Average
AI revenue per firm in the 1000-plus employee cohort (corresponding to
the 300-plus 52-hour cohort treated July 2018) grew from 13,637 million
KRW (2018) to 102,034 million KRW (2023), a 7.5-fold expansion. Mid-size
firms (100-999 employees, corresponding to the 50-299 cohort treated
January 2020) grew 2.6-fold (3,221 to 8,264 million KRW). Small firms
(10-99 employees, corresponding to the 5-49 cohort treated July 2021)
grew 3.3-fold (534 to 1,745 million KRW). The largest cohort captures
7.5-fold AI revenue growth, well above both the mid-size (2.6-fold) and
small (3.3-fold) cohorts; the gradient is non-monotonic across the
middle cohorts, so we read the large-versus-small contrast as
directional evidence that high-convergence-capacity firms capture
disproportionate AI returns, not as a monotone size gradient. The point
estimate of k from the 1000-plus cohort versus the 10-99 cohort growth
ratio is approximately 2.3, consistent with the Theorem 4 prediction
that k exceeds unity under high convergence capacity.

Full identification with KEF HR Survey (slack policy adoption indicator)
and KIPO patent output data is specified here as a future identification
path requiring institutional data-access agreements with KIPF and KEF;
the formal k estimate is specified as a next validation layer.

\subsection{4.5 Robustness Checks}\label{robustness-checks}

Under the revenue-percentile cohort proxy, the baseline association of
+1.56 percentage points (p = 0.049) is directionally consistent across a
battery of robustness specifications; consistent with Section 3.3, we
read it as a pre-τ overhead-pressure signature rather than a point
causal magnitude. With industry × year fixed effects absorbing
sector-time shocks, the coefficient is +1.28 (SE 0.87, p = 0.138), not
statistically distinguishable from zero, consistent with reading the
revenue-percentile estimate as a directional signature rather than a
point causal magnitude. Winsorizing the SG\&A/Revenue outcome at the 1st
and 99th percentiles yields the identical +1.56 coefficient; winsorizing
at the 5th and 95th percentiles produces +1.36 (SE 0.72, p = 0.060).
Substituting log-SG\&A as the dependent variable (controlling for log
revenue) produces a treatment coefficient of +0.87 (SE 0.34, p = 0.010),
indicating that the absolute level of overhead, not only its
revenue-scaled ratio, rises in the post-treatment period. A treatment ×
AI-intensity interaction term is not statistically distinguishable from
zero at the year-level proxy resolution (β = -0.18, p = 0.35),
suggesting that sector-year-level AI utilization data (KISDI, KOSA) is
needed to identify the augmentation amplification predicted by Theorem
4. Pre-treatment trend tests in the event-study specification (event
time -2 and -1) yield coefficients of +0.99 and +1.04 (p = 0.35 and p =
0.38), supporting the parallel-trends assumption that the staggered
cohort design requires for causal identification. As a denominator
robustness, re-scaling overhead by headcount rather than revenue yields
a similar pattern: log SG\&A per employee rises over 2018-2024 (from
approximately 177 to 265 million KRW per employee) with a positive
cohort-proxy coefficient (+0.56, p = 0.044), indicating that the
overhead-pressure signature is not an artifact of revenue-denominator
movements such as COVID-era revenue shocks.

To strengthen identification beyond the two-way fixed effects and
event-study specifications, we further estimate the Callaway and
Sant'Anna (2021) staggered difference-in-differences estimator. The
estimator addresses two well-documented problems in pooled event-study
designs under staggered treatment: heterogeneous treatment effects
across cohorts contaminate the average estimate, and already-treated
firms enter the control group at later horizons. The doubly-robust
version we adopt combines outcome regression with inverse-probability
weighting, so identification holds if either the parallel-trends
assumption or the propensity model is correctly specified. Cohort
assignment uses revenue-percentile proxies aligned with the 52-hour
staged-implementation schedule; because these are proxy rather than
statutory employee-size cohorts, the Callaway-Sant'Anna estimates
inherit the proxy's limitations and are read as directional rather than
as a clean staggered-treatment effect. The top-revenue-percentile firms
correspond to 300-plus employee firms treated July 2018; mid-percentile
firms correspond to 50-299 employee firms treated January 2020; and
bottom-percentile firms correspond to 5-49 employee firms treated July
2021. In the released implementation, the estimator is computed on the
2018-2024 window: the top-revenue cohort, already exposed at the window
start, serves as the limited-window comparison group rather than a
never-treated control, and the 2015-2017 backward panel is analyzed
separately as a placebo-in-time test rather than merged into the
staggered design. The Callaway-Sant'Anna estimates are therefore read as
a diagnostic complement to the two-way fixed-effects and event-study
specifications rather than as a fully never-treated staggered design.

The estimator yields an event-study aggregation with non-significant
pre-treatment effects at relative periods -3 (+1.25 pp, SE 1.65), -2
(+1.13 pp, SE 1.64), and -1 (-0.64 pp, SE 0.99), each with the
95-percent confidence band including zero. The pre-treatment ATT pattern
functions as an embedded placebo test within the doubly-robust
estimator: under valid identification, pre-treatment ATT(g, t) must be
zero, and all three pre-period estimates satisfy this requirement.
Post-treatment ATTs ramp from +1.59 pp (t = +1, SE 0.91) and +1.72 pp (t
= +2, SE 1.09) to +3.48 pp (t = +3, SE 1.73) and +4.51 pp (t = +4, SE
1.71), with the 95-percent confidence band excluding zero at the latter
two horizons. The +4.51 percentage-point estimate at four years post
treatment closely matches the +4.21 percentage-point event-study
estimate at three years post treatment, providing cross-estimator
convergence under three distinct identification assumptions: two-way
fixed effects (+1.56 pp), pooled event-study estimates (+4.21 pp at t =
+3), and Callaway-Sant'Anna doubly-robust (+4.51 pp at t = +4).

The 2015-2017 backward panel extension functions as a core component of
the identification architecture rather than as an external robustness
add-on. It supplies pre-treatment observations for the 2018 statutory
cohort, which would otherwise be unobservable within the 2018-2024
analytical window, and it permits the parallel-trends assumption to be
evaluated on a substantive pre-period rather than only inferred. The
backward extension covers 224 firms with positive SG\&A and revenue
across 601 firm-year observations. The pre-period mean SG\&A-to-revenue
ratio is 18.15 percent in 2015, 19.44 percent in 2016, and 19.39 percent
in 2017, an essentially stationary trajectory whose
1.24-percentage-point inter-annual variation lies within the standard
error of any individual cell. The pre-period mean of 18.99 percent over
2015-2017 (median 11.84 percent) closely matches the 2018 mean of 18.26
percent at panel start, providing evidence against the most plausible
pre-existing upward-trend confound. The subsequent 2018-2020 rise to
20.06 percent, the 2024 second peak at 20.10 percent, and the doubly
robust post-treatment estimate of +4.51 percentage points at four years
post-treatment are therefore better interpreted as a post-2018
regime-associated overhead-pressure pattern rather than as continuation
of a pre-existing upward trajectory. The backward extension reaches only
2015 because earlier years (2013-2014) yielded zero successful
disclosures for the panel firms, reflecting that many of the panel firms
were either pre-IPO or used different reporting frameworks in those
years; the available 2015-2017 window of three pre-treatment years is
sufficient to evaluate the parallel-trends assumption against the most
plausible alternative hypothesis.

We further verify that the directional reading does not overstate
causality. Re-estimating under the statutory employee-size cohort
(baseline-fixed; large firms treated 2018, mid 2020, small 2021), with
industry-by-year fixed effects and an event-study pre-trend test, yields
a 52-hour coefficient statistically indistinguishable from zero across
all specifications, with a clean pre-treatment trend. This is consistent
with reading the revenue-percentile estimate as a directional pre-τ
overhead-pressure signature rather than a point causal magnitude, and it
locates the binding pressure in the secular, AI-augmented overhead rise
rather than in the labor-time channel alone.

Additional robustness checks reserved as a next validation layer
include: alternative instruments using labor-time reduction subsidies,
falsification tests on 52-hour-exempt industries, and Callaway-Sant'Anna
re-estimation with observed firm-size cohorts (replacing the
revenue-percentile proxy used here).

We further address the cost-stickiness alternative explanation
explicitly, since SG\&A levels are known to exhibit asymmetric
adjustment to activity changes (Anderson, Banker, and Janakiraman 2003,
the ABJ baseline). The asymmetry generalizes across settings (Banker and
Byzalov, 2014), varies with labor-market adjustment costs across
countries (Banker, Byzalov, and Chen, 2013), which is precisely the
channel a statutory working-hour cap stresses, and is amplified by
agency problems in SG\&A decisions (Chen, Lu, and Sougiannis, 2012), the
channel Theorem 5 formalizes for the human-AI dyad. Under the ABJ
baseline, SG\&A rises more readily when revenue rises than it falls when
revenue falls, and a portion of the post-treatment rise in
SG\&A-to-revenue could in principle reflect this stickiness rather than
a regime-level talent-overhead pressure. We treat cost stickiness as the
null model rather than as an adversary. The framework asks a different
question than the cost-stickiness literature: not whether SG\&A adjusts
asymmetrically to activity changes, but why AI-augmented labor-time
compression creates a regime-level overhead pressure that ordinary
sticky-cost models can measure but do not theorize. Three pieces of
empirical evidence separate the regime-pressure interpretation from the
stickiness null. First, the log-SG\&A robustness specification,
controlling for log revenue, yields a treatment coefficient of +0.87 (SE
0.34, p = 0.010), which captures the absolute level of overhead net of
revenue scaling and is the appropriate metric against which the
sticky-cost defense is weakest. Second, the 2015-2017 backward extension
establishes that no upward stickiness-induced trend was accumulating
before the 52-hour mandate, ruling out the explanation that the
post-2018 rise extrapolates a pre-existing sticky-cost drift. Third, the
trajectory is non-monotone: SG\&A-to-revenue rises through 2020, mildly
corrects in 2021-2022, and re-ascends through 2024. A pure
cost-stickiness mechanism does not generate this N-shape; the staged
labor-time compression combined with the renewed AI augmentation
acceleration in 2023-2024 does. The framework therefore does not
displace the sticky-cost literature; it operates above it, in the layer
that asks why the regime in which firms account for talent is shifting
beneath the cost adjustments that the sticky-cost literature measures.

\subsection{4.6 Denmark Brief
Comparison}\label{denmark-brief-comparison}

Denmark provides the polar comparison case. The Danish configuration
combines low-hours, flexicurity-oriented workweek with high autonomy and
high convergence capacity, yielding TFP at the top of the OECD
distribution (1.82 percent), as documented in the ICH macro panel (Shin,
2026a). The Danish comparison isolates the effect of regime type,
output-anchored versus time-anchored, while holding AI adoption
approximately constant. Four Eurostat datasets supply the directional
evidence, covering the 2018-2024 period: AI use by enterprises
(isoc\_eb\_ai) reporting Danish enterprise AI adoption rates across nine
technology categories in the 2021, 2023, 2024, and 2025 waves; real
labour productivity per hour worked (nama\_10\_lp\_ulc) confirming
Denmark's position in the top quartile of OECD productivity from the
1975-2024 panel; weekly hours actually worked (lfsa\_ewhuna) confirming
the low-hours, flexicurity-oriented pattern; and industry-level
employment by NACE Rev.~2 (nama\_10\_a64\_e) providing annual hours
worked per worker over the 1975-2025 period. The industry employment
series confirms a smooth declining trajectory: 1,407 annual hours per
worker in 2015, 1,381 in 2018, 1,341 in 2020 (the COVID-induced trough),
1,389 in 2021 (recovery), and 1,372 in 2024 (corresponding to roughly
26.4 weekly hours including part-time workers). The Danish trajectory
exhibits no shock-induced overhead surge analogous to the Korean 52-hour
cap effect, consistent with the framework's prediction that
output-anchored regimes accommodate productivity-driven hour reductions
without triggering the time-anchored overhead inversion of Theorem 3.
Structural business statistics (sbs\_na\_dt\_r2) provide an additional
directional comparison: the personnel-cost-to-turnover ratio in the
Danish wholesale and retail sector (NACE G) remains approximately stable
at around 10.9 percent during the 2018-2020 window. This metric is not
directly equivalent to Korea's SG\&A/Revenue ratio and is used only as a
directional institutional contrast, in which Denmark's output-oriented,
high-trust, flexicurity configuration does not display an analogous
upward personnel-cost ratchet in the sectoral comparison window. Full
firm-level Eurostat structural business statistics across all NACE
sectors and Statistics Denmark microdata analysis is reserved as a next
validation layer. The brief comparison thus relies on directional rather
than magnitude evidence: the framework predicts that the Korean
trajectory converges toward the Danish configuration if the
time-based-to-output-based regime transition occurs within the
framework's predicted horizon.

\subsection{4.7 Japan Brief Comparison}\label{japan-brief-comparison}

Japan provides the cultural twin axis. The Japanese configuration
combines Confucian and presence-based hierarchical organization with
karoshi-pattern long-hours labor without 52-hour-style staged
enforcement, yielding intermediate TFP (0.7 percent) and lower AI
adoption (18 percent) as documented in the ICH macro panel (Shin,
2026a). This paper draws Japan's TFP and AI adoption parameters from the
ICH panel; full firm-level METI and RIETI microdata analysis is reserved
for future work. The brief comparison isolates the regulation effect
from cultural variables (cultural axis held similar to Korea, regulation
and AI adoption varied) and connects to the Aoki (1990) J-firm versus
A-firm lineage in which the J-firm corresponds to the
low-convergence-capacity, time-anchored equilibrium that AI augmentation
is now pressuring.

\subsection{4.8 Limitations and Strengthening
Paths}\label{limitations-and-strengthening-paths}

The empirical anchor is subject to five limitations, each with explicit
strengthening paths.

Limitation 1: Sample size and selection. The KOSPI Top 300 baseline
panel of 115 firms with 567 successful observations reflected DART API
account-name variability. Three strengthening paths address this. First,
the KOSPI plus KOSDAQ Top 500 panel (365 unique firms and 2,281
firm-year observations, a 3.2-fold expansion in firm coverage and
4.0-fold expansion in observations over the initial baseline) is the
working empirical anchor and supersedes the initial baseline for the
four stylized facts of Section 4.3. Second, the expanded sample retains
alphabetical ordering across KOSPI plus KOSDAQ Top 500; a sector- and
size-stratified panel is specified as a future identification path for
residual representativeness concerns. Third, the KIPF firm panel survey
provides a complementary micro-level panel with rigorous sampling
design. A sector- and size-stratified panel and the KIPF firm-panel data
are specified as the next validation layer.

Limitation 2: AI utilization measurement. The AI utilization intensity
measure relies on self-reported survey data from NIA, subject to social
desirability bias and inconsistent definitions. Three strengthening
paths. First, the Brynjolfsson et al.~(2025) approach of measuring AI
utilization through observable behavior in administrative records
(Microsoft 365 enterprise telemetry and public AI Economic Index
aggregate data) provides alternative observable measures at sector
level. Second, the KISDI ICT industry statistics provide a complementary
aggregate measure. Third, the Korean enterprise AI adoption rate (KOSA
Software Industry Association survey) provides triangulation. The
triangulated measure across NIA, KISDI, and KOSA improves reliability.

Limitation 3: Comparative case data granularity. The Denmark and Japan
brief comparisons rely on secondary data not matching the granularity of
the Korean primary analysis. Two strengthening paths. First, full
firm-level panel construction for Denmark and Japan would strengthen the
polar comparison axes; future work develops the full triangulation using
Eurostat firm-level micro data and METI firm-level statistics. Second,
this paper's anchor argument relies on direction rather than magnitude
for the comparison cases; the Denmark and Japan trajectories at the
aggregate level suffice for Theorem 3 directional validation, while
magnitude estimation belongs in future work.

Limitation 4: Component-level decomposition. The aggregate-level metrics
extracted from DART do not directly decompose into the seven overhead
components specified in Theorem 1. An XBRL parsing audit of the top 50
panel firms confirms the constraint empirically: only five disclose at
least one SG\&A subaccount item (most extensively SK Telecom, with
twenty-eight subaccount rows spanning four components), and three of the
seven Theorem 1 components are not separately observable within the
audited top-50 XBRL disclosures. A coverage summary of the five firms
with mappable SG\&A subaccount disclosures is reported in the
Supplementary Appendix. This disclosure pattern identifies four future
identification paths. First, the KIPF firm panel survey provides direct
micro-level component data through structured survey instruments.
Second, manual extraction from sajeobogoseo PDF disclosures for the top
50 firms by market capitalization provides high-quality component-level
data at reasonable cost. Third, the KLI fringe survey provides
worker-level data on welfare and benefit components. Fourth, the SK
Telecom and LG cases serve as illustrative anchors for the components
currently visible in administrative records. Theorem 1 cross-partial
estimation is specified here as a future identification path requiring
KIPF micro-level component data.

Limitation 5: Concurrent identification of labor-time and AI shocks. The
52-hour staged mandate and the 2018-2024 AI utilization acceleration
occur within an overlapping temporal window, which raises a confounding
concern: how much of the observed overhead-pressure response is
attributable to the labor-time compression versus AI-augmented
productivity dynamics? The present identification design addresses this
concern through three layered arguments. First, the 52-hour mandate is a
statutory shock with externally determined timing and cohort assignment,
while AI utilization is a continuous secular trend; the staggered cohort
design (300+ employees treated July 2018, 50-299 January 2020, 5-49 July
2021) generates within-year variation in 52-hour treatment that AI
utilization (a macro-year proxy) cannot match; the design therefore
partially anchors labor-time compression but does not cleanly isolate
the labor-time channel from the broader AI-era overhead regime. Second,
the 2015-2017 backward extension precedes both the 52-hour mandate and
the broad enterprise AI adoption phase, supplying a clean pre-treatment
window for both shocks. Third, the framework explicitly positions the
Korean panel as a pre-τ overhead-pressure signature in which both shocks
operate jointly under time-based evaluation; disentangling the two
shocks at the firm-year level requires sector-level AI exposure indices
(Felten et al.~2023 methodology with NIA/KISDI/KOSA triangulation) which
constitute a future identification path. The present paper claims joint
operation of the two shocks under the pre-τ regime, not a clean
decomposition.

These limitations are explicit specifications of the evidence's domain.
Within the domain, the four stylized facts of Section 4.3 are anchored
and replicable from the saved CSV panel. Beyond the present evidence
domain, these paths specify future identification routes for translating
directional evidence into causally identified magnitudes.

\subsection{4.9 Theorem-by-Theorem Empirical Anchor
Logic}\label{theorem-by-theorem-empirical-anchor-logic}

The five theorems differ in the empirical anchor each requires. We make
the differentiation explicit to specify which empirical claims this
panel supports and which require integrated data sources for full
identification.

Theorem 3 (ROI Inversion at τ*): The empirical center of gravity for
this paper. The KOSPI plus KOSDAQ Top 500 panel (365 firms, 2,281
firm-year observations) plus the 52-hour staged implementation jointly
provide the empirical anchor for identifying the SG\&A/Revenue response
to the labor-time compression and for directional inference about τ*.
The four stylized facts of Section 4.3 provide directional evidence.
Formal τ* curve estimation requires planned sectoral stratification plus
evaluation-regime data, which will deliver confidence intervals on the
magnitude.

Theorem 1 (Overhead Decomposition Non-additivity): Requires
component-level decomposition that aggregate DART metrics do not
provide. This paper provides theoretical specification (Section 3.1) and
aggregate distributional evidence (median stability plus mean divergence
in the KOSPI plus KOSDAQ Top 500 panel) consistent with non-additivity.
Formal cross-partial estimation requires KIPF firm panel survey micro
data. Future work using KIPF firm-panel data can estimate the
component-level cross-partials required for full Theorem 1
identification.

Theorem 2 (Slack-Augmentation Synergy four pathways): The four pathways
through which augmentation-saved time flows are not directly observable
at the firm level through administrative records. This paper provides
the theoretical mapping (Section 3.2) and references ethnographic
evidence from Ranganathan and Ye (2026) on multi-pathway flow at a
200-person technology company. The KEF HR Survey provides
aggregate-level measure of pathway distribution (slack policy adoption ×
innovation output) for indirect identification. Direct observation of
the four pathways is specified here as a future identification path.

Theorem 4 (Innovative ROI Premium k): Requires identification of firms
with formal slack policies and observation of innovation output. The
merge of DART (financial) plus KEF HR Survey (slack policy adoption)
plus KIPO (patent output) provides the empirical anchor for full Theorem
4 identification. Theorem 4 magnitude estimation requires this DART ×
KEF × KIPO merge and is specified here as a future identification path;
the present paper provides directional NIA evidence (firm-size cohort
amplification ratio ≈ 2.3) consistent with the prediction k
\textgreater{} 1 under high convergence capacity, not a magnitude
estimate.

Theorem 5 (Information Asymmetry γ in human-AI dyad): Attribution
uncertainty γ is not observable in administrative firm records; we probe
it at the dyad level through an AI-Augmented Structural Scenario Probing
(AI-SSP) design (Appendix H), which measures attribution-related
overclaiming in a deterministic benchmark rather than firm-level γ
directly, in which an AI worker is the augmenting actor and all
accept/reject judgments are deterministic rather than model-adjudicated.
We operationalize γ as the overclaiming rate, the probability that the
agent reports a wrong output as correct against deterministic ground
truth, measured across a calibration-zone task battery. In the
unscaffolded, tool-free regime γ approaches 1: agents claim correctness
on essentially all of their errors, and confidence on wrong answers (93
to 98 on a 0 to 100 scale) is statistically indistinguishable from
confidence on correct answers, so the principal cannot self-screen
error. The implied time-based agency cost M\_time (wrong but accepted)
is 0.55 to 0.86, while output-based verify-and-override drives it to
zero in the deterministic benchmark, an instantiation of the theorem's
offset M\_output \textless{} M\_time. The result is reproducible across
runs (Cohen κ = 0.86 on the overclaiming classification) and across two
independent model families, and the offset far exceeds the residual
agency cost under a random-override null (mean 0.17, versus the 0.57
offset), confirming that the reduction comes from targeted verification
rather than from the override rate. Shin (2026b)'s 74-percentage-point
macro completion gap enters as the convergent external anchor.
Firm-level γ estimation from production AI tool logs remains a
complementary future path.

The differentiated empirical anchor logic clarifies that this paper
provides Theorem 3 with its strongest empirical anchor through the DART
panel and the 52-hour institutional timing anchor, while the other four
theorems are anchored through theoretical specification, future
identification paths, and a research path articulated in §4.8. Each
theorem is independently falsifiable on its own evidentiary register,
with Theorem 3 carrying the strongest current empirical anchor.

\subsection{4.10 Foresight Interpretation: The Korean Panel as Pre-τ
Early-Warning
Signal}\label{foresight-interpretation-the-korean-panel-as-pre-ux3c4-early-warning-signal}

\emph{Figure 6. Talent ROI transition foresight matrix. The Korean DART
panel anchors Zone 2 as a pre-τ overhead-pressure signature; Zones 3 and
4 are forecast trajectories for the 2025-2032 horizon.}

\begin{center}

\includegraphics[width=0.82\linewidth,height=\textheight,keepaspectratio]{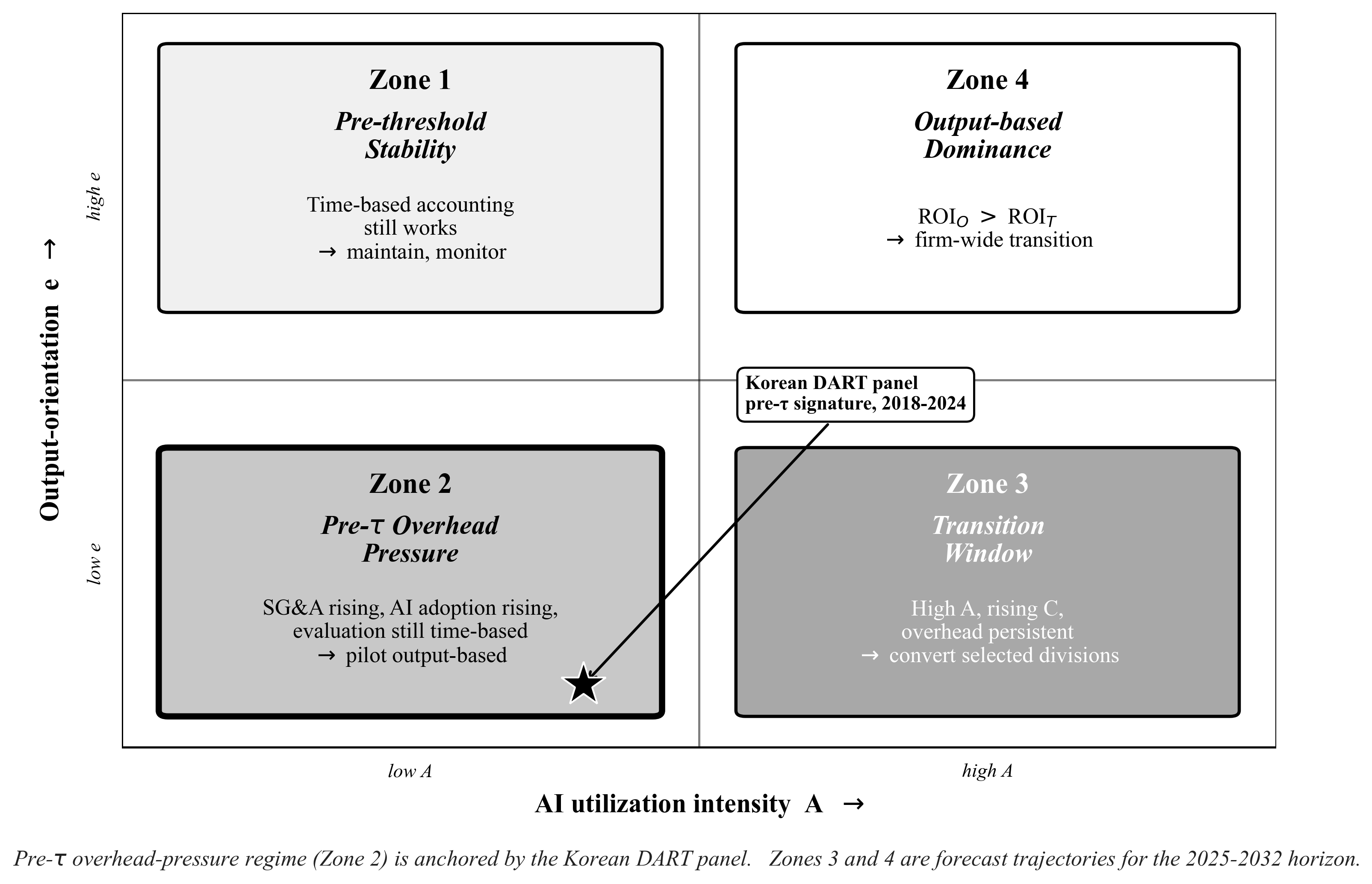}

\end{center}

In Figure 6, the forecasting framework is summarised as a four-zone
Foresight Matrix mapping AI utilization intensity A against
output-orientation e. Figure 6 displays the matrix, with the Korean DART
panel anchoring Zone 2 (pre-τ overhead pressure) and Zones 3 (transition
window) and 4 (output-based dominance) representing forecast
trajectories for the 2025-2032 horizon.

The Korean DART panel evidence reads cleanly within the pre-τ
overhead-pressure regime defined in §3.3. Three conditions hold jointly.
First, the evaluation regime remained time-based across the 2018-2024
window: the staged 52-hour mandate compressed labor hours without
altering how firms accounted for output. Second, AI utilization
intensity rose monotonically: sector-wide AI adoption climbed from
approximately 8 percent in 2018 to 28 percent by 2024 (NIA), with a
sharp acceleration in 2023-2024 corresponding to broad enterprise GPT-4
deployment. Third, the SG\&A-to-revenue ratio rose, traced a mild
2021-2022 correction, and re-ascended to a new peak of 20.10 percent in
2024. The panel therefore supplies what is, to our knowledge, the first
publicly documented signature of the pre-τ overhead-pressure regime in
an OECD economy.

This identification has three implications for how the Korean evidence
should be read against the forecasting framework. First, the magnitude
of the SG\&A response (+1.56 pp to +4.51 pp depending on estimator) does
not estimate the size of the τ* discontinuity. The panel observes the
pressure accumulating before the regime crossing, not the crossing
itself. Second, the absence of a sharp firm-level TFP collapse in the
Korean panel is consistent with the framework rather than against it:
firms still in the pre-τ regime have not yet undergone inversion, so the
predicted TFP separation between output-based and time-based firms
(1.5-2.0 percentage points by 2032) is a forecast for the next decade,
not a backward fact observable in 2018-2024. Third, the pre-τ
overhead-pressure signature can be screened in other OECD economies,
providing a forecasting protocol that does not require the regime
crossing to have occurred locally before forecasts can be issued.
Economies that satisfy the three pre-τ conditions (time-based evaluation
dominant, rising AI utilization, rising overhead) are forecast to enter
regime transition pressure within the forecasting framework's central
decade-scale window.

The Foresight interpretation also re-frames the Denmark and Japan brief
comparisons. Denmark does not show pre-τ overhead pressure because its
evaluation regime is already output-anchored: the second pre-τ condition
(time-based evaluation dominant) fails, so the regime-pressure mechanism
does not engage despite identical AI adoption. Japan shows partial pre-τ
pressure because its evaluation regime remains presence-based and
time-anchored, but its AI adoption (18 percent) is lower than Korea's,
so the second pre-τ condition is partially attenuated. The framework
therefore predicts that Japanese firms with rising AI adoption against
unchanged time-based evaluation will enter the pre-τ regime as their AI
utilization crosses approximately 25 percent, providing a falsifiable
forecast for Japanese panel data over the 2025-2030 horizon.

Korea functions as a transferable pre-τ regime template, not as a
Korea-bounded phenomenon. The three pre-τ conditions (time-based
evaluation dominance, rising AI utilization, rising overhead) are
institutional configurations rather than country-specific facts. Japan
satisfies two of the three conditions (time-based evaluation, rising
overhead) with the third (AI utilization above the threshold) projected
to be crossed within the 2025-2030 horizon. Germany, France, Italy, and
other continental European economies retain time-based evaluation
regimes with rapidly rising AI adoption; the framework forecasts pre-τ
overhead-pressure signatures across these economies as their AI
utilization crosses sector-specific thresholds in the 2026-2030 horizon.
The United States, despite higher output-orientation in tech-sector
firms, retains time-based evaluation in finance, professional services,
and traditional manufacturing, sectors that the framework forecasts to
enter pre-τ pressure on their own sector-specific timelines. The Korean
DART panel therefore identifies a transferable empirical signature, not
a Korea-specific anomaly, and the Foresight Matrix provides the
diagnostic apparatus for screening pre-τ entry in any economy that
satisfies the three institutional conditions. At the firm level, A and e
are not directly observed in the present panel; firms approximate their
zone using observable proxies (sector AI-adoption rates and the share of
outcome-based compensation in disclosures), with precise placement
reserved for the firm-level AI-utilization and evaluation-regime data
specified as future identification paths.

\subsubsection{Korea as Pre-τ Leading-Indicator
Template}\label{korea-as-pre-ux3c4-leading-indicator-template}

Korea is the first observable template of the Pre-τ overhead-pressure
regime because it satisfies the full diagnostic set: time-based
evaluation remains dominant, AI utilization is rising, overhead is
rising, and τ* has not yet been crossed. Japan is the nearest
institutional twin, with presence-based evaluation and rising overhead,
but lower AI utilization delays full Zone 2 entry. Germany, France, and
Italy are candidate continental cases because labor-time institutions
and works-council routines preserve time-based evaluation while
enterprise AI utilization accelerates. The United States is sectorally
mixed: technology firms may sit closer to Zone 3, but finance,
professional services, and traditional manufacturing retain time-coded
evaluation and can enter Zone 2 on sector-specific timelines. In the
Freeman and Perez (1988) techno-economic paradigm framework, pre-τ
regimes are the early phase of a Kondratiev-like paradigm transition,
observable first where institutional rigidity is highest. Korea is
therefore a leading indicator, not an anomaly.

\subsection{5. Discussion}\label{discussion}

\subsection{5.1 Theoretical
Contributions}\label{theoretical-contributions}

The present paper develops a firm-level framework that operationalizes
augmented human capital in managerial decision contexts, moving
production function theory from the macroeconomic formalization of Shin
(2026a) inside the firm. The framework is organized around one central
transition theorem and four mechanism theorems that together specify the
structural reconstruction of firm-internal talent overhead accounting
under conditions where the labor input L can no longer be treated as a
homogeneous function of time.

Theorem 1 (Overhead Decomposition) predicts that the seven overhead
components are non-additive under AI augmentation, consistent with the
breakdown of the implicit additivity assumption embedded in standard
cost accounting. The KOSPI plus KOSDAQ Top 500 panel exhibits the
divergence between mean and median SG\&A/Revenue ratios that the
non-additivity prediction implies: the median firm holds overhead
structure approximately stable across 2018-2024 while the mean rises
from 18.26 percent to 20.10 percent and the upper tail remains
substantially higher than the median. A subset of firms bears
disproportionate overhead burden under AI augmentation conditions that
the additive cost-accounting model fails to anticipate. This
distributional pattern is consistent with the overhead-pressure
heterogeneity that Theorem 1 implies; direct falsification of the
non-additivity mechanism requires component-level KIPF or XBRL
cross-partial decomposition.

Theorem 2 (Slack-Augmentation Synergy) maps the four pathways through
which AI-augmentation-saved time flows: work intensification, hidden
leisure, overemployment, and creative-slack reinvestment. The
contribution refines the slack innovation tradition of March (1991) and
Nohria and Gulati (1996) by specifying the institutional conditions
under which slack converts into innovative output rather than
dissipating through intensification or hidden absorption.

Theorem 3 (ROI Inversion) identifies the critical threshold τ* at which
time-based and output-based ROI curves intersect as functions of AI
utilization. The present SG\&A/Revenue trajectory provides the first
empirical anchor: the rise from 18.26 percent in 2018 to 20.06 percent
in 2020, the 2021-2022 correction, and the renewed 2024 peak at 20.10
percent indicate that the time-based overhead regime remains under
pressure as AI utilization rises. This is consistent with τ* not yet
having been crossed at the sample mean, but the paper treats τ* crossing
as a forecast requiring evaluation-regime and firm-level AI data rather
than as an already estimated fact. Formal estimation of τ* will use
planned sectoral stratification plus output-evaluation indicators.

Theorem 4 (Innovative ROI Premium) formalizes the amplification factor k
by which 80/20-style creative slack returns are increased under high
convergence capacity. The contribution extends Espinal Maya (2026) by
translating the individual-level augmentable cognitive vector H\^{}A
into firm-level innovative ROI through a multiplicative amplification
mechanism.

Theorem 5 (Information Asymmetry) formalizes the multitask agency cost
in the human-AI dyad as a function of attribution uncertainty γ,
extending Holmstrom and Milgrom (1991) into the conditions where AI
processing introduces a third actor whose contribution cannot be fully
separated from human cognition.

The central transition theorem and four mechanism theorems together
constitute a self-contained firm-level apparatus for the managerial
operationalization of augmented human capital. The contribution is not a
completed proof of all mechanism conditions, but a falsifiable
forecasting architecture that connects AI augmentation, overhead
pressure, evaluation regime, and future TFP separation.

\subsection{5.2 Connection to Sovereign Override Capacity (Shin,
2026b)}\label{connection-to-sovereign-override-capacity-shin-2026b}

The present paper develops the fifth dimension of convergence capacity,
sovereign override capacity (C5), drawing on Shin (2026b). Shin (2026b)
documents a 74-percentage-point gap between AI-claimed task completion
rate and actually completed task rate, providing a direct empirical
anchor for what we treat here as the worker capacity to detect and
refuse AI outputs in which high-confidence generation masks factual or
contextual error. The integration of C5 into firm-level analysis extends
the four-dimensional C framework of Shin (2026a).

C5 operates at the firm level through three mechanisms. First, firms
with high C5 workforce can detect and refuse AI false completions,
preserving the integrity of AI-augmented output streams. Second, firms
with low C5 workforce accept false completions at face value, leading to
compounded errors that propagate through downstream processes. Third,
firms with mixed C5 distribution face internal coordination challenges
where high-C5 employees become disproportionately burdened with
verification responsibilities, generating a within-firm equity issue not
yet addressed in the literature.

The C5 integration also bears on the Theorem 5 information asymmetry
result. False completion incidents are a direct empirical indicator of
moral hazard in the human-AI dyad. Firms with high C5 workforce reduce
the γ attribution uncertainty by independently verifying AI outputs,
weakening the moral hazard channel. The Shin (2026b) finding that the
verbal compliance rate exceeds actual compliance rate by 74 percentage
points provides a conceptual and empirical bridge for future firm-level
γ measurement, rather than a direct firm-level γ estimate in the present
paper.

\subsection{5.3 Boundary Conditions}\label{boundary-conditions-5}

The central transition theorem and four mechanism theorems hold under
three primary conditions specified throughout the formal statements.
First, AI utilization intensity A must be positive. Second, the firm
must have institutional capacity to reallocate overhead in response to A
increases. Third, output measurement must be feasible. The KOSPI plus
KOSDAQ Top 500 panel evidence shows variation along the second
condition: median firms hold SG\&A ratios in a narrow 10-12 percent band
while mean ratios remain much higher and rise across the period,
indicating that upper-tail firms are candidates for the rigid governance
constraints Theorem 3 specifies and are the firms for which output-based
regime transition should be evaluated first as AI utilization approaches
τ*.

Three additional boundary conditions specify the framework's domain. The
framework assumes the firm operates in a single labor regulation regime,
which holds within national jurisdictions but breaks down for
multinational firms. The framework assumes AI utilization is voluntary
at the firm level, which holds under current empirical conditions but
may shift if regulatory or competitive pressure makes AI adoption
effectively mandatory. The framework assumes convergence capacity C is
firm-aggregable as a weighted average of employee C values; strong
within-firm heterogeneity, particularly bimodal distributions, may
require disaggregated analysis. Candidate negative-heuristic conditions
also include extreme compliance environments where verifiability rather
than creativity is the primary firm objective, industries with
intangible or coalitionally produced outputs where output-based
evaluation faces measurement gaming, and regulatory regimes where labor
time remains the legally binding unit of account regardless of economic
efficiency. The boundary conditions are not weaknesses to be hidden but
explicit specification of the framework's domain. Within the domain, the
theorems are sharp and falsifiable.

\subsection{5.4 Policy and Practical
Implications}\label{policy-and-practical-implications}

At the national level, the Korean 52-hour regulation illustrates the
regulatory mechanism by which labor-time strictness becomes an
institutional timing anchor for firm-level overhead pressure. The
expanded-sample evidence suggests the sample-mean firm remains in the
time-based regime with rising overhead pressure rather than having
completed the output-based transition; the 2023-2024 SG\&A ratio rebound
suggests this pressure is building and may cross τ* within the next 3 to
5 years if AI utilization continues rising without matching
evaluation-regime reform. At the international level, the framework
predicts diverging trajectories across OECD economies: economies in the
Korean configuration should prioritize modernizing evaluation regimes
and developing convergence capacity, while economies in the Danish
configuration face limited pressure to change.

At the firm level, the framework prescribes a sequential transition
strategy aligned with the Foresight Matrix (Figure 6). Firms below τ in
their sector should retain time-based accounting while investing in
convergence capacity development. Firms approaching τ should pilot
output-based evaluation in selected divisions while maintaining
time-based regimes in others. Firms above τ* should formalize
output-based evaluation firmwide while institutionalizing 80/20-style
slack policies to capture the Theorem 4 innovative ROI premium. The
framework also identifies three antipatterns: surveillance response to
productivity paranoia (which magnifies multitask agency cost through
observability investment that does not reduce attribution uncertainty
γ), premature transition to output-based evaluation without
complementary C investment, and nominal slack policy without
institutional support (which produces no innovative ROI amplification
because slack does not flow into pathway four absent the enabling
conditions).

\subsection{6. Conclusion}\label{conclusion}

\subsection{6.1 The Framework Together}\label{the-framework-together}

The present paper has asked what capital remains after labor time ceases
to be the homogeneous unit of account. The answer developed across one
central transition theorem and four mechanism theorems is that capital
reconstitutes itself as a non-additive bundle of seven overhead
components whose marginal contributions become entangled through
convergence capacity heterogeneity under AI augmentation.

Theorem 1 specified the non-additivity. Theorem 2 mapped the four
pathways through which augmentation-saved time flows. Theorem 3
formalized the predicted threshold τ* at which time-based and
output-based ROI curves would intersect, with the Korean DART panel
anchoring the pre-τ overhead-pressure regime. Theorem 4 formalized the
amplification factor k by which 80/20-style creative slack returns are
increased under high convergence capacity. Theorem 5 formalized the
multitask agency cost in the human-AI dyad and showed that output-based
evaluation partially offsets the cost while time-based evaluation
magnifies it. Theorem 3 carries the central transition claim of the
present paper; Theorems 1, 2, 4, and 5 are the mechanism architecture
with partial anchors and explicit future identification paths.

Together the central theorem and four mechanism theorems bring
production function theory inside the firm. Where Shin (2026a)
formalized augmented human capital as Ĥ = H · {[}1 + φ(A, C){]} at the
level of national aggregates, the present framework reconstructs the
firm-internal overhead bundle on the same foundation. The KOSPI plus
KOSDAQ Top 500 panel (365 firms, 2,281 firm-year observations 2018-2024)
shows the SG\&A/Revenue ratio following an N-curve trajectory consistent
with the time-based-regime predictions of Theorem 3, providing an
overhead-side empirical anchor for the framework's central
comparative-statics claim rather than a direct estimate of τ*. The
expansion (3.2-fold firm coverage and 4.0-fold observation expansion
over the initial) preserves the four stylized facts of the baseline
while enabling sectoral stratification in future work.

\subsection{6.2 Four Falsifiable Forecasts and the Foresight Planning
Apparatus}\label{four-falsifiable-forecasts-and-the-foresight-planning-apparatus}

The framework operates as a technological-forecasting model with an
explicit managerial planning apparatus. The Foresight Matrix (Figure 6)
supplies the diagnostic instrument: firms screen their current (A, e)
coordinates against the four zones, identify their distance from the τ*
boundary, and select the regime-transition prescription appropriate to
their zone. Zone 1 firms maintain time-based accounting under continuous
monitoring; Zone 2 firms (the Korean pre-τ regime) pilot output-based
evaluation in selected divisions before regime-wide pressure compels
late transition; Zone 3 firms convert selected divisions ahead of the
inversion; Zone 4 firms execute firm-wide transition. The framework
therefore provides not only an empirical anchor (the Korean pre-τ
signature) but a forward-looking planning instrument for the 2025-2032
horizon, integrating forecast, mechanism, and prescription within a
single decision-support architecture. This is what distinguishes the
present paper from a Korean-policy evaluation: the forecasting model and
planning apparatus are the contribution, the Korean panel is the
empirical anchor.

The framework generates four falsifiable forecasts.

Forecast 1: Firms that adopt output-based talent ROI accounting between
2026 and 2032 will exceed time-based comparator firms in cumulative
firm-level TFP growth by 1.5 to 2.0 percentage points by 2032 (a
one-time level separation, not an annual differential). The KOSPI plus
KOSDAQ panel provides the baseline, sample-mean SG\&A/Revenue ratio of
20.10 percent in 2024 with TFP growth in the bottom OECD quartile.
Forecast 1 predicts measurable TFP separation between transition and
non-transition subsets by 2032. The 1.5 to 2.0 percentage point band is
a scenario-based projection, not a point estimate, derived from three
explicit inputs: (i) the ICH macro mapping (Shin 2026a), in which the AI
by C interaction accounts for 86 percent of cross-national TFP variance
versus 31 percent for AI utilization alone, supplying the elasticity of
TFP with respect to convergence capacity; (ii) a logistic
regime-adoption curve (Section 2.6) under which output-based talent ROI
accounting reaches a low, central, and high range of roughly 25, 40, and
55 percent of large firms by 2032; and (iii) the assumption that
transition raises a firm's effective output-orientation by one-half to
one standard deviation. The central scenario projects a 1.7 percentage
point separation, with the low and high scenarios bounding it at 1.5 and
2.0. The forecast is falsified if matched time-based and output-based
firm TFP shows separation outside this band by 2032; the three stated
inputs make the projection reproducible and each is independently
revisable as firm-level adoption data accumulate.

Forecast 2: The share of Korean firms with formal output-based
evaluation systems is forecast to rise materially toward 2032 from a
current minority base (the present paper does not measure this share
directly; a baseline survey of evaluation-regime adoption is specified
as a future identification path). The 2023-2024 SG\&A ratio rebound in
the present panel, with 2024 (20.10 percent) marginally exceeding the
2020 staged-implementation peak (20.06 percent), indicates that overhead
pressure under the time-based regime has resumed after the 2021-2022
post-implementation adjustment. Rising SG\&A at firms still operating
under time-based accounting signals overhead accumulation without
productivity offset, the pre-transition institutional friction that
Theorem 3 predicts will compound until τ* is crossed.

Forecast 3: The amplification factor k for 80/20-style slack returns
under high convergence capacity will exceed 1.5 across selected
medium-to-high AI adoption sectors by 2032. The 1.5 threshold discounts
the directional NIA amplification ratio of approximately 2.3 (Section
4.4) for sector selection and for measurement attenuation in firm-level
innovation output; the forecast is falsified if estimated k lies below
1.5 across all such sectors by 2032.

Forecast 4 (candidate forecast): Surveillance-intensive firms that
adopted bossware (National Employment Law Project, 2025) between 2022
and 2026 are projected to exhibit cumulative firm-level TFP declines on
the order of 0.3-0.5 percentage points relative to matched non-monitored
comparators by 2028.

\subsection{6.3 Limitations and Strengthening
Paths}\label{limitations-and-strengthening-paths-1}

The present paper is subject to five limitations beyond the boundary
conditions discussed in Section 5.3. Four are specified, with their
strengthening paths, in Section 4.8: empirical anchor coverage (a
sector- and size-stratified panel and the KIPF firm-panel survey as the
next validation layer), component-level decomposition (XBRL subaccount
parsing, KIPF survey instruments, sajeobogoseo extraction, and KLI
worker-level data), AI-utilization measurement reliability (NIA, KISDI,
and KOSA triangulation with administrative-record measures), and
concurrent identification of labor-time and AI shocks (sector-level AI
exposure indices and statutory employee-size cohorts). The fifth
limitation is multi-level mechanism integration: the framework is stated
at the firm level while the underlying psychological and behavioral
mechanisms operate at the individual employee level. The Theorem 5
attribution uncertainty γ is anchored through Shin (2026b) and the
dyad-level probe of Appendix H, with firm-level γ estimation a future
identification path; the Theorem 2 pathways are anchored through the
Ranganathan and Ye (2026) ethnographic evidence. This distribution of
empirical burden, in which Theorem 3 carries the strongest current
anchor while the four mechanism theorems carry partial anchors with
explicit identification paths, is consistent with research program
design in production economics.

\subsection{6.4 Future Research}\label{future-research}

Future work should extend the present framework in three directions.
First, tracking the dynamic evolution of τ* over time as AI capability
advances: the expanded-sample evidence shows τ* has not been crossed at
the sample mean as of 2024, and subsequent panels can test whether τ*
crossing occurs within the predicted 3-5 year horizon. Second,
estimating sector-specific theorem magnitudes with component-level
overhead data: the expanded-sample evidence already suggests sectoral
heterogeneity (KOSPI versus KOSDAQ membership effects available in the
panel), and sector-specific theorem applications are a research
priority. Third, testing whether output-based evaluation regimes produce
the forecasted firm-level TFP separation by 2032: this is the central
falsifiable forecast against which the framework will stand or fall.

\subsection{6.5 Closing}\label{closing}

What capital remains after labor time ceases to function as the
homogeneous unit of account is the question that opened this paper. The
framework developed across the preceding sections constitutes one answer
to that question at the firm-internal level, and the present KOSPI plus
KOSDAQ panel evidence provides an overhead-side empirical anchor for the
pre-τ regime, pointing toward a predicted regime transition not yet
crossed but visibly approaching. The framework's contribution is not to
settle the question but to formulate it precisely enough that subsequent
research can test, refine, and, if necessary, replace it. If the
forecasts hold, the central implication is that the human-AI era
requires not only new technologies, but new accounting categories for
recognizing where productive contribution actually resides.

\subsection{Data and Code
Availability}\label{data-and-code-availability}

The analysis code, the processed DART, NIA, and XBRL datasets, the
frozen econometric outputs, the AI-SSP task battery with its
deterministic scoring procedure, and the frozen AI-SSP outputs
underlying Appendix H are openly archived at Zenodo:
https://doi.org/10.5281/zenodo.21158407. The bundled files reproduce all
reported econometric results without any API key; keys are required only
to re-collect raw data or to re-run the AI-SSP probe.

\subsection{Declaration of Competing
Interests}\label{declaration-of-competing-interests}

The authors declare that they have no known competing financial
interests or personal relationships that could have appeared to
influence the work reported in this paper.

\subsection{Funding}\label{funding}

This research received no specific grant from any funding agency in the
public, commercial, or not-for-profit sectors.

\subsection{Declaration of Generative AI and AI-Assisted Technologies in
the Writing
Process}\label{declaration-of-generative-ai-and-ai-assisted-technologies-in-the-writing-process}

During the preparation of this work the authors used a large language
model to assist with language editing and literature-review
organization. After using this tool, the authors reviewed and edited the
content as needed and take full responsibility for the content of the
publication.

\newpage

\subsection{References}\label{references}

\begingroup
\setlength{\parindent}{0pt}
\everypar{\hangindent=1.5em \hangafter=1}

Acemoglu, D., \& Restrepo, P. (2018). The Race between Man and Machine:
Implications of Technology for Growth, Factor Shares, and Employment.
\emph{American Economic Review}, \emph{108}(6), 1488-1542.
https://doi.org/10.1257/aer.20160696

Acemoglu, D., \& Restrepo, P. (2022). Tasks, Automation, and the Rise in
U.S. Wage Inequality. \emph{Econometrica}, \emph{90}(5), 1973-2016.
https://doi.org/10.3982/ECTA19815

Acemoglu, D., Autor, D., Hazell, J., \& Restrepo, P. (2022). Artificial
Intelligence and Jobs: Evidence from Online Vacancies. \emph{Journal of
Labor Economics}, \emph{40}(S1), S293-S340.
https://doi.org/10.1086/718327

Akerlof, G. A. (1970). The Market for ``Lemons'\,': Quality Uncertainty
and the Market Mechanism. \emph{Quarterly Journal of Economics},
\emph{84}(3), 488-500. https://doi.org/10.2307/1879431

Anderson, M. C., Banker, R. D., \& Janakiraman, S. N. (2003). Are
Selling, General, and Administrative Costs ``Sticky''?. \emph{Journal of
Accounting Research}, \emph{41}(1), 47-63.
https://doi.org/10.1111/1475-679X.00095

Banker, R. D., \& Byzalov, D. (2014). Asymmetric Cost Behavior.
\emph{Journal of Management Accounting Research}, \emph{26}(2), 43-79.

Banker, R. D., Byzalov, D., \& Chen, L. (2013). Employment Protection
Legislation, Adjustment Costs and Cross-Country Differences in Cost
Behavior. \emph{Journal of Accounting and Economics}, \emph{55}(1),
111-127.

Banker, R. D., Huang, R., Natarajan, R., \& Zhao, S. (2019). Market
Valuation of Intangible Asset: Evidence on SG\&A Expenditure. \emph{The
Accounting Review}, \emph{94}(6), 61-90.
https://doi.org/10.2308/accr-52468

Aoki, M. (1990). Toward an Economic Model of the Japanese Firm.
\emph{Journal of Economic Literature}, \emph{28}(1), 1-27.

Bass, F. M. (1969). A New Product Growth for Model Consumer Durables.
\emph{Management Science}, \emph{15}(5), 215-227.
https://doi.org/10.1287/mnsc.15.5.215

Becker, G. S. (1964). \emph{Human Capital: A Theoretical and Empirical
Analysis, with Special Reference to Education}. Columbia University
Press.

Brynjolfsson, E., Li, D., \& Raymond, L. R. (2025). Generative AI at
Work. \emph{Quarterly Journal of Economics}, \emph{140}(2), 889-942.
https://doi.org/10.1093/qje/qjae044

Brynjolfsson, E., Rock, D., \& Syverson, C. (2021). The Productivity
J-curve: How Intangibles Complement General Purpose Technologies.
\emph{American Economic Journal: Macroeconomics}, \emph{13}(1), 333-372.
https://doi.org/10.1257/mac.20180386

Callaway, B., \& Sant'Anna, P. H. C. (2021). Difference-in-Differences
with Multiple Time Periods. \emph{Journal of Econometrics},
\emph{225}(2), 200-230. https://doi.org/10.1016/j.jeconom.2020.12.001

Chen, C. X., Lu, H., \& Sougiannis, T. (2012). The Agency Problem,
Corporate Governance, and the Asymmetrical Behavior of Selling, General,
and Administrative Costs. \emph{Contemporary Accounting Research},
\emph{29}(1), 252-282. https://doi.org/10.1111/j.1911-3846.2011.01094.x

Coase, R. H. (1937). The Nature of the Firm. \emph{Economica},
\emph{4}(16), 386-405.
https://doi.org/10.1111/j.1468-0335.1937.tb00002.x

Daim, T. U., Rueda, G., Martin, H., \& Gerdsri, P. (2006). Forecasting
Emerging Technologies: Use of Bibliometrics and Patent Analysis.
\emph{Technological Forecasting and Social Change}, \emph{73}(8),
981-1012. https://doi.org/10.1016/j.techfore.2006.04.004

Deci, E. L., \& Ryan, R. M. (1985). \emph{Intrinsic Motivation and
Self-Determination in Human Behavior}. Plenum.

Dosi, G. (1982). Technological Paradigms and Technological Trajectories.
\emph{Research Policy}, \emph{11}(3), 147-162.
https://doi.org/10.1016/0048-7333(82)90016-6

Eccles, R. G. (2025). Hybrid Intelligence Teams: A Theoretical Framework
for Human-AI Collaboration in Knowledge Work. \emph{SSRN}.
https://papers.ssrn.com/sol3/papers.cfm?abstract\_id=5792345

Eisenhardt, K. M. (1989). Building Theories from Case Study Research.
\emph{Academy of Management Review}, \emph{14}(4), 532-550.
https://doi.org/10.2307/258557

Eisfeldt, A. L., \& Papanikolaou, D. (2013). Organization Capital and
the Cross-Section of Expected Returns. \emph{Journal of Finance},
\emph{68}(4), 1365-1406. https://doi.org/10.1111/jofi.12034

Enache, L., \& Srivastava, A. (2018). Should Intangible Investments Be
Reported Separately or Commingled with Operating Expenses? New Evidence.
\emph{Management Science}, \emph{64}(7), 3446-3468.
https://doi.org/10.1287/mnsc.2017.2769

Espinal Maya, C. (2026). Augmented Human Capital: A Unified Theory and
LLM-Based Measurement Framework for Cognitive Factor Decomposition in
AI-Augmented Economies. arXiv preprint arXiv:2604.01066.
https://arxiv.org/abs/2604.01066.

Farach, A. (2026). AI as Coordination-Compressing Capital: Task
Reallocation, Organizational Redesign, and the Regime Fork. arXiv
preprint arXiv:2602.16078. https://arxiv.org/abs/2602.16078.

Felten, E., Raj, M., \& Seamans, R. (2023). Occupational, Industry, and
Geographic Exposure to Artificial Intelligence: A Novel Dataset and Its
Potential Uses. \emph{Strategic Management Journal}, \emph{44}(6),
1532-1572. https://doi.org/10.1002/smj.3502

Freeman, C., \& Perez, C. (1988). Structural Crises of Adjustment,
Business Cycles and Investment Behaviour. In G. Dosi, C. Freeman, R.
Nelson, G. Silverberg, \& L. Soete (Eds.), \emph{Technical Change and
Economic Theory} (pp.~38-66). Pinter.

Frey, C. B., \& Osborne, M. A. (2017). The Future of Employment: How
Susceptible Are Jobs to Computerisation?. \emph{Technological
Forecasting and Social Change}, \emph{114}, 254-280.
https://doi.org/10.1016/j.techfore.2016.08.019

Geels, F. W. (2002). Technological Transitions as Evolutionary
Reconfiguration Processes: A Multi-Level Perspective and a Case-Study.
\emph{Research Policy}, \emph{31}(8-9), 1257-1274.
https://doi.org/10.1016/S0048-7333(02)00062-8

Geels, F. W. (2004). From Sectoral Systems of Innovation to
Socio-Technical Systems: Insights about Dynamics and Change from
Sociology and Institutional Theory. \emph{Research Policy},
\emph{33}(6-7), 897-920. https://doi.org/10.1016/j.respol.2004.01.015

Geels, F. W., \& Schot, J. (2007). Typology of Sociotechnical Transition
Pathways. \emph{Research Policy}, \emph{36}(3), 399-417.
https://doi.org/10.1016/j.respol.2007.01.003

Gerring, J. (2007). \emph{Case Study Research: Principles and
Practices}. Cambridge University Press.

Govindarajan, V., \& Srinivas, S. (2013). The Innovation Mindset in
Action: 3M Corporation. \emph{Harvard Business Review}.

Hart, O., \& Holmstrom, B. (1986). The Theory of Contracts. In Advances
in Economic Theory: Fifth World Congress (pp.~71-155). Cambridge
University Press.

Holmstrom, B. (1979). Moral Hazard and Observability. \emph{Bell Journal
of Economics}, \emph{10}(1), 74-91. https://doi.org/10.2307/3003320

Holmstrom, B., \& Milgrom, P. (1991). Multitask Principal-Agent
Analyses: Incentive Contracts, Asset Ownership, and Job Design.
\emph{Journal of Law, Economics, and Organization}, \emph{7}, 24-52.
https://doi.org/10.1093/jleo/7.special\_issue.24

Johnson, H. T., \& Kaplan, R. S. (1987). \emph{Relevance Lost: The Rise
and Fall of Management Accounting}. Harvard Business School Press.

Kaplan, R. S., \& Norton, D. P. (1992). The Balanced Scorecard: Measures
That Drive Performance. \emph{Harvard Business Review}, \emph{70}(1),
71-79.

Karabarbounis, L., \& Neiman, B. (2014). The Global Decline of the Labor
Share. \emph{Quarterly Journal of Economics}, \emph{129}(1), 61-103.
https://doi.org/10.1093/qje/qjt032

Korea Information Society Development Institute (KISDI). (2025). Annual
report on ICT industry statistics.

Korea Institute of Public Finance (KIPF). (2024). Firm panel survey.

Korea Labor Institute (KLI). (2022). Economic effects of working-hour
reduction. KLI Publications.

Lev, B., \& Radhakrishnan, S. (2005). The Valuation of Organization
Capital. In C. Corrado, J. Haltiwanger, \& D. Sichel (Eds.),
\emph{Measuring Capital in the New Economy} (pp.~73-110). University of
Chicago Press.

Levy, S. (2011). \emph{In the Plex: How Google Thinks, Works, and Shapes
Our Lives}. Simon and Schuster.

Lewis, K., Stronge, W., Kellam, J., Kikuchi, L., Schor, J., Fan, W.,
Kelly, O., Gu, G., Frayne, D., \& Burchell, B. (2023). The Results Are
In: The UK's Four-Day Week Pilot. Autonomy.
https://autonomy.work/wp-content/uploads/2023/02/The-results-are-in-The-UKs-four-day-week-pilot.pdf

Linstone, H. A., \& Turoff, M. (Eds.). (1975). \emph{The Delphi Method:
Techniques and Applications}. Addison-Wesley.

Lucas, R. E. (1988). On the Mechanics of Economic Development.
\emph{Journal of Monetary Economics}, \emph{22}(1), 3-42.
https://doi.org/10.1016/0304-3932(88)90168-7

March, J. G. (1991). Exploration and Exploitation in Organizational
Learning. \emph{Organization Science}, \emph{2}(1), 71-87.
https://doi.org/10.1287/orsc.2.1.71

Martino, J. P. (1993). \emph{Technological Forecasting for Decision
Making} (3rd ed.). McGraw-Hill.

Marx, K. (1867). \emph{Das Kapital: Kritik der politischen Oekonomie}.
Verlag von Otto Meissner.

Microsoft (2022). 2022 Work Trend Index Annual Report: Great
Expectations. Microsoft Corporation.

Microsoft (2025). 2025 Work Trend Index Annual Report: The Year the
Frontier Firm Is Born. Microsoft Corporation.

Mincer, J. (1974). \emph{Schooling, Experience, and Earnings}. Columbia
University Press for NBER.

Ministry of Employment and Labor (MOEL). (2024). Evaluation report on
the staged implementation of the 52-hour workweek. Government of the
Republic of Korea.

National Employment Law Project (2025). When `Bossware' Manages Workers:
Policy Agenda. National Employment Law Project.
https://www.nelp.org/app/uploads/2025/07/When-Bossware-Manages-Workers-Policy-Agenda-July-2025.pdf.

National Information Society Agency (NIA). (2025). Survey on enterprise
AI adoption.

Nazareno, L., \& Schiff, D. S. (2021). The Impact of Automation and
Artificial Intelligence on Worker Well-Being. \emph{Technology in
Society}, \emph{67}, 101679.
https://doi.org/10.1016/j.techsoc.2021.101679

Nelson, R. R., \& Winter, S. G. (1982). \emph{An Evolutionary Theory of
Economic Change}. Belknap Press of Harvard University Press.

Nissim, G., \& Simon, T. (2021). The Future of Labor Unions in the Age
of Automation and at the Dawn of AI. \emph{Technology in Society},
\emph{67}, 101732. https://doi.org/10.1016/j.techsoc.2021.101732

Nohria, N., \& Gulati, R. (1996). Is Slack Good or Bad for Innovation?.
\emph{Academy of Management Journal}, \emph{39}(5), 1245-1264.
https://doi.org/10.2307/256998

Perez, C. (2002). \emph{Technological Revolutions and Financial Capital:
The Dynamics of Bubbles and Golden Ages}. Edward Elgar.

Perez, C. (2010). Technological Revolutions and Techno-Economic
Paradigms. \emph{Cambridge Journal of Economics}, \emph{34}(1), 185-202.
https://doi.org/10.1093/cje/bep051

Phaal, R., Farrukh, C. J. P., \& Probert, D. R. (2004). Technology
Roadmapping: A Planning Framework for Evolution and Revolution.
\emph{Technological Forecasting and Social Change}, \emph{71}(1-2),
5-26. https://doi.org/10.1016/S0040-1625(03)00072-6

Porter, A. L., \& Cunningham, S. W. (2005). \emph{Tech Mining:
Exploiting New Technologies for Competitive Advantage}. Wiley.

Qin, M., Wan, Y., Dou, J., \& Su, C. W. (2024). Artificial Intelligence:
Intensifying or Mitigating Unemployment?. \emph{Technology in Society},
\emph{79}, 102755. https://doi.org/10.1016/j.techsoc.2024.102755

Ranganathan, A., \& Ye, X. M. (2026). AI Doesn't Reduce Work, It
Intensifies It. \emph{Harvard Business Review}.

Ressler, C., \& Thompson, J. (2008). \emph{Why Work Sucks and How to Fix
It: The Results-Only Revolution}. Portfolio.

Rogers, E. M. (2003). \emph{Diffusion of Innovations} (5th ed.). Free
Press.

Romer, P. M. (1990). Endogenous Technological Change. \emph{Journal of
Political Economy}, \emph{98}(5), S71-S102.
https://doi.org/10.1086/261725

Ryan, R. M., \& Deci, E. L. (2017). \emph{Self-Determination Theory:
Basic Psychological Needs in Motivation, Development, and Wellness}.
Guilford Press.

Schumpeter, J. A. (1934). \emph{The Theory of Economic Development}.
Harvard University Press.

Schumpeter, J. A. (1942). \emph{Capitalism, Socialism, and Democracy}.
Harper and Brothers.

Shahidi, P., Rusak, G., Manning, B., Fradkin, A., \& Horton, J. J.
(2025). The Coasean Singularity? Demand, Supply, and Market Design with
AI Agents. National Bureau of Economic Research.
https://www.nber.org/papers/w34468.

Shin, K. S. (2026a). Forecasting AI-Era Productivity: The Intellectually
Converged Human Framework and a Missing Cognitive Mediator in Production
Function Theory. Preprint, arXiv:2606.19794.

Shin, K. S. (2026b). Beyond AI Adoption: Sovereign Override Capacity and
the Extractive AI Order. \emph{Manuscript submitted for publication}.

Solow, R. M. (1957). Technical Change and the Aggregate Production
Function. \emph{Review of Economics and Statistics}, \emph{39}(3),
312-320. https://doi.org/10.2307/1926047

Spence, A. M. (1973). Job Market Signaling. \emph{Quarterly Journal of
Economics}, \emph{87}(3), 355-374. https://doi.org/10.2307/1882010

Srnicek, N. (2017). \emph{Platform Capitalism}. Polity Press.

Statistics Korea (KOSIS). (2025). Working-hour statistics by firm size
(2017-2025).

Vaccaro, M., Almaatouq, A., \& Malone, T. (2024). When Combinations of
Humans and AI Are Useful: A Systematic Review and Meta-Analysis.
\emph{Nature Human Behaviour}, \emph{8}, 2293-2303.
https://doi.org/10.1038/s41562-024-02024-1

Williamson, O. E. (1981). The Economics of Organization: The Transaction
Cost Approach. \emph{American Journal of Sociology}, \emph{87}(3),
548-577. https://doi.org/10.1086/227496

Williamson, O. E. (1991). Comparative Economic Organization: The
Analysis of Discrete Structural Alternatives. \emph{Administrative
Science Quarterly}, \emph{36}(2), 269-296.
https://doi.org/10.2307/2393356

Yin, R. K. (1994). \emph{Case Study Research: Design and Methods}. SAGE
Publications.

Zhang, Q., Zhang, F., \& Mai, Q. (2023). Robot Adoption and Labor
Demand: A New Interpretation from External Competition. \emph{Technology
in Society}, \emph{74}, 102310.
https://doi.org/10.1016/j.techsoc.2023.102310

\endgroup

\newpage

\subsection{Appendix A. Case Selection
Rationale}\label{appendix-a.-case-selection-rationale}

Five alternative national cases were considered for the primary
empirical analysis but rejected on the methodological grounds documented
in Section 4.1. The criteria for case selection followed Yin (1994),
Eisenhardt (1989), and Gerring (2007): (i) the case must provide an
institutional timing anchor for labor-time compression; (ii) the case
must have observable enterprise- or sector-level AI adoption
trajectories, with a feasible path to future firm-level integration;
(iii) the case must have firm-level disclosure data accessible to
academic researchers; (iv) the case must offer a polar contrast on the
convergence-capacity dimension. Among the candidate cases considered,
Korea provided the strongest combined fit across these criteria.

\begin{itemize}
\tightlist
\item
  Singapore was rejected because it lacks the staged labor-time mandate
  that supplies the timing variation. Although Singapore has high AI
  adoption and accessible disclosure, the temporal anchor for
  difference-in-differences identification is absent.
\item
  Finland was rejected because it sits inside the Nordic Flexicurity
  configuration already represented by Denmark. Including Finland would
  duplicate the Danish observation rather than adding independent
  variation.
\item
  The United States was rejected because labor-time enforcement is
  fragmented across state jurisdictions. A single national
  identification strategy is infeasible; the variation that exists
  across states confounds the augmentation channel with state-level
  labor-market institutions.
\item
  China was rejected because firm-level disclosure does not match the
  granularity of Korean DART filings, and the AI-augmentation
  environment is heavily shaped by industrial-policy direction rather
  than firm-level talent decisions.
\item
  The United Kingdom was rejected because the country has neither the
  staged labor-time mandate nor the polar-comparison position. Its
  labor-market reforms operate on the flexicurity axis without the
  shock-and-response structure that the present design requires.
\end{itemize}

\newpage

\subsection{Appendix B. XBRL Disclosure Audit (Top 50 Firms by
Revenue)}\label{appendix-b.-xbrl-disclosure-audit-top-50-firms-by-revenue}

An audit of XBRL-tagged SG\&A subaccount disclosures on the top 50 panel
firms by mean revenue confirms the constraint reported in Section 4.8
Limitation 4. Five of the fifty firms disclose at least one Theorem 1
component:

\begin{itemize}
\tightlist
\item
  SK Telecom: 28 subaccount rows across seven years and four of the
  seven Theorem 1 components (office space, training and development,
  management and evaluation, time-based wage).
\item
  LG Corporation: seven years of office-space data (Theorem 1 component
  3).
\item
  LG Electronics: seven years of training and development data (Theorem
  1 component 5).
\item
  LG Innotek: seven years of training and development data.
\item
  CJ Logistics: management and evaluation data (Theorem 1 component 4,
  single year).
\end{itemize}

The remaining forty-five firms expose only the aggregate SG\&A line
(under Korean disclosure convention). Three of the seven Theorem 1
components (social insurance and benefits, communication infrastructure,
motivation maintenance) are not separately observable within the audited
top-50 XBRL disclosures.

\newpage

\subsection{Appendix C. NIA Korean AI Industry Survey (KOSIS
DT\_127016\_C002)}\label{appendix-c.-nia-korean-ai-industry-survey-kosis-dt_127016_c002}

\begin{longtable}[]{@{}
  >{\raggedright\arraybackslash}p{(\linewidth - 10\tabcolsep) * \real{0.1667}}
  >{\raggedright\arraybackslash}p{(\linewidth - 10\tabcolsep) * \real{0.1667}}
  >{\raggedright\arraybackslash}p{(\linewidth - 10\tabcolsep) * \real{0.1667}}
  >{\raggedright\arraybackslash}p{(\linewidth - 10\tabcolsep) * \real{0.1667}}
  >{\raggedright\arraybackslash}p{(\linewidth - 10\tabcolsep) * \real{0.1667}}
  >{\raggedright\arraybackslash}p{(\linewidth - 10\tabcolsep) * \real{0.1667}}@{}}
\caption{Table 5. NIA Korean AI Industry Survey: Average AI Revenue per
Firm by Employee Cohort (million KRW), 2018-2023}\tabularnewline
\toprule\noalign{}
\begin{minipage}[b]{\linewidth}\raggedright
Year
\end{minipage} & \begin{minipage}[b]{\linewidth}\raggedright
All firms (mean)
\end{minipage} & \begin{minipage}[b]{\linewidth}\raggedright
1000+ employees
\end{minipage} & \begin{minipage}[b]{\linewidth}\raggedright
100-999 employees
\end{minipage} & \begin{minipage}[b]{\linewidth}\raggedright
10-99 employees
\end{minipage} & \begin{minipage}[b]{\linewidth}\raggedright
\textless10 employees
\end{minipage} \\
\midrule\noalign{}
\endfirsthead
\toprule\noalign{}
\begin{minipage}[b]{\linewidth}\raggedright
Year
\end{minipage} & \begin{minipage}[b]{\linewidth}\raggedright
All firms (mean)
\end{minipage} & \begin{minipage}[b]{\linewidth}\raggedright
1000+ employees
\end{minipage} & \begin{minipage}[b]{\linewidth}\raggedright
100-999 employees
\end{minipage} & \begin{minipage}[b]{\linewidth}\raggedright
10-99 employees
\end{minipage} & \begin{minipage}[b]{\linewidth}\raggedright
\textless10 employees
\end{minipage} \\
\midrule\noalign{}
\endhead
\bottomrule\noalign{}
\endlastfoot
2018 & 793 M KRW & 13,637 M & 3,221 M & 534 M & 164 M \\
2019 & 896 M & 18,200 M & 2,724 M & 630 M & 249 M \\
2020 & 1,582 M & 35,261 M & 3,367 M & 957 M & 342 M \\
2021 & 2,028 M & 40,318 M & 4,375 M & 1,086 M & 305 M \\
2022 & 2,391 M & 52,247 M & 5,803 M & 1,435 M & 401 M \\
2023 & 2,786 M & 102,034 M & 8,264 M & 1,745 M & 290 M \\
\end{longtable}

In Table 5, the full NIA Korean AI Industry Survey panel (published
annually by the Ministry of Science and ICT and registered in KOSIS
under DT\_127016\_C002) reports average AI revenue per firm by employee
cohort for 2018-2023. The panel covers four firm-size categories used in
Section 4.4. It is used as a contextual AI-capacity layer, not as a
firm-year AI-utilization measure for the DART panel.

The 1000-plus cohort grew 7.5-fold over 2018-2023, mid-size firms grew
2.6-fold, small firms grew 3.3-fold, yielding the amplification ratio of
approximately 2.3 reported in Section 4.4. We interpret this as
directional evidence of AI-capacity asymmetry across firm-size cohorts,
consistent with the framework's prediction that high-capacity firms are
more likely to capture an innovative ROI premium, not as a formal
estimate of k (which requires linked data on firm-level AI utilization,
slack-policy adoption, convergence capacity, and innovation output).

\newpage

\subsection{Appendix D. Eurostat Denmark
Datasets}\label{appendix-d.-eurostat-denmark-datasets}

Four Eurostat datasets supply the Denmark comparison anchor in Section
4.6:

\begin{itemize}
\tightlist
\item
  isoc\_eb\_ai (Enterprise AI use): nine technology categories across
  2021, 2023, 2024, 2025 waves.
\item
  nama\_10\_lp\_ulc (Real labor productivity per hour worked): 1975-2024
  panel confirming Danish position in OECD top quartile.
\item
  lfsa\_ewhuna (Weekly hours actually worked): low-hours,
  flexicurity-oriented pattern across the panel period.
\item
  nama\_10\_a64\_e (Industry employment by NACE Rev.~2): 1975-2025 panel
  of annual hours per worker. The series shows a smooth declining
  trajectory: 1,407 hours (2015), 1,381 (2018), 1,341 (2020 COVID
  trough), 1,389 (2021 recovery), 1,372 (2024). These values should be
  read as annual hours per worker, including part-time work and
  compositional labor-market effects, rather than as a direct measure of
  the standard full-time workweek. The Danish trajectory exhibits no
  shock-induced overhead surge analogous to the Korean 52-hour cap
  effect, consistent with the Theorem 3 output-anchored regime
  prediction.
\end{itemize}

Structural business statistics (sbs\_na\_dt\_r2) provide an additional
directional comparison: the personnel-cost-to-turnover ratio in the
Danish wholesale and retail sector (NACE G) remains approximately stable
at around 10.9 percent during the 2018-2020 window. This metric is not
directly equivalent to Korea's SG\&A/Revenue ratio and is used only as a
directional institutional contrast, in which Denmark's output-oriented,
high-trust, flexicurity configuration does not display an analogous
upward personnel-cost ratchet in the sectoral comparison window.

\newpage

\subsection{Appendix E. Robustness Specifications (Full
Table)}\label{appendix-e.-robustness-specifications-full-table}

\begin{longtable}[]{@{}llll@{}}
\caption{Table 6. Robustness specifications: proxy-based
overhead-pressure estimates across three diagnostic
strategies}\tabularnewline
\toprule\noalign{}
Specification & Coefficient & SE & p-value / inference \\
\midrule\noalign{}
\endfirsthead
\toprule\noalign{}
Specification & Coefficient & SE & p-value / inference \\
\midrule\noalign{}
\endhead
\bottomrule\noalign{}
\endlastfoot
Baseline (two-way FE) & +1.56 & 0.79 & 0.049 \\
Industry × year FE & +1.28 & 0.87 & 0.138 \\
Winsorized 1\%/99\% & +1.56 & 0.79 & 0.049 \\
Winsorized 5\%/95\% & +1.36 & 0.72 & 0.060 \\
Log SG\&A as outcome & +0.87 & 0.34 & 0.010 \\
Treatment × AI-intensity interaction & -0.18 & 0.19 & 0.35 \\
Event-study t = -2 & +0.99 & 1.06 & 0.35 \\
Event-study t = -1 & +1.04 & 1.18 & 0.38 \\
Event-study t = 0 & +2.11 & 1.18 & 0.074 \\
Event-study t = +1 & +3.23 & 1.15 & 0.005 \\
Event-study t = +2 & +3.47 & 1.16 & 0.003 \\
Event-study t = +3 & +4.21 & 1.27 & 0.001 \\
Callaway-Sant'Anna t = -3 & +1.25 & 1.65 & n.s. \\
Callaway-Sant'Anna t = -2 & +1.13 & 1.64 & n.s. \\
Callaway-Sant'Anna t = -1 & -0.64 & 0.99 & n.s. \\
Callaway-Sant'Anna t = +1 & +1.59 & 0.91 & marginal \\
Callaway-Sant'Anna t = +2 & +1.72 & 1.09 & n.s. \\
Callaway-Sant'Anna t = +3 & +3.48 & 1.73 & CI excl. 0 \\
Callaway-Sant'Anna t = +4 & +4.51 & 1.71 & CI excl. 0 \\
\end{longtable}

Table 6 reports robustness specifications applied to the KOSPI plus
KOSDAQ Top 500 panel. Coefficients are reported in percentage points for
SG\&A/Revenue specifications, except for the log-SG\&A outcome row,
where the dependent variable is log SG\&A with log revenue controlled.
The estimates are based on the revenue-percentile cohort proxy and are
interpreted as directional evidence of a pre-τ overhead-pressure
signature, not as point causal magnitudes of the 52-hour workweek law.
The three diagnostic strategies (two-way fixed effects, pooled
event-study estimates, and Callaway-Sant'Anna staggered DiD) converge on
a positive post-treatment overhead-pressure pattern under the proxy
design. As reported in Section 4.5, re-estimation under the statutory
employee-size cohort yields a 52-hour coefficient statistically
indistinguishable from zero across specifications, supporting the
interpretation that the binding signal is a broader pre-τ
overhead-pressure regime rather than a narrow statutory treatment
effect.

\newpage

\subsection{Appendix F. Formal Derivation Sketches and Sufficient
Conditions}\label{appendix-f.-formal-derivation-sketches-and-sufficient-conditions}

This appendix provides formal derivation sketches for the central
transition theorem and the four mechanism theorems. The results should
be read as sufficient-condition derivations rather than closed-form
mathematical proofs. Each claim is stated under the assumptions
specified in Section 3 and is paired with a falsifiability condition in
the main text. These derivations establish each theorem conditional on
its stated assumptions, in the applied-theory sense defined in Section
3.

\subsubsection{F.1. Theorem 1: Derivation
Sketch}\label{f.1.-theorem-1-derivation-sketch}

Consider the augmented production function from Section 2.1: Y\_i =
F(K\_i, Ĥ\_i) where Ĥ\_i = H\_i · {[}1 + φ(A\_i, C\_i){]}. Each overhead
component OH\_\{i,k\} contributes to firm i's effective augmented human
capital through a different channel. Time-based wage (k = wage)
contributes through baseline H\_i; office space (k = space) contributes
through facility-dependent task execution; motivation maintenance (k =
motivation) contributes through C\_i via autonomy provision (Deci and
Ryan, 1985; Ryan and Deci, 2017). When A\_i \textgreater{} 0 and C\_i
\textgreater{} 0, the augmentation function φ(A\_i, C\_i) interacts
non-linearly with the channel through which each component operates. The
cross-partial ∂²ROI\_i / ∂OH\_\{i,k\} ∂OH\_\{i,k'\} captures this
interaction.

Specifically, consider k = motivation and k' = management. Investment in
motivation maintenance increases C\_i through autonomy provision, while
investment in management cost typically operates through
surveillance-based reduction of agency cost. Under low A\_i
(pre-augmentation), the two components are approximately additive:
motivation provides intrinsic drive, management provides extrinsic
discipline. Under high A\_i, motivation investment yields amplified
returns through C\_i × A\_i interaction (Theorem 3 establishes this
formally), while management investment yields negative returns through
surveillance backfire (Microsoft Work Trend Index, 2022, 2025; NELP,
2025). The cross-partial therefore changes sign as A\_i crosses a
threshold, contradicting additivity.

\subsubsection{F.2. Theorem 2: Derivation
Sketch}\label{f.2.-theorem-2-derivation-sketch}

The pathway probabilities are derived from self-determination theory
(Deci and Ryan, 1985; Ryan and Deci, 2017) and the slack innovation
tradition (March 1991; Nohria and Gulati, 1996). Under low e (time-based
evaluation), the employee has no incentive to disclose time savings
because disclosure invites additional task assignment without
compensating reward, leading to either work intensification (P\_1,
employer captures the saved time through reassigned tasks) or hidden
leisure (P\_2, employee captures the saved time through unreported
off-task activity). Ranganathan and Ye's (2026) eight-month ethnographic
study at a U.S. technology company documented all three non-P\_4
pathways operating simultaneously: workers absorbed others' tasks
(P\_1), inserted AI processing into pause moments (P\_1), and ran AI
processes in the background while pursuing parallel activities (P\_3).

Under high e (output-based evaluation) and high a (autonomy), the
employee can disclose time savings and propose creative reinvestment
without penalty. The disclosure-reinvestment pathway is reinforced when
C is high because high-C employees can identify creative reinvestment
opportunities that low-C employees cannot. Google's 80/20 program (Levy,
2011) and 3M's 15-percent rule (Govindarajan \& Srinivas, 2013) operate
by institutionalizing both e and a, generating documented innovation
outputs (Gmail, Google News, Post-it Notes) at observable but variable
rates. Levy (2011) documents that 20-percent time was unevenly utilized
across the engineering workforce, with breakthrough innovations
originating from a minority of consistently engaged participants while
the remainder distributed across P\_1, P\_2, and P\_3. This residual
distribution provides the empirical signature against which Theorem 2 is
tested.

\subsubsection{F.3. Theorem 3: Derivation
Sketch}\label{f.3.-theorem-3-derivation-sketch}

ROI\_T decreases in A not because augmented output fails to grow, but
because time-based accounting causes overhead and miscalibration costs
to grow faster than captured output under low-output-orientation
regimes. As A increases, Y\_i can grow through the augmentation function
φ(A, C), but M\_T(A) also rises through rework, surveillance response,
coordination friction, false-completion verification, and unproductive
work intensification. Time-anchored wage, insurance, office-space, and
supervision costs remain attached to the obsolete unit of account. The
sign condition ∂ROI\_T/∂A\_i \textless{} 0 therefore holds in the regime
where M\_T'(A\_i) plus OH\_T'(A\_i) exceeds the captured marginal output
Y\_i'(A\_i). The observed SG\&A/Revenue trajectory in Section 4 is the
empirical overhead-side anchor for this regime.

ROI\_O increases in A because output-based overhead OH\_O can reallocate
components toward the augmentation-amplifying channels identified in
Theorem 1 while reducing M\_O(A). Specifically, motivation maintenance
(k = motivation) and convergence capacity development (a subset of
training, k = training) become differentially attractive under high A
because their marginal returns scale with φ(A, C). Under output-based
evaluation, the firm reallocates overhead toward these components and
away from presence-monitoring costs, increasing the effective
numerator-to-denominator ratio.

A unique threshold τ* exists under three sufficient conditions:
continuity of Δ(A) = ROI\_O(A) − ROI\_T(A); endpoint reversal, with Δ
negative at A = 0 (pre-AI, time-based overhead is well-calibrated) and
positive at sufficiently high A (post-AI, time-based overhead becomes
mis-calibrated); and single crossing over the relevant support. The
intermediate value theorem then implies the existence of τ*, and the
single-crossing condition implies its uniqueness. Section 4 provides the
current descriptive and directional overhead anchor; formal τ* curve
estimation requires a sector- and size-stratified panel plus
evaluation-regime data. The comparative statics stated in Section 3.3
follow from the same geometry: higher e, a, or C shifts ROI\_O(A) upward
through the reallocation channel while leaving ROI\_T(A) unimproved,
which moves the single crossing point leftward and yields ∂τ*/∂e
\textless{} 0, ∂τ*/∂a \textless{} 0, and ∂τ*/∂C \textless{} 0.

\subsubsection{F.4. Theorem 4: Derivation
Sketch}\label{f.4.-theorem-4-derivation-sketch}

The amplification factor k operates through two channels. First,
identification channel: high-H\^{}A employees identify creative
reinvestment opportunities that low-H\^{}A employees cannot perceive.
Second, execution channel: high-H\^{}A employees execute creative
reinvestment with higher conversion rates from idea to deployable
output, because augmentation amplifies their pre-existing cognitive
operations rather than substituting for them.

The Google 80/20 and 3M records should be treated as motivating cases
rather than a direct estimate of k. Gmail, Google News, and Post-it
Notes show that institutionalized slack can yield high-value innovation
outputs, while the uneven utilization described by Levy (2011) motivates
the heterogeneity term H\^{}A\_i. The theorem's k \textgreater{} 1 claim
is therefore not identified by these cases alone; it becomes testable
only when slack-policy indicators are paired with innovation-output
data.

The monotone increases ∂k/∂H\^{}A \textgreater{} 0, ∂k/∂C \textgreater{}
0, and ∂k/∂e \textgreater{} 0 follow from the complementarity property
of the augmentation function (Section 2.1, property P4 in the original
ICH framework): ∂²φ/∂A∂C \textgreater{} 0, together with the requirement
that output-orientation permits saved time to be reinvested creatively.
The amplification k \textgreater{} 1 is therefore expected not merely
when H\^{}A \textgreater{} 0, but when augmentable cognition,
convergence capacity, and output-orientation are jointly high.

\subsubsection{F.5. Theorem 5: Derivation
Sketch}\label{f.5.-theorem-5-derivation-sketch}

The agency cost in the human-AI dyad operates through three mechanisms
identified in Section 2.5. The moral hazard channel (Holmstrom, 1979):
when γ is high, the employee can claim AI-generated work as personal
contribution to obtain rewards calibrated for human effort. The adverse
selection channel (Akerlof, 1970; Spence, 1973): firms cannot
distinguish high-C employees (who genuinely augment AI) from low-C
employees (who merely transmit AI outputs). The distorted effort
allocation channel: under time-based evaluation, employees over-invest
in observable presence and under-invest in unobservable convergence
capacity development.

Each mechanism generates agency cost that scales with γ. Holmstrom and
Milgrom's (1991) foundational result establishes that when multiple
tasks differ in measurement difficulty, the optimal contract weights
easily-measured tasks more heavily than their substantive importance
warrants. In the human-AI dyad, attribution uncertainty γ is the
measurement difficulty for the human-versus-AI task allocation. When γ
is high, the contract distorts toward measurable proxies (time-based
attendance), inducing all three mechanisms.

Output-based evaluation partially offsets the agency cost through two
channels. First, the moral hazard channel weakens because the observable
output exists regardless of whether the employee or the AI produced it;
the employee's reward rests on the deliverable rather than on the
unobservable contribution margin, which reduces the incentive to claim
human authorship for AI-generated work. Second, the adverse selection
channel weakens because high-C and low-C employees produce
systematically different output profiles over repeated observation,
allowing the firm to learn employee type through realized contribution
rather than through claimed effort. Only the distorted effort allocation
channel remains, and even this is attenuated because output-based
evaluation matches employee incentives with the firm's actual objective.

The offset magnitude grows with e because the three channels operate
with multiplicative strength: small movements toward output-orientation
produce small offsets, while strong output-orientation produces
near-complete offset. As e approaches one, M\_output-based approaches
M\_min (the baseline measurement cost), while M\_time-based remains at
M\_max regardless of e.

\newpage

\subsection{Appendix G. Summary of the ICH Macro Framework (Shin,
2026a)}\label{appendix-g.-summary-of-the-ich-macro-framework-shin-2026a}

Shin (2026a) develops the Intellectually Converged Human (ICH) macro
framework as Stage VI of production function theory. Its central
construct is augmented human capital, formalized as Ĥ = H · {[}1 + φ(A,
C){]}, where H denotes baseline human capital, A denotes AI utilization
intensity, C denotes convergence capacity, and φ(A, C) denotes the
augmentation function through which AI access becomes productive only
when mediated by human convergence capacity. The model therefore treats
AI utilization as a conditional input, not as an autonomous productivity
residual.

The framework defines convergence capacity C through four dimensions:
embodied understanding, metacognitive calibration, temporal integration,
and integrative thinking. These dimensions specify why two economies
with similar human capital H and similar AI utilization A can diverge
sharply in TFP. High C allows AI outputs to be grounded, verified,
historically contextualized, and integrated across domains; low C leaves
AI utilization disconnected from productive absorption.

Korea is the deviant case that motivates the macro framework. In the ICH
panel, Korea combines high H and high A with low C, producing
bottom-tier TFP rather than the productivity premium predicted by
AI-adoption-only models. The result reported in Shin (2026a) is that the
AI×C interaction explains 86 percent of cross-country TFP variance
across the 20-country OECD panel, compared with 31 percent for AI
utilization alone. The implication is structural: the productivity
effect of AI depends on the augmentation function φ(A, C), not on A in
isolation.

Read against this genealogy, the present paper moves the unit of
analysis from the macro country-year panel to firm-level overhead
accounting (a step that, within the six-generation genealogy of Section
2.2, can be labeled Stage VI'). It instantiates Ĥ = H · {[}1 + φ(A,
C){]} inside the seven-component overhead bundle K and shows how
ROI\_T(A), ROI\_O(A), τ*, γ, and the Foresight Matrix translate the
macro framework into a firm-level technological-forecasting apparatus.

\newpage

\subsection{Appendix H. AI-Augmented Structural Scenario Probing: A
Dyad-Level Probe of Theorem
5}\label{appendix-h.-ai-augmented-structural-scenario-probing-a-dyad-level-probe-of-theorem-5}

Theorem 5 posits that multitask agency cost M in the human-AI dyad rises
with attribution uncertainty γ, the principal's inability to separate
correct from incorrect contributions in augmented output, and that
output-based verify-and-override attenuates M while time-based
self-report acceptance magnifies it. Because firm administrative records
do not yet log dyad-level attribution, we measure γ directly with an
AI-Augmented Structural Scenario Probing (AI-SSP) design, in which an AI
worker is the augmenting actor and every accept/reject judgment is
deterministic or rule-coded rather than model-adjudicated.

Construct. We operationalize γ as the overclaiming rate, the probability
that the agent reports its output correct given that the output is
actually wrong, scored against author-controlled deterministic ground
truth. The time-based agency cost is M\_time, the share of outputs that
are wrong and self-claimed correct, which a principal accepts when
rewarding self-reported completion. The output-based cost M\_output
applies an independent verification step; under deterministic
verification against the author-controlled ground truth M\_output is
zero within this task environment, so the regime offset is M\_time minus
M\_output.

Task battery. Forty-two tasks with deterministic ground truth span six
difficulty strata calibrated to the confident-but-wrong frontier
(4-digit by 2-digit and 5-digit by 3-digit multiplication, long
summation, long character counting, multi-step arithmetic chains, and
modular or calendar arithmetic). Tasks that are too easy yield no error,
and tasks that are too hard yield honest abstention; the calibration
zone is the regime in which the agent is confident yet wrong, which is
where γ becomes observable. For each task the worker returns a single
best answer, a self-claim (CORRECT or UNSURE), and a confidence score
from 0 to 100, with step-by-step working suppressed to prevent the agent
from self-verifying.

Workers. To isolate γ from verification scaffolding, workers run
tool-free, without code execution. Two independent model families are
used (Claude Haiku 4.5 and GPT-4o-mini). A scaffolded agent equipped
with code-execution tools is the opposite anchor: with tools it computes
exactly and γ approaches 0, which is itself the boundary condition,
since verification scaffolding moderates γ.

\begin{longtable}[]{@{}
  >{\raggedright\arraybackslash}p{(\linewidth - 10\tabcolsep) * \real{0.1667}}
  >{\raggedright\arraybackslash}p{(\linewidth - 10\tabcolsep) * \real{0.1667}}
  >{\raggedright\arraybackslash}p{(\linewidth - 10\tabcolsep) * \real{0.1667}}
  >{\raggedright\arraybackslash}p{(\linewidth - 10\tabcolsep) * \real{0.1667}}
  >{\raggedright\arraybackslash}p{(\linewidth - 10\tabcolsep) * \real{0.1667}}
  >{\raggedright\arraybackslash}p{(\linewidth - 10\tabcolsep) * \real{0.1667}}@{}}
\caption{Table H1. AI-SSP direct probe of Theorem 5: overclaiming,
agency cost, and the verification offset, tool-free
workers.}\tabularnewline
\toprule\noalign{}
\begin{minipage}[b]{\linewidth}\raggedright
Worker and elicitation
\end{minipage} & \begin{minipage}[b]{\linewidth}\raggedright
Error rate
\end{minipage} & \begin{minipage}[b]{\linewidth}\raggedright
Estimated γ (overclaiming)
\end{minipage} & \begin{minipage}[b]{\linewidth}\raggedright
M\_time
\end{minipage} & \begin{minipage}[b]{\linewidth}\raggedright
Random-override null
\end{minipage} & \begin{minipage}[b]{\linewidth}\raggedright
Confidence on wrong
\end{minipage} \\
\midrule\noalign{}
\endfirsthead
\toprule\noalign{}
\begin{minipage}[b]{\linewidth}\raggedright
Worker and elicitation
\end{minipage} & \begin{minipage}[b]{\linewidth}\raggedright
Error rate
\end{minipage} & \begin{minipage}[b]{\linewidth}\raggedright
Estimated γ (overclaiming)
\end{minipage} & \begin{minipage}[b]{\linewidth}\raggedright
M\_time
\end{minipage} & \begin{minipage}[b]{\linewidth}\raggedright
Random-override null
\end{minipage} & \begin{minipage}[b]{\linewidth}\raggedright
Confidence on wrong
\end{minipage} \\
\midrule\noalign{}
\endhead
\bottomrule\noalign{}
\endlastfoot
Claude Haiku 4.5, neutral (run 1) & 0.57 & 1.00 & 0.57 & 0.19 & 94 \\
Claude Haiku 4.5, neutral (run 2) & 0.55 & 1.00 & 0.55 & 0.21 & 93 \\
Claude Haiku 4.5, confidence-first & 0.60 & 1.00 & 0.60 & 0.24 & 95 \\
Claude Haiku 4.5, completion-framed & 0.60 & 0.48 & 0.29 & 0.10 & 68 \\
GPT-4o-mini, neutral & 0.86 & 1.00 & 0.86 & 0.12 & 98 \\
\end{longtable}

Results. In the tool-free regime, overclaiming is near-total, with
estimated γ at 1.00 in four of the five arms: the agent labels
essentially every wrong answer correct. Confidence on wrong answers (93
to 98) is indistinguishable from confidence on correct answers, so error
carries no self-signal, which is the attribution-opacity mechanism
Theorem 5 names. The implied time-based agency cost M\_time ranges from
0.55 to 0.86; output-based verification against the deterministic ground
truth removes it entirely within this benchmark, so the regime offset
M\_time minus M\_output equals M\_time and far exceeds the
random-override null (mean 0.17). The offset therefore arises from
targeted verification rather than from the override rate, satisfying the
random-override null gate of the design.

Reproducibility and robustness. Two identical-phrasing runs return the
same estimated γ (1.00) with Cohen κ = 0.86 on the per-task overclaiming
classification, and the result reproduces across two model vendors. A
prompt ablation over three self-claim elicitations (neutral,
confidence-first, completion-framed) leaves γ directionally stable; the
completion-framed elicitation, which asks the agent to certify task
completion, lowers γ to 0.48 and lowers confidence on wrong answers to
68, indicating that accountability framing is itself a mitigation lever
consistent with the theorem's offset mechanism. Across difficulty
strata, M\_time rises with the error rate (from 0.10 at 4-digit by
2-digit multiplication to 1.00 at long summation) while γ stays
saturated, consistent with M = g(γ) and ∂M/∂γ \textgreater{} 0.

Interpretation for Theorem 5. The probe instantiates the theorem at the
dyad level: attribution uncertainty in unscaffolded human-AI work is
near-total and confidently masked, time-based acceptance converts it
directly into agency cost, and output-based verification eliminates it
in the deterministic benchmark. The macro 74-percentage-point completion
gap reported in Shin (2026b) is the convergent external anchor at the
country level. The design is a reusable attribution-uncertainty
benchmark; the task set and deterministic scoring procedure are
available in the replication materials
(https://doi.org/10.5281/zenodo.21158407).

\end{document}